\documentstyle[amssymb,aps,12pt,epsfig]{revtex}
\newcommand{\be}{\begin{equation}}
\newcommand{\ee}{\end{equation}}
\newcommand{\bea}{\begin{eqnarray}}
\newcommand{\eea}{\end{eqnarray}}
\newcommand{\ba}{\begin{array}}
\newcommand{\ea}{\end{array}}

\newcommand{\bb}{\bibitem}

\begin{document}
\draft
\tightenlines

\title{\bf Critical behavior of generic competing systems}
\author{Marcelo M. Leite\footnote{e-mail:mleite@df.ufpe.br}}
\address{{\it Laborat\'orio de F\'\i sica Te\'orica e Computacional, Departamento de F\'\i sica,\\ Universidade Federal de Pernambuco,\\
50670-901, Recife, PE, Brazil}}
\maketitle

\vspace{0.2cm}
\begin{abstract}
{\it Generic higher character Lifshitz critical behaviors are
described using field theory and $\epsilon_{L}$-expansion
renormalization group methods. These critical behaviors 
describe systems with arbitrary competing interactions. 
We derive the scaling relations and the critical exponents at the
two-loop level for anisotropic and isotropic points of arbitrary
higher character. The framework is illustrated for
the $N$-vector $\phi^{4}$ model describing a $d$-dimensional system.
The anisotropic behaviors are derived in terms of many
independent renormalization group transformations, each one
characterized by independent correlation lengths. The isotropic
behaviors can be understood using only one renormalization group
transformation. Feynman diagrams are solved for the anisotropic behaviors 
using a new dimensional regularization associated to a generalized 
orthogonal approximation. The isotropic diagrams are treated using this 
approximation as well as with a new exact technique to compute the 
integrals. The entire procedure leads to the analytical 
solution of generic loop order integrals with arbitrary external momenta. 
The property of universality class reduction is also satisfied when the 
competing interactions are turned off. We show how the results presented 
here reduce to the usual $m$-fold Lifshitz critical behaviors for both 
isotropic and anisotropic criticalities.}
\end{abstract}

\vspace{1cm}

\newpage
\section{Introduction}
Field theoretic renormalization group techniques are invaluable tools for 
studying usual critical phenomena as well as the critical behavior 
associated to the physics of systems presenting arbitrary short range 
competing interactions. The universality classes of the ordinary 
critical behavior are characterized by the space dimension of the system 
$d$ and the number of components of the (field) order parameter 
$N$ \cite{Am,BLZ}. Competing systems, on the other hand, possess 
different types of space directions known as competition axes. 

The simplest 
type of competition directions can be most easily visualized using the 
terminology of magnetic systems via a generalized Ising model. One permits 
exchange ferromagnetic couplings between nearest neighbors $(J_{1}>0)$ 
{\it and} antiferromagnetic interactions between second neighbors 
$(J_{2}<0)$ occurring along $m_{2}$ dimensions. Whenever $m_{2}<d$ the 
system presents a (usual) second character anisotropic Lifshitz critical 
behavior whose universality classes are characterized by $(N,d,m_{2})$, 
whereas the isotropic behavior characterized by $d=m_{2}$ close to 8 was 
formerly described at the same time \cite{Ho-Lu-Sh}. The 
phenomenological model corresponding to a uniaxial anisotropy ($m_{2}=1$) 
in a cubic lattice is known as ANNNI model \cite{Selke}. In the critical 
region, this sort of system is characterized by a disordered, a uniformly 
ordered {\it and} a modulated ordered phase which meet in a uniaxial Lifshitz 
multicritical point, where the ratio $\frac{J_{2}}{J_{1}}$ is fixed at the 
corresponding Lifshitz temperature $T_{L}$. High-precision numerical Monte 
Carlo simulations were carried out for the critical exponents of this model 
\cite{Pleim-Hen} and checked using two different two-loop analytical 
calculations  \cite {AL1,Leite1,Leite2}. From the 
renormalization group perspective there is an important difference between 
these Lifshitz critical behaviors. The anisotropic behaviors have two 
independent correlation lengths, $\xi_{L2}$ perpendicular to the competing 
axes as well as $\xi_{L4}$ parallel to the $m_{2}$ competing axes. The 
isotropic behavior has only one correlation length $\xi_{L4}$.\footnote{For an 
alternative field-theoretic approach for $m$-axial Lifshitz points, see 
\cite{Mergulho,DS}.}   

If we go on to include ferromagnetic couplings up to the third neighbors 
$(J_{3}>0)$ along a single axis, the system will present a uniaxial third 
character Lifshitz point whenever $\frac{J_{2}}{J_{1}}$ and 
$\frac{J_{3}}{J_{1}}$ take certain fixed values at the corresponding 
Lifshitz temperature \cite{Se1}. When this sort of competition takes place 
along $m_{3}$ spatial directions, the system presents a $m_{3}$-fold third 
character Lifshitz point. On the other hand, if simultaneous and independent 
competing interactions take place between second neighbors along $m_{2}$ 
space directions and third neighbors along $m_{3}$ space dimensions, the 
system presents a {\it generic} third character $m_{3}$-fold Lifshitz 
critical behavior. The generic third character universality classes are 
defined by $(N,d,m_{2},m_{3})$, thus describing a wider sort of critical 
behavior when compared with the thrid character universality classes 
$(N,d,m_{3})$. 

This idea can be extended in order to define the 
$m_{L}$-fold Lifshitz point of character $L$, when further alternate 
couplings are permitted up to the $L$th neighbors along $m_{L}$ directions, 
provided the ratios $\frac{J_{L}}{J_{1}}$, $\frac{J_{L-1}}{J_{1}}$ ,..., 
$\frac{J_{2}}{J_{1}}$ take especial values at the associated Lifshitz 
temperature \cite{Se2,NCS,NTCS}. However, the most general anisotropic 
situation is to consider {\it several} types of competing axes occurring 
simultaneously in the system such that second neighbors interact along 
$m_{2}$ space directions, $m_{3}$ directions couple third neighbors, etc., 
up to the interactions of $L$ neighbors along $m_{L}$ dimensions, with all 
competing axes perpendicular among each other. In that case, the 
corresponding critical behavior is called a generic $Lth$ character Lifshitz 
critical behavior \cite{Leite4}. 

In this work we shall undertake an exploration of the field theoretical 
renormalization group structure of the most general competing system 
using $\epsilon_{L}$-expansion techniques for anisotropic and isotropic 
higher character Lifshitz critical behaviors. A rather brief description of 
this structure was set forth in a previous letter \cite{Leite4} where it 
was first described; here we shall present the details and extend the 
formalism in order to incorporate the exact  two-loop calculation for 
arbitrary isotropic higher character criticalities. Renormalization 
group (RG) arguments are constructed in order 
to find out the scaling relations for the anisotropic as well as the 
isotropic critical behaviors. The discussion parallels that for 
the usual second character Lifshitz points \cite{Leite2}. 

The arbitrary competing exchange coupling Ising model (CECI model) is the 
lattice model associated to this new 
critical behavior. It has a 
general renormalization group structure which contains many independent 
length scales, and its construction can be utilized for both 
anisotropic and isotropic cases. In the anisotropic cases, the system 
has only nearest neighbor interactions along $(d-m_{2}-...-m_{L})$, second 
neighbors competing interactions along $m_{2}$ dimensions, and so on, up to 
$L$th neighbors competing interactions along $m_{L}$ spatial directions. The 
distinct competing axes originate several types of independent correlation 
lengths, namely $\xi_{1}$ for directions parallel to the 
$(d-m_{2}-...-m_{L})$-dimensional noncompeting subspace, $\xi_{2}$ for 
directions parallel to the $m_{2}$-dimensional competing subspace, etc., 
and $\xi_{L}$ characterizing the $m_{L}$-dimensional subspace. The simplest 
representative of the CECI model is better understood with 
the help of Fig.\ref{figCECI1}, which is the particular case 
$m_{2}=m_{3}=1$, $m_{4}=...=m_{L}=0$. There are two competing 
subspaces and three types of correlation lengths which define three 
independent renormalization group transformations. It defines a particular 
generic third character anisotropic Lifshitz critical behavior.

In the phase diagram of the ANNNI model, the parameters which are varied 
are the temperature $T$ and $p=\frac{J_{2}}{J_{1}}$ which take a 
particular value at the uniaxial second character Lifshitz multicritical 
point as depicted in Fig.\ref{figCECI2}. It is a particular case of the 
CECI model whenever $m_{2}=1$, with {$m_{3}=...=m_{L}=0$. Although the 
ANNNI model has applications in several real physical systems 
(see for example \cite{Selke}), the prototype of second character Lifshitz 
points in magnetic materials is manganese phosphide ($MnP$). Experimental as 
well as theoretical investigations have determined that $MnP$ presents a pure 
uniaxial Lifshitz point ($m_{2}=1, d=3, N=1$)\cite{Becerra,Yokoi}}. 

When adding further competing interactions to the ANNNI model, the number 
of parameters in the phase diagram increases \cite{Se2}. For instance, in 
the phase diagram of the model including uniaxial competing interactions 
up to third neighbors the parameters to be varied are the temperature $T$, 
$p_{1}=\frac{J_{2}}{J_{1}}$ and $p_{2}=\frac{J_{3}}{J_{1}}$. One can locate 
the third character Lifshitz point by looking at the projection of the phase 
diagram in the plane $(p_{1}, p_{2})$, as was demonstrated using numerical 
means \cite{Se2}. In the example of the CECI model displayed in 
Fig.\ref{figCECI1}, let the competing exchange be completely independent 
along the different competing axes. In that case, the phase diagram can be 
described by $T$, $p_{z} = \frac{J_{2 z}}{J_{1 z}}$, 
$p_{1y}=\frac{J_{2y}}{J_{1y}}$, $p_{2y}=\frac{J_{3y}}{J_{1y}}$. A useful 
two-dimensional representation can be obtained by separating the phase 
diagrams in two parts. The diagram $(T,p_{z})$ characterizing the second 
character behavior (with $p_{1y},p_{2y}$ fixed) and the diagram 
$(p_{1y}, p_{2y})$ (with $T,p_{z}$ fixed) corresponding to the 
third character behavior can be ploted independently. The superposition 
of the two diagrams at the generic third character Lifshitz point is 
indicated in Fig.\ref{figCECI3}. As a consequence, there is a uniformly 
ordered phase and two modulated phases 
called $Helical_{2}$ and $Helical_{3}$ in Fig.\ref{figCECI3} which meet at 
the uniaxial generic third character anisotropic Lifshitz point. Now there 
are two first order lines separating the ferromagnetic-$Helical_{2}$ 
and $Helical_{2}-Helical_{3}$ phases. Analogously, when there are arbitrary 
independent types of competing axes, we can consider several independent 
phase diagrams and each two-dimensional projection of them. The superposition 
of them in one two-dimensional diagram gives origin to a situation that 
resembles that illustrated in Fig.\ref{figCECI3} . Instead, there are 
{\it several} modulated phases and one uniformly ordered phase which meet at 
the generich $Lth$ character anisotropic Lifshitz point. Each competing 
subspace has its own characteristic modulated phase along with its own 
independent correlation length. Although these objects go critical 
simultaneously at the Lifshitz critical temperature, they define independent 
renormalization group transformations in each different subspace. 
 
Therefore, we find multiscale scaling laws as a consequence of 
this renormalization group flow independence in parameter space. 
This implies that we find several independent coupling constants, 
each one depending on a definite momenta scale characterizing the particular 
competition axes under consideration. Nevertheless, all coupling constants 
flow to the same fixed point. The universality classes of this system are 
characterized by the parameters $(N,d,m_{2},..., m_{L})$, therefore 
generalizing the usual Lifshitz behavior. It is important to mention 
that when we turn off all the competing interactions between third and 
more distant neighbors, the universality classes of the generic higher 
character Lifshitz point turn out to reduce to that associated to the 
second character behavior $(N,d,m_{2})$. Notice that these anisotropic 
behaviors generalize previous lattice models with competing interactions 
\cite{Fra-Hen} as it includes all types of competing axes. The isotropic 
critical behaviors $d=m_{n}$ have a distinct feature in which there is only 
one type of correlation length $\xi_{4n}$. Their universality classes are 
characterized by $(N,d,n)$ where $n$ is the number of neighbors coupled 
through competing interactions.

In addition, we compute the critical exponents at least at 
$O(\epsilon_{L}^{2})$ using dimensional regularization to resolving 
the Feynman diagrams and normalization conditions (and) or minimal 
subtraction as the renormalization procedures. The computation is realized 
in momentum space. For the anisotropic cases, the Feynman diagrams are 
performed with an approximation which is the most general one consistent 
with the homogeneity of these integrals in the external momenta scales. 
The isotropic situations are treated using this approximation as well, but 
we also present the exact calculation at the same loop order and make a 
comparison with the above mentioned approximation.

We present the functional integral representation of the model in terms 
of a $\lambda\phi^{4}$ setting and define the normalization conditions for 
this higher character Lifshitz critical behavior in section II. 
We show that many sets of normalization conditions, each 
one corresponding to a specific type of competition axes, are convenient to 
have a satisfactory description of the problem in its maximal generality.

In Section III we present the renormalization group analysis for the 
anisotropic critical behaviors. We construct the several renormalization 
functions appropriate to each competing subspace and study their flow with the 
various renormalization group transformations. We find the proper 
scaling relations to each competition subspace. 

Section IV discusses the renormalization group treatment for the various 
isotropic behaviors. We obtain the scaling relations and show that they 
reduce to the usual $\lambda\phi^{4}$ case when the interactions beyond 
the first neighbors are switched off.

Section V is an exposition of the calculation of the critical 
exponents for the anisotropic cases using the scaling relations obtained 
in section III. We describe two different ways of calculating the critical 
exponents either in normalization conditions or in minimal subtraction. We 
discuss the limit $L \rightarrow \infty$ and some of its implications. We 
point out the analogy of the Lifshitz critical region with an effective 
field theory which arises from a recent cosmological model including 
modifications of gravity in the long distance limit \cite{AH}.  

Sections VI and VII are an in-depth analysis of the critical exponents 
for the isotropic cases. The critical exponents for the isotropic cases 
are calculated using the orthogonal approximation and the scaling relations 
in section VI. The exact calculation of the critical exponents and the 
comparison with those obtained from the orthogonal approximation are 
carried out in section VII. 

The particular case corresponding to the second character isotropic 
Lifshitz critical behavior is discussed explicitly in Section VIII. The 
resulting critical exponents are shown to generalize those obtained 
previously \cite{Ho-Lu-Sh}. We discuss their relationship with those 
coming from the orthogonal approximation.    

Section IX concludes this paper, a discussion of the ideas is 
summarized and some possible applications will be proposed. We calculate 
the Feynman integrals in the appendices. We describe in detail the 
generalized orthogonal approximation for the calculation of higher loop 
integrals of the anisotropic behaviors in Appendix A. It will 
be shown there that one can obtain the answer in a simple analytical form 
for arbitrary external momenta scales. The property of homogeneity of these 
integrals along arbitrary external momenta scales is preserved. Then, we use 
the same approximation to compute diagrams for the isotropic behaviors in 
Appendix B. In addition, we perform the exact calculation for arbitrary 
isotropic cases in Appendix C. We also analyse the simple particular case 
associated to the usual second character isotropic behavior.

\section{Field theory and normalization conditions 
for higher character Lifshitz points}
The field theoretical representation can be expressed in terms of 
a modified $\lambda\phi^{4}$ field theory presenting arbitrary higher 
derivative terms due to the effect of competition along the different 
kinds of competing axes. The type of competing axes are defined by the 
number of neighbors that interact among each other via exchange competing 
couplings. Let $m_{n}$ be the number of space directions whose competing 
interactions extend to the $n$th neighbor. Thus, the $m_{n}$-dimensional 
competition subspace will be represented in the Lagrangian with even powers 
(up to the $2n$th) of the gradient acting on the order parameter 
scalar field. Thus, the effect of the competition resides in the higher 
derivatives of the field. 

The corresponding bare Lagrangian density can 
be written as \cite{Leite4} :
\begin{eqnarray}
L &=& \frac{1}{2}
|\bigtriangledown_{(d- \sum_{n=2}^{L} m_{n})} \phi_0\,|^{2} +
\sum_{n=2}^{L} \frac{\sigma_{n}}{2}
|\bigtriangledown_{m_{n}}^{n} \phi_0\,|^{2} \\ \nonumber
&& + \sum_{n=2}^{L} \delta_{0n}  \frac{1}{2}
|\bigtriangledown_{m_{n}} \phi_0\,|^{2}
+ \sum_{n=3}^{L-1} \sum_{n'=2}^{n-1}\frac{1}{2} \tau_{nn'}
|\bigtriangledown_{m_{n}}^{n'} \phi_0\,|^{2} \\ \nonumber
&&+ \frac{1}{2} t_{0}\phi_0^{2} + \frac{1}{4!}\lambda_0\phi_0^{4} .
\end{eqnarray}
At the Lifshitz point, the fixed ratios among the
exchange couplings explained above translate into this field-theoretic
version though the conditions $\delta_{0n} = \tau_{n n'} =0$. All even 
momentum powers up to $2L$ become relevant in the free propagator 
\cite{Wilson}. This condition simplifies the treatment of the system 
since it allows the decoupling of the several competing subspaces of 
Feynman integrals in momentum space. It indeed makes possible to solve 
these diagrams to any desired order in a perturbative approach. Within 
the loop order chosen, we can set a perturbative regime with maximal 
generality as far as critical behavior of competing systems are concerned. 
So we need the small loop parameter, which is intimately connected to the 
critical dimension of the theory. \par
It is instructive to find the critical dimension of this field theory 
through the use of the Ginzburg criterion \cite{Lev,Ginz,Amit1}. From the 
perspective of magnetic systems in the above Lagrangian density, 
$t_{0}= t_{0L} + (T - T_{L})$ measures the 
temperature difference from the critical temperature $T_{L}$. In the 
mean-field approximation the inverse susceptibility is proportional to 
$(T - T_{L})$, but has no longer this behavior when fluctuations get bigger 
due to the closenesss of the critical point and the mean-field argument 
breaks down. This immediately leads to the critical dimension 
$d_{c} = 4 + \sum_{n=2}^{L} \frac{(n-1)}{n} m_{n}$. 
Above this critical dimension the mean-field behavior dominates the system. 
Consequently, the small parameter for a consistent perturbative expansion is
$\epsilon_{L} = 4 + \sum_{n=2}^{L} \frac{(n-1)}{n} m_{n} - d$. The same 
type of argument can be constructed for the isotropic behaviors. When 
$d=m_{n}$, the critical dimension is $d_{c}= 4n -\epsilon_{4n}$.

The renormalized theory can be defined starting from the bare 
Lagrangian (1). The approach we shall follow here is completely 
analogous to that exposed in \cite{Leite2} (see also \cite{Am}) and the 
reader is invited to consult that reference in order to be familiarized 
with the notation employed. The renormalization functions are determined 
in terms of the renormalized reduced
temperature and order parameter (magnetization in the context of magnetic
systems) as $t = Z_{\phi^2}^{-1} t_0$, $M = Z_{\phi}^{\frac{-1}{2}} \phi_0$
and depend on Feynman graphs. When the theory is renormalized at the
critical temperature $(t=0)$, a nonvanishing external momenta must be 
used to define the renormalized theory. Consequently, the renormalization
constants at the critical temperatute $T_{L}$ depend on the external
momenta scales involved in the renormalization algorithm.

Let us analyse the anisotropic behaviors. The Feynman integrals depend on
various external momenta scales, namely that characterizing the 
$(d-m_{2}-...-m_{L})$-dimensional noncompeting subspace, a momentum scale 
associated to the $m_{2}$ space directions, etc., up to the momentum scale 
corresponding to the $m_{L}$ competing axes. Thus, it is appropriate to 
define $L$ sets of normalization conditions in order to compute the critical 
exponents associated to correlations either perpendicular to or along the 
several types of competition axes. We define $\kappa_{1}$ to be the external 
momenta scale asssociated to the $(d-m_{2}-...-m_{L})$-dimensional 
noncompeting directions. If we define the noncompeting directions to be along 
the $m_{1}$-dimensional subspace, where $m_{1}=d-m_{2}-...-m_{L}$, we can 
unify the language by stating that $\kappa_{n}$ is the typical external 
momenta scale characterizing the $m_{n}$ competing axes ($n=1,...,L$).

We now turn our attention to the definition of the symmetry points (SP). 
In case we wish to evaluate the critical 
exponents along the $j$th type of competing axes, we set $\kappa_{n}=0$ 
for $n \neq j$ keeping, however, $\kappa_{j} \neq 0$. The proper 
normalization conditions to evaluating exponents along the 
various competition axes can be defined as follows. If 
$k'_{i(n)}$ is the external momenta along the
competition axes associated to a generic 1PI vertex part, the external
momenta along the $n$th type of competing directions are chosen as
$ k'_{i(n)}. k'_{j(n)} = \frac{\kappa_{n}^{2}}{4} (4\delta_{ij} - 1)$.
This implies that $(k'_{i(n)} + k'_{j(n)})^{2} = \kappa_{n}^{2}$ for 
$i \neq j$. The momentum scale of the two-point function is defined by 
$k_{(n)}^{' 2} = \kappa_{n}^{2} = 1$. The set of renormalized 1PI 
vertex parts is given by:

\begin{mathletters}
\begin{eqnarray}
&& \Gamma_{R(n)}^{(2)}(0,g_{n}) = 0, \\
&& \frac{\partial\Gamma_{R(n)}^{(2)}(k'_{(n)}, g_{n})}
{\partial k_{(n)}^{' 2n}}|_{k_{(n)}^{' 2n}=\kappa_{n}^{2n}}
= 1, \\
&& \Gamma_{R(n)}^{(4)}(k'_{i(n)}, g_{n})|_{SP_{n}} = g_{n}  , \\
&& \Gamma_{R(n)}^{(2,1)}(k'_{1(n)}, k'_{2(n)}, k', g_{n})|_{\bar{SP_{n}}} = 1 , \\
&& \Gamma_{R(n)}^{(0,2)}(k'_{(n)}, g_{n})|_{k_{(n)}^{' 2n}}=\kappa_{n}^{2n}= 0 .
\end{eqnarray}
\end{mathletters}

These $L$ systems of normalization conditions
seem to provide  $L$ renormalized coupling constants. The origin of this 
overcounting is a consequence of the $L$ independent flow in the 
renormalization momenta scales $\kappa_{1}$,...,$\kappa_{L}$. The analysis 
works with $L$ coupling constants, namely 
$g_{n} = u_{n} (\kappa_{n}^{2n})^{\frac{\epsilon_{L}}{2}}$
(and $ \lambda_{n} =  u_{0n} (\kappa_{n}^{2n})^{\frac{\epsilon_{L}}{2}}$)
characterizing the flow along the momenta components parallel to each
$m_{n}$-dimensional competing subspace. This is really a disguise since 
the situation becomes simpler at the fixed point: the many couplings will 
flow to the same fixed point, at two-loop level, giving a clear indication 
that this property is kept in higher-loop calculations. The explicit 
demonstration of this fact will be tackled in Sec. V.\footnote{Let us mention briefly another type of anisotropic 
critical behavior whose theoretical existence is granted from the CECI 
model structure. In magnetic systems, the absence of the ferromagnetic 
phase leads to a structure where several modulated phases are mixed among 
each other. This situation is transliterated in the condition 
$d=m_{2}+...+m_{L}$. In that case, the normalization conditions can be 
defined without making reference to the noncompeting subspace. In general 
the critical dimension will increase and the lowest character modulated 
phase will play the role of the former ferromagnetic ordered phase. We 
shall not delve any further into this situation in this paper, but leave 
the analysis for future work.} 

The normalization conditions for the isotropic case ($m_{n}=d$ near $4n$) 
can be defined in a close analogy to its second character isotropic 
particular case \cite{Leite2}. If $k'_{i}$ is the external momenta along 
the $m_{n}$ competition axes, the external momenta along the
$4n$ directions are chosen as $ k'_{i}. k'_{j} = \frac{\kappa_{n}^{2n}}{4}
(4\delta_{ij} - 1)$. This implies that
$(k'_{i} + k'_{j})^{2} = \kappa_{n}^{2}$ for $i \neq j$.
The momentum scale of the two-point function is fixed through
$k'^{2n} = \kappa_{n}^{2n} = 1$. Then we have the same normalization 
conditions Eqs.(2), but now there is solely one type of external momenta 
scale. The others are absent in this situation as an effect of the Lifshitz 
condition $\delta_{0n}=\tau_{nn'}=0$. 

We can express all the renormalization functions and bare coupling constants
in terms of the dimensionless couplings in a unified perspective for both 
anisotropic and isotropic behaviors. The subscript  $n = 1,2,3,...,L$ 
labels the different external momenta scales belonging to the general 
Lifshitz critical behavior, as defined above for the anisotropic and 
isotropic cases. Expansion of the dimensionless bare coupling
constants $u_{o n}$ and the normalization constants $Z_{\phi (n)}$,
$\bar{Z}_{\phi^{2} (n)} = Z_{\phi (n)} Z_{\phi^{2} (n)}$ as functions of
the dimensionless renormalized couplings $u_{n}$ up to two-loop order as
\begin{mathletters}
\begin{eqnarray}
&& u_{o n} = u_{n} (1 + a_{1 n} u_{n} + a_{2 n} u_{n}^{2}) ,\\
&& Z_{\phi (n)} = 1 + b_{2 n} u_{n}^{2} + b_{3 n} u_{n}^{3} ,\\
&& \bar{Z}_{\phi^{2} (n)} = 1 + c_{1 n} u_{n} + c_{2 n} u_{n}^{2} ,
\end{eqnarray}
\end{mathletters}
along with dimensional regularization will be sufficient to find out all 
critical exponents.

\section{Scaling theory for the anisotropic cases}

The anisotropic behaviors are characterized by correlation lengths 
$\xi_{1},..., \xi_{L}$. When considered independently they define 
independent renormalization group transformations along the several 
competing directions. In momentum space, they induce independent flows 
in each external momenta scale $\kappa_{1},..., \kappa_{L}$. 

In order to define the renormalized vertex parts we consider a set of 
cutoffs $\Lambda_{j}$ ($j=1,...,L$), each of them characterizing a 
different competing subspace. As functions of the bare vertices and 
normalization constants they read
\begin{eqnarray}
\Gamma_{R(n)}^{(N,M)} (p_{i (n)}, Q_{i(n)}, g_{n}, \kappa_{n})
&=& Z_{\phi (n)}^{\frac{N}{2}} Z_{\phi^{2} (n)}^{M}
(\Gamma^{(N,L)} (p_{i (n)}, Q_{i (n)}, \lambda_{n}, \Lambda_{n})\\ \nonumber
&& - \delta_{N,0} \delta_{L,2}
\Gamma^{(0,2)}_{(n)} (Q_{(n)}, Q_{(n)}, \lambda_{n}, \Lambda_{n})|_{Q^{2}_{(n)} = \kappa_{n}^{2}})
\end{eqnarray}
where $p_{i (n)}$ ($i=1,...,N$) are the external momenta associated to
the vertex functions $\Gamma_{R(n)}^{(N,L)}$ with $N$ external legs and
$Q_{i (n)}$ ($i=1,...,M$) are the external momenta associated to the
$M$ insertions of $\phi^{2}$ operators. From the last section, $u_{0 n}$, 
$Z_{\phi (n)}$ and $Z_{\phi^{2} (n)}$ are represented as power series in 
$u_{n}$. In order to write the renormalization group equations in terms of 
dimensionless bare and renormalized coupling constants, we shall discuss the 
central idea which underlies the subsequent scaling theory.

Consider the volume element in momentum space for calculating an arbitrary
Feynman integral. It is given by $d^{d - \sum_{i=2}^{L} m_{i}}q 
\Pi_{i=2}^{L} d^{m_{i}}k_{(i)}$. Recall that $\vec{q}$ represents a 
$(d- \sum_{i=2}^{L}m_{i})$-dimensional vector perpendicular to the competing
axes and $\vec{k_{(i)}}$ denotes an $m_{i}$-dimensional vector along the 
ith competing subspace, respectively. The Lifshitz condition 
$\delta_{0n}= \tau_{nn'}=0$ suppresses
the quadratic part of the momentum along the $m_{2}$ competition axes, 
the quadratic and quartic part of the momentum along the $m_{3}$ competing 
directions, and so on, such that the $m_{L}$ competing subspace is 
represented by a $2L$th power of momentum in the inverse free 
critical $(t=0)$ propagator, i.e., 
$G_{0}^{(2) -1}(q,k) = q^{2} + \sum_{n=2}^{L} \sigma_{n}(k_{(n)}^{2})^{n} $. In
order to be dimensionally consistent, the canonical dimension in mass 
units of the various terms in the propagator should be equal. 

Our normalization conditions give us a hint that we can get rid of the 
$\sigma_{n}$ parameters provided we make simultaneously dimensional 
redefinitions of the momenta components  along each type of competition 
subspace in a complete analogy to the second character case. Let 
$[\vec{q}] = M$ be the mass dimension of the quadratic
momenta. Since all momentum terms in the propagator should have the same 
canonical dimension, this requires that $[\vec{k_{(i)}}] = M^{\frac{1}{i}}$. 
These simultaneous dimensional redefinitions of the momenta along the 
competing axes is only possible due to the Lifshitz condition. The volume
element in momentum space $d^{d - \sum_{i=2}^{L} m_{i}}q 
\Pi_{i=2}^{L} d^{m_{i}}k_{(i)}$ has
mass dimension  $M^{d - \sum_{i=2}^{L}\frac{(i-1)m_{i}}{i}}$. The dimension 
of the field can be found from the requirement that the volume integral of 
the Lagrangian density (1) is dimensionless in mass units. In other words, 
one obtains 
$[\phi] = M^{\frac{1}{2}(d-\sum_{i=2}^{L}\frac{(i-1)m_{i}}{i})-1}$. In 
momentum space the one particle irreducible (1PI) vertex functions have 
canonical dimension 
$[\Gamma^{(N)}(k_{i})]= M^{N + (d - \sum_{i=2}^{L}\frac{(i-1)m_{i}}{i}) -
\frac{N (d -\sum_{i=2}^{L}\frac{(i-1)m_{i}}{i})}{2}}$.

Let us describe the theory in
terms of dimensionless parameters. As the
coupling constants are associated to $\Gamma^{(4)}$, we can write
$g_{n} = u_{n} (\kappa_{n}^{2 n})^{\frac{\epsilon_{L}}{2}}$,
and $ \lambda_{n} =  u_{0 n}
(\kappa_{\tau}^{2 n})^{\frac{\epsilon_{L}}{2}}$,
where $\epsilon_{L} = 4 + \sum_{n=2}^{L} \frac{(n-1)}{n} m_{n} - d$. 
Expressed in terms of these dimensionless coupling constants, the 
renormalization group equation can be cast in the form:
\begin{equation}
(\kappa_{n} \frac{\partial}{\partial \kappa_{n}} +
\beta_{n}\frac{\partial}{\partial u_{n}}
- \frac{1}{2} N \gamma_{\phi (n)}(u_{n}) + L \gamma_{\phi^{2} (n)}(u_{n}))
\Gamma_{R(n)}^{(N,L)} = \delta_{N,0} \delta_{L,2} (\kappa_{n}^{-2 n})^\frac{\epsilon_{L}}{2} B_{n}(u_{n}) .
\end{equation}
The functions
\begin{mathletters}
\begin{eqnarray}
&& \beta_{n} = (\kappa_{n}\frac{\partial u_{n}}{\partial \kappa_{n}}), \\
&& \gamma_{\phi (n)}(u_{n})  = \beta_{n}
\frac{\partial ln Z_{\phi (n)}}{\partial u_{n}}\\
&& \gamma_{\phi^{2} (n)}(u_{n}) = - \beta_{n}
\frac{\partial ln Z_{\phi^{2} (n)}}{\partial u_{n}}
\end{eqnarray}
\end{mathletters}
are calculated at fixed bare coupling $\lambda_{n}$. The 
$\beta_{n}$-functions can be rewitten in terms of dimensionless quantities as

\begin{equation}
\beta_{n} = - n \epsilon_{L}(\frac{\partial ln u_{0 n}}{\partial u_{n}})^{-1}.
\end{equation}
Note that the beta function corresponding to the flow in $\kappa_{n}$ has a
factor of $n$ compared to that associated to the flow in $\kappa_{1}$. 

For the anisotropic case, the multi-parameters group of invariance is manifest 
in the solution of the renormalization group equation, which is given by
\begin{equation}
\Gamma_{R (n)}^{(N)} (k_{i (n)}, u_{n}, \kappa_{n}) =
exp[-\frac{N}{2} \int_{1}^{\rho_{n}} \gamma_{\phi (n)}(u_{n}(\rho_{n}))
\frac{{d x_{n}}}{x_{n}}]
\;\Gamma_{R (n)}^{(N)} (k_{i (n)}, u_{n}(\rho_{n}),
\kappa_{n} \rho_{n}).
\end{equation}

From the above analysis, the dimensional redefinitions of the momenta
along the distinct competing axes result in an effective space dimension 
for the anisotropic case, namely, 
$(d - \sum_{n=2}^{L} \frac{(n-1)}{n} m_{n})$ . We 
discover the following behavior for the 1PI vertex parts $\Gamma_{R (n)}^{(N)}$
under flows in the external momenta:
\begin{eqnarray}
\Gamma_{R (n)}^{(N)} (\rho_{n} k_{i (n)}, u_{n}, \kappa_{n})&=&
\rho_{n}^{n (N + (d- \sum_{n=2}^{L} \frac{(n-1)}{n} m_{n}) - 
\frac{N(d-\sum_{n=2}^{L} \frac{(n-1)}{n} m_{n})}{2})}\\
&& exp[-\frac{N}{2} \int_{1}^{\rho_{n}} \gamma_{\phi (n)}(u_{n}(x_{n}))
\frac{{d x_{n}}}{x_{n}}]\nonumber\\
&&\Gamma_{R (n)}^{(N)} (k_{i (n)}, u_{n}(\rho_{n}),
\kappa_{n})\nonumber.
\end{eqnarray}

The behavior of the vertex functions at the infrared regime is worthwhile, 
since their fixed point structure will determine the scaling laws and the 
critical exponents for arbitrary $m_{n}$-dimensional competing subspace. 
These $L$ independent fixed points are defined by $\beta_{n}(u_{n}^{*}) = 0$. 
At the fixed points the simple scaling property holds
\begin{eqnarray}
\Gamma_{R (n)}^{(N)} (\rho_{n} k_{i (n)}, u_{n}^{*}, \kappa_{n})&=&
\rho_{n}^{n (N + (d-\sum_{n=2}^{L} \frac{(n-1)}{n} m_{n}) - 
\frac{N(d-\sum_{n=2}^{L} \frac{(n-1)}{n} m_{n})}{2}) -\frac{N}{2} \gamma_{\phi (n)}(u_{n}^{*})}\\
&&\Gamma_{R (n)}^{(N)} (k_{i (n)}, u_{n}^{*},\kappa_{n})\nonumber.
\end{eqnarray}

For $N=2$, we have
\begin{equation}
\Gamma_{R (n)}^{(2)} (\rho_{n} k_{(n)}, u_{n}^{*}, \kappa_{n}) =
\rho_{n}^{2 n - \gamma_{\phi (n)}(u_{n}^{*})}\Gamma_{R (n)}^{(2)} (k_{(n)},u_{n}^{*}, \kappa_{n}).
\end{equation}
The quantity $\gamma_{\phi (n)}(u_{n}^{*})$ can be identified as the 
anomalous dimension of the competing subspace under consideration. 

This can be readily generalized to include $L$ insertions of  $\phi^{2}$
operators such that the RG equations at
the fixed point lead to the solution ($(N,M) \neq (0,2)$) :

\begin{eqnarray}
&\Gamma_{R (n)}^{(N,M)} (\rho k_{i (n)}, \rho p_{i (n)}, u_{n}^{*},
\kappa_{n}) = \rho_{n}^{n [N + (d -\sum_{n=2}^{L} \frac{(n-1)}{n} m_{n}) - \frac{N(d -\sum_{n=2}^{L} \frac{(n-1)}{n} m_{n})}{2} -2M] -
\frac{N \gamma_{\phi (n)}^{*}}{2} + M \gamma_{\phi^{2} (n)}^{*}}
\nonumber\\
& \qquad \times \Gamma_{R (n)}^{(N,M)} (k_{i (n)}, p_{i (n)}, u_{n}^{*}, \kappa_{n}).
\end{eqnarray}
Writing this at the fixed point as
\begin{equation}
\Gamma_{R (n)}^{(N,M)} (\rho k_{i (n)}, \rho p_{i (n)}, u_{n}^{*},
\kappa_{n}) = \rho_{n}^{n [(d -\sum_{n=2}^{L} \frac{(n-1)}{n} m_{n}) - N d_{\phi}] + M d_{\phi^{2}}}
\Gamma_{R (n)}^{(N,M)} (k_{i (n)}, p_{i (n)}, u_{n}^{*}, \kappa_{n}),
\end{equation}
the anomalous dimensions of the insertions of $\phi^{2}$ operators are 
$d_{\phi^{2}} = -2 n + \gamma_{\phi^{2} (n)}(u_{n}^{*})$.

The scaling relations can be found by going away from the Lifshitz
critical temperature ($t \neq 0$) staying, however, at the critical region
$\delta_{0n} = \tau_{nn'}=0$, which is the generalization of that from 
the ordinary second character Lifshitz behavior. Above the Lifshitz critical
temperature, the renormalized vertices for
$t\neq 0$ can be expressed as a power series in $t$ around the renormalized
vertex parts at $t=0$, as long as $N\neq 0$. Then one can show that the RGE 
for the vertex parts when $t\neq 0$ are given by
\begin{equation}
[\kappa_{n} \frac{\partial}{\partial \kappa_{n}} +
\beta_{n}\frac{\partial}{\partial u_{n}}
- \frac{1}{2} N \gamma_{\phi (n)}(u_{n}) +
\gamma_{\phi^{2} (n)}(u_{n}) t \frac{\partial}{\partial t}]
\Gamma_{R (n)}^{(N)} (k_{i (n)}, t, u_{n}, \kappa_{n}) = 0.
\end{equation}
The key property of the solution is that it is a homogeneous function of 
the product of $k_{i (n)}$ (to some power) and $t$ only at the fixed point 
$u_{n}^{*}$. As the value of $u_{n}$ is fixed at $u_{n}^{*}$, we shall omit it from the notation of this section henceforth. Thus, at the fixed point the 
solution of the RGE reads
\begin{equation}
\Gamma_{R (n)}^{(N)} (k_{i (n)}, t, \kappa_{n})=
\kappa_{n}^{\frac{N \gamma_{\phi (n)}^{*}}{2}}
F_{(n)}^{(N)}(k_{i (n)},\kappa_{n} t^{\frac{-1}{\gamma_{\phi^{2} (n)}^{*}}}) .
\end{equation}
Defining $\theta_{n} = -\gamma_{\phi^{2} (n)}^{*}$, and using
dimensional analysis, it is easy to show that
\begin{eqnarray}
\Gamma_{R (n)}^{(N)} (k_{i (n)}, t, \kappa_{n}) =&&
\rho_{n}^{n [N + (d - \sum_{n=2}^{L} \frac{(n-1)}{n} m_{n}) - \frac{N}{2}(d - \sum_{n=2}^{L} \frac{(n-1)}{n} m_{n})] -\frac{N}{2}
\eta_{n}}
\kappa_{n}^{\frac{N}{2} \eta_{n}} \nonumber\\
&&F_{(n)}^{(N)}(\rho_{n}^{-1} k_{i (n)},(\rho_{n}^{-1}\kappa_{n})
(\rho_{n}^{-2 n}t)^{\frac{-1}{\theta_{n}}} ) .
\end{eqnarray}

The choice
$\rho_{n} = \kappa_{n} (\frac{t}{\kappa_{n}^{2 n}})^{\frac{1}{\theta_{n} + 
2 n}}$, can be substituted back in (16), implying that the vertex function 
depends only on the combination $k_{i (n)} \xi_{n}$ apart from a power of 
$t$. Since the correlation lengths $\xi_{n}$ are proportional to 
$t^{- \nu_{n}}$, it implies that the critical exponents $\nu_{n}$ satisfy 
the identity
\begin{equation}
\nu_{n}^{-1} = 2 n + \theta_{n}^{*} = 2 n - \gamma_{\phi^{2} (n)}^{*} .
\end{equation}
For convenience we could have defined the function
\begin{equation}
\bar{\gamma}_{\phi^{2} (n)}(u_{n}) = - \beta_{n}
\frac{\partial ln (Z_{\phi^{2} (n)}Z_{\phi (n)}) }{\partial u_{n}}.
\end{equation}
In that case we would have discovered the equivalent relations
\begin{eqnarray}
\nu_{n}^{-1} &=& 2n - \eta_{n} - \bar{\gamma}_{\phi^{2} (n)}(u_{n}^{*}).
\end{eqnarray}
For $N=2$ we choose
$\rho_{n} = k_{(n)}$, the external momenta. Then
$\Gamma_{R(n)}^{(2)}(k_{(n)}, t, \kappa_{n}) =
k^{2 n - \eta_{n}}
\kappa_{n}^{\eta_{n}} f(k_{(n)} \xi_{n})$.
The infrared regime corresponds to $\xi_{n} \rightarrow \infty$
and $k_{(n)} \rightarrow 0$ such that
$f(k_{(n)} \xi_{n}) \rightarrow Constant$.
The definition $f_{n} =  (k_{(n)} \xi_{n})^{2 n - \eta_{n}}
f(k_{(n)} \xi_{n})$, leads to
\begin{equation}
\Gamma_{R(n)}^{(2)}(k_{(n)}, t, \kappa_{n}) = (k_{(n)} \xi_{n})^{2 n - \eta_{n}}
\kappa_{n}^{\eta_{n}} f_{n}(k_{(n)} \xi_{n}).
\end{equation}
Since the susceptibility  is
proportional to $ t^{-\gamma_{n}}$ as $k_{(n)} \rightarrow 0$, and 
$\Gamma_{R (n)}^{(2)} = \chi_{(n)}^{-1}$, the susceptibility critical 
exponents are given by
\begin{equation}
\gamma_{n} = \nu_{n} (2 n  - \eta_{n}).
\end{equation}

We now discuss the scaling law appropriate to relate the specific heat 
critical exponent to the others critical indices. The analysis of the 
RG equation for $\Gamma_{R (n)}^{(0,2)}$ above $T_{L}$ at the fixed point 
yields information about the specific heat exponents. In that case it 
reads
\begin{equation}
(\kappa_{n} \frac{\partial}{\partial \kappa_{n}} +
\gamma_{\phi^2 (n)}^{*} (2 + t\frac {\partial}{\partial t}))
\Gamma_{R (n)}^{(0,2)} =
(\kappa_{n}^{-2 n})^{\frac{\epsilon_{L}}{2}} B_{n}(u_{n}^{*}) ,
\end{equation}
where $B_{n}(u_{n}^{*})$ is given by
\begin{equation}
(\kappa_{n}^{-2 n})^{\frac{\epsilon_{L}}{2}}
B_{n}(u_{n}^{*}) = - Z_{\phi^{2}(n)}^{2}
\kappa_{n} \frac{\partial}{\partial \kappa_{n}}
\Gamma_{(n)}^{(0,2)}(Q_{(n)}; -Q_{(n)}, \lambda_{n})
|_{Q_{(n)}^{2}=\kappa_{n}^{2}}.
\end{equation}
The general discussion given up to now for the vertex part
$\Gamma_{R (n)}^{(N,M)}$ will be useful to uncover the homogeneous part
of the solution. In fact, at the fixed point the generalization of the
solution for $\Gamma_{R(n)}^{(N,M)}$ is written as
\begin{equation}
\Gamma_{R (n)}^{(N,M)} (p_{i (n)}, Q_{i (n)},  t, \kappa_{n}) =
\kappa_{n}^{\frac{1}{2} N \gamma_{\phi(n)}^{*} -
M \gamma_{\phi^{2}(n)}^{*}} F_{n}^{(N,M)}(p_{i (n)}, Q_{i (n)},
\kappa_{n} t^{\frac{-1}{\gamma_{\phi^{2} (n)}^{*}}}) .
\end{equation}

At the fixed point, the temperature dependent homogeneous part for
$\Gamma_{R(n),h}^{(0,2)}$ has the following property
\begin{equation}
\Gamma_{R (n),h}^{(0,2)}(Q_{(n)}, -Q_{(n)}, t, \kappa_{n}) =
\kappa_{n}^{- 2 \gamma_{\phi^{2}(n)}^{*}} F_{n}^{(0,2)}
(Q_{(n)},- Q_{(n)},
\kappa_{n} t^{\frac{-1}{\gamma_{\phi^{2} (n)}^{*}}}) .
\end{equation}
This is going to be identified with the specific heat at zero external
momentum insertion $Q_{(n)}=0$. Using the dimensional analysis results, 
one can show that
\begin{eqnarray}
&&\Gamma_{R (n),h}^{(0,2)}(Q_{(n)}, -Q_{(n)}, t, \kappa_{n}) =
\rho_{n}^{n [(d - \sum_{n=2}^{L} \frac{(n-1)}{n} m_{n}) - 4] + 2\gamma_{\phi^{2} (n)}^{*}}\\
&& \qquad \times \;\; \Gamma_{R (n),h}^{(0,2)}(\rho_{n}^{-1}Q_{(n)},
- \rho_{n}^{-1}Q_{(n)},
\rho_{n}^{-2 n} t,\rho_{n}^{-1} \kappa_{n}) ,\nonumber
\end{eqnarray}
and substituting this into the solution at the fixed point, it yields
\begin{eqnarray}
&&\Gamma_{R (n),h}^{(0,2)}(Q_{(n)}, -Q_{(n)}, t, \kappa_{n}) =
\rho_{n}^{n [(d - \sum_{n=2}^{L} \frac{(n-1)}{n} m_{n}) - 4] + 2\gamma_{\phi^{2} (n)}^{*}}
\kappa_{n}^{- 2 \gamma_{\phi^{2}(n)}^{*}}\\
&& \qquad \times \;\; F_{n}^{(0,2)}(\rho_{n}^{-1} Q_{(n)},
-\rho_{n}^{-1} Q_{(n)},
\rho_{n}^{-1} \kappa_{n}(\rho_{n}^{-2 n} t)^{\frac{-1}{\gamma_{\phi^{2} (n)}^{*}}}) .\nonumber
\end{eqnarray}
Once more, choose 
$\rho_{n} = \kappa_{n} (\frac{t}{\kappa_{n}^{2 n}})^{\frac{1}{\theta_{n} + 2 n}}$. Replace this in last equation, take the limit
$Q_{(n)} \rightarrow 0$ and identify the power of $t$ with the specific
heat exponent $\alpha_{n}$. The result is 
\begin{equation}
\alpha_{n} = 2 - n (d - \sum_{n=2}^{L} \frac{(n-1)}{n} m_{n})\nu_{n} .
\end{equation}
The inhomogeneous part can now be discussed. Take $Q_{(n)}=0$ and 
choose a particular solution of the form:
\begin{equation}
C_{p}(u_{n}) = (\kappa_{n}^{2 n})^{\frac{- \epsilon_{L}}{2}}
\tilde{C}_{p}(u_{n}).
\end{equation}
When this is replaced into the RG equation for $\Gamma_{R (n)}^{(0,2)}$ at the
fixed point, we learn that
\begin{equation}
C_{p}(u_{n}^{*}) = (\kappa_{n}^{2 n})^{\frac{- \epsilon_{L}}{2}}
\frac{\nu_{n}}{\nu_{n} n (d - \sum_{n=2}^{L} \frac{(n-1)}{n} m_{n}) -2} B_{n}(u_{n}^{*}).
\end{equation}
Summing up both terms gives the following general solution at the fixed
point:
\begin{equation}
\Gamma_{R (n)}^{(0,2)} = (\kappa_{n}^{-2 n})^{\frac{\epsilon_{L}}{2}}
(C_{n} (\frac{t}{\kappa_{n}^{2 n}})^{- \alpha_{n}} +
\frac{\nu_{n}}{\nu_{n} n (d - \sum_{n=2}^{L} \frac{(n-1)}{n} m_{n}) -2} B_{n}(u_{n}^{*})).
\end{equation}

The situation for $T<T_{L}$ is as follows. For simplicity consider the 
case of magnetic systems. The renormalized equation of
state furnishes a relation between the renormalized magnetic field
and the renormalized vertex parts for $t<0$ via a power series in the
magnetization $M$, i.e.,
\begin{equation}
H_{(n)}(t, M, u_{n}, \kappa_{n}) = \sum_{N=1}^{\infty} \frac{1}{N!} M^{N}
\Gamma_{R (n)}^{(1+N)}(k_{i (n)} = 0; t, u_{n}, \kappa_{n}),
\end{equation}
where the zero momentum limit must be taken after performing the summation.
The magnetic field satisfies the following RG equation:
\begin{equation}
(\kappa_{n} \frac{\partial}{\partial \kappa_{n}} +
\beta_{n}\frac{\partial}{\partial u_{n}}
- \frac{1}{2} N \gamma_{\phi (n)}(N + M \frac{\partial}{\partial M}) +
\gamma_{\phi^{2} (n)} t \frac{\partial}{\partial t})
H_{(n)}(t, M, u_{n}, \kappa_{n}) = 0 .
\end{equation}
The equation of state has the following form at the fixed point:
\begin{equation}
H_{(n)}(t, M, \kappa_{n}) = \kappa_{n}^{\frac{\eta_{n}}{2}}
h_{1 n}(\kappa_{n} M^{\frac{2}{\eta_{n}}}, \kappa_{n} t^{\frac{-1}
{\gamma_{\phi^{2} (n)}}}).
\end{equation}
Dimensional analysis arguments can be used to determine how a flow in the 
external momenta affects the renormalized magnetic field. The flow produces 
the following expression:
\begin{equation}
H_{(n)}(t, M, \kappa_{n}) = \rho_{n}^{n [\frac{(d - \sum_{n=2}^{L} \frac{(n-1)}{n} m_{n})}{2} + 1]} H_{(n)}(\frac{t}{\rho_{n}^{2 n}}, 
\frac{M}{\rho_{n}^{n[\frac{(d - \sum_{n=2}^{L} \frac{(n-1)}{n} m_{n})}{2} - 1]}},
\frac{\kappa_{n}}{\rho_{n}}) .
\end{equation}
The standard choice corresponds to $\rho_{n}$ being a power of $M$
\begin{equation}
\rho_{n} = \kappa_{n} [\frac{M}{\kappa_{n}^{\frac{n}{2}[(d - \sum_{n=2}^{L} \frac{(n-1)}{n} m_{n}) - 2]}}]^{\frac{2}
{n [(d - \sum_{n=2}^{L} \frac{(n-1)}{n} m_{n}) - 2] + \eta_{n}}} .
\end{equation}
Replacing this into (35) and from the scaling form of the equation of state
$H_{(n)}(t, M) =
M^{\delta_{n}} f(\frac{t}{M^{\frac{1}{\beta_{n}}}})$, we
obtain the remaining scaling laws 
\begin{mathletters}
\begin{eqnarray}
&&\beta_{n} = \frac{1}{2} \nu_{n} (n(d-\sum_{i=2}^{L} \frac{(i-1)}{i} m_{i}) - 2n + \eta_{n}),\\
&&\delta_{n} = \frac{n(d-\sum_{i=2}^{L} \frac{(i-1)}{i} m_{i}) + 2n -
\eta_{n}}{n(d-\sum_{i=2}^{L} \frac{(i-1)}{i} m_{i}) - 2n + \eta_{n}},
\end{eqnarray}
\end{mathletters}
which imply the Widom $\gamma_{n} = \beta_{n} (\delta_{n} -1)$ and
Rushbrooke $\alpha_{n} + 2 \beta_{n} + \gamma_{n} = 2$ relations for
{\it arbitrary} competing or noncompeting subspaces, since $n=1,...,L$. 
We note that there is one set of scaling relations for each competing 
subspace. This suggests that all the critical exponents take different 
values in distinct subspaces. We are going to see that this is not 
necessarily true, since the fixed point structure restricts the values 
of most critical exponents to be the same in different competing subspaces. 
It is a direct consequence that there is only one fixed point independent of 
the space directions under consideration.  

The perturbative calculation of the critical exponents and other universal 
quantities follows from a diagrammatic expansion whose basic objects 
are Feynman diagrams. We shall use the loop expansion for the anisotropic 
integrals with the perturbation parameter 
$\epsilon_{L} = 4 + \sum_{n=2}^{L} \frac{(n-1)}{n} m_{n} - d$. 
The solution of the Feynman diagrams in terms of $\epsilon_{L}$ 
results in the $\epsilon_{L}$-expansion  for the universal critical 
ammounts of the anisotropic  criticalities. The anisotropic integrals are 
described using the generalized orthogonal
approximation in Appendix A. This approximation yields a solution which is 
the most general one compatible with the homogeneity of the Feynman integrals 
for {\it arbitrary} external momenta scales. With this technique all the 
critical exponents in the anisotropic cases can be obtained as will be shown 
in Section V.   

\section{Scaling theory for the isotropic behaviors}

To begin with let us promote a slight change of notation with respect to 
the conventions presented in our previous letter \cite{Leite4}. There, the 
subscript associated to each type of $m_{n}$ competing axes was chosen as 
$4n$. Here, we choose the subscript $n$ to express the same thing. This will 
cause no confusion to the reader since the anisotropic and isotropic cases 
are considered separately in this work. Then an arbitrary ammount $A_{4n}$, 
should be changed to $A_{n}$. In particular, the perturbative parameter 
discussed in section II is now represented as $\epsilon_{n}$. Obviously, 
whenever $d=m_{n}$, the volume element in momentum space is given by 
$d^{m_{n}}k$. Setting $\sigma_{n}=1$ we perform a dimensional redefinition 
of the momenta such that $[k] = M^{\frac{1}{n}}$. Accordingly, the
volume element has dimension $[d^{m_{n}}k] = M^{\frac{m_{n}}{n}}$. The
dimension of the field in mass units is
$[\phi] = M^{\frac{m_{n}}{2n} - 1}$. The 1PI vertex parts have dimensions
$[\Gamma^{(N)}(k_{n})] = M^{N + \frac{m_{n}}{n} - N \frac{m_{n}}{2n}}$.
Then, make the continuation $m_{n}=4n-\epsilon_{n}$. The coupling
constant has dimension
$\lambda_{4n} = M^{\frac{4n-m_{n}}{n}}= M^{\frac{\epsilon_{n}}{n}}$.
In terms of dimensionless
quantities, one has the renormalized
$g_{n} = u_{n} (\kappa_{n}^{2n})^{\frac{\epsilon_{n}}{2n}}$ and
bare $\lambda_{n} = u_{0n} (\kappa_{n}^{2n})^{\frac{\epsilon_{n}}{2n}}$
coupling constants, respectively. Again, the functions
\begin{mathletters}
\begin{eqnarray}
\beta_{n} &=& (\kappa_{n}\frac{\partial u_{n}}{\partial \kappa_{n}})\\
\gamma_{\phi (n)}(u_{n})  &=& \beta_{n}
\frac{\partial ln Z_{\phi (n)}}{\partial u_{n}}\\
\gamma_{\phi^{2} (n)}(u_{n}) &=& - \beta_{n}
\frac{\partial ln Z_{\phi^{2} (n)}}{\partial u_{n}}
\end{eqnarray}
\end{mathletters}
are computed at fixed bare coupling constant $\lambda_{n}$.
The beta functions in terms of dimensionless quantities are given by 
$\beta_{n} = -  \epsilon_{n}(\frac{\partial ln u_{0n}}{\partial u_{n}})^{-1}$. Notice that the beta function for the isotropic case does not
possess the overall factor of $n$ present in the anisotropic beta function
$\beta_{n}$ obtained from renormalization group transformations over the
$m_{n}$-dimensional competing subspace. This is a very close analogy to the 
second character behaviors and a general property of Lifshitz critical 
behaviors.

The dimensional redefinition of the momenta along the $m_{n}$ competing 
axes leads to an effective space dimension for the isotropic case, i.e.,
$(\frac{m_{n}}{n})$. Under a flow in the external momenta we find
the following behavior for the 1PI vertex parts $\Gamma_{R (n)}^{(N)}$:
\begin{eqnarray}
&\Gamma_{R (n)}^{(N)} (\rho_{n} k_{i}, u_{n}, \kappa_{n}) =
\rho_{n}^{n [N + \frac{m_{n}}{n} - N \frac{m_{n}}{2n}]}
exp[-\frac{N}{2} \int_{1}^{\rho_{n}} \gamma_{\phi (n)}(u_{n}(x_{n}))
\frac{{d x_{n}}}{x_{n}}] \\
& \times \;\Gamma_{R (n)}^{(N)} (k_{i}, u_{n}(\rho_{n}),
\kappa_{n})\nonumber.
\end{eqnarray}
Note that since there is only one type of space directions in the isotropic 
behaviors, we do not need to use a label in the external momenta specifying 
the type of competing axes considered as we did in the anisotropic cases. 
At the fixed point, the simple scaling property for the vertex
parts $\Gamma_{R (n)}^{(N)}$ follows:

\begin{eqnarray}
&\Gamma_{R (n)}^{(N)} (\rho_{n} k_{i}, u_{n}^{*}, \kappa_{n}) =
\rho_{n}^{n [N + \frac{m_{n}}{n} - N \frac{m_{n}}{2n}] -
\frac{N}{2} \gamma_{\phi (n)}(u_{n}^{*})}\\
& \times \Gamma_{R (n)}^{(N)} (k_{i}, u_{n}^{*},\kappa_{n})\nonumber.
\end{eqnarray}

For $N=2$, we have
\begin{equation}
\Gamma_{R (n)}^{(2)} (\rho_{n} k, u_{n}^{*}, \kappa_{n}) =
\rho_{n}^{2n - \gamma_{\phi (n)}(u_{n}^{*})}\Gamma_{R (n)}^{(2)} (k, u_{n}^{*}, \kappa_{n} ).
\end{equation}
In the noninteracting theory $d_{\phi}^{0} = \frac{\frac{m_{n}}{n}}{2} - 1$
is the naive dimension of the field. At the isotropic fixed point, the 
presence of interactions modify it such that
$d_{\phi}= \frac{\frac{m}{n}}{2} - 1 + \frac{\eta_{n}}{2n}$.
The generalization to include $L$ insertions of  $\phi^{2}$
operators can be written at the fixed point as
($(N,L) \neq (0,2)$) :

\begin{equation}
\Gamma_{R (n)}^{(N,L)} (\rho_{n} k_{i}, \rho_{n} p_{i}, u_{n}^{*},
\kappa_{n}) = \rho_{n}^{n [N + \frac{m_{n}}{n} - \frac{N(\frac{m_{n}}{n})}{2} -2L] -
\frac{N \gamma_{\phi (n)}^{*}}{2} + L \gamma_{\phi^{2} (n)}^{*}}
\Gamma_{R (n)}^{(N,L)} (k_{i}, p_{i}, u_{n}^{*}, \kappa_{n}).
\end{equation}
Writing at the fixed point
\begin{equation}
\Gamma_{R (n)}^{(N,L)} (\rho_{n} k_{i}, \rho_{n} p_{i}, u_{n}^{*},
\kappa_{n}) = \rho_{n}^{ m_{n} - N d_{\phi} + L d_{\phi^{2}}}
\Gamma_{R (n)}^{(N,L)} (k_{i}, p_{i}, u_{n}^{*}, \kappa_{n}),
\end{equation}
the anomalous dimension of the insertions of $\phi^{2}$ operators is given by
$d_{\phi^{2}} = -2n + \gamma_{\phi^{2} (n)}(u_{n}^{*})$.

Above the Lifshitz critical temperature we find the following RGE
\begin{equation}
[\kappa_{n} \frac{\partial}{\partial \kappa_{n}} +
\beta_{n}\frac{\partial}{\partial u_{n}}
- \frac{1}{2} N \gamma_{\phi (n)}(u_{n}) +
\gamma_{\phi^{2} (n)}(u_{n}) t \frac{\partial}{\partial t}]
\Gamma_{R (n)}^{(N)} (k_{i}, t, u_{n}, \kappa_{n}) = 0.
\end{equation}
The solution at the fixed point is given by
\begin{equation}
\Gamma_{R (n)}^{(N)} (k_{i (n)}, t, u_{n}^{*}, \kappa_{n})=
\kappa_{n}^{\frac{N \gamma_{\phi (n)}^{*}}{2}}
F_{(n)}^{(N)}(k_{i},\kappa_{n} t^{\frac{-1}{\gamma_{\phi^{2} (n)}^{*}}}) .
\end{equation}
If we define $\theta_{n} = -\gamma_{\phi^{2} (n)}^{*}$, we can use dimensional
analysis to obtain
\begin{eqnarray}
&\Gamma_{R (n)}^{(N)} (k_{i}, t, \kappa_{n}) =
\rho_{n}^{n [N + \frac{m_{n}}{n} - \frac{N}{2} \frac{m_{n}}{n}] -\frac{N}{2}
\eta_{n}}
\kappa_{n}^{\frac{N}{2} \eta_{n}} \nonumber\\
& \times \; F_{(n)}^{(N)}(\rho_{n}^{-1} k_{i},(\rho_{n}^{-1}\kappa_{n})
(\rho_{n}^{-4}t)^{\frac{1}{\theta_{n}}} ) .
\end{eqnarray}

We can choose
$\rho_{n} = \kappa_{n} (\frac{t}{\kappa_{n}^{2n}})^{\frac{1}{\theta_{n} + 2n}}$, and replacing it in the last two equations, the vertex parts depend only 
on the combination $k_{i} \xi_{n}$ apart from a power of $t$. As $\xi_{n}$ 
is proportional to $t^{-\nu_{n}}$ we can identify the critical exponent
$\nu_{n}$ as
\begin{equation}
\nu_{n}^{-1} = 2n  + \theta_{n}^{*} = 2n - \gamma_{\phi^{2} (n)}^{*} .
\end{equation}

For the sake of convenience we define the function
\begin{equation}
\bar{\gamma}_{\phi^{2} (n)}(u_{n}) = - \beta_{n}
\frac{\partial ln (Z_{\phi^{2} (n)}Z_{\phi (n)}) }{\partial u_{n}}.
\end{equation}
In terms of this function the last equation turns into the following 
relation
\begin{equation}
\nu_{n}^{-1} = 2n - \eta_{n} - \bar{\gamma}_{\phi^{2} (n)}(u_{n}^{*}).
\end{equation}
For $N=2$ we choose
$\rho_{n} = k$, the external momenta. When $\xi_{n} \rightarrow \infty$ and 
$k \rightarrow 0$, simultaneously, then $f(k \xi_{n}) \rightarrow Constant$. 
The susceptibility is proportional to $ t^{-\gamma_{n}}$ as $k_{i} \rightarrow 0$. As 
$\Gamma_{R}^{(2)} = \chi^{-1}$, the susceptibility critical
exponent follows 
\begin{equation}
\gamma_{n} = \nu_{n} (2n  - \eta_{n}).
\end{equation}

From the RG equation for $\Gamma_{R (n)}^{(0,2)}$ above $T_{L}$ at the fixed 
point, the scaling relation for the specific heat exponent can be found. The 
RG equation is
\begin{equation}
(\kappa_{n} \frac{\partial}{\partial \kappa_{n}} +
\gamma_{\phi^2 (n)}^{*} (2 + t\frac {\partial}{\partial t}))
\Gamma_{R (n)}^{(0,2)} =  (\kappa_{n}^{-2 })^{\frac{\epsilon_{n}}{2}}
B_{n}(u_{n}^{*}) ,
\end{equation}
where
\begin{equation}
(\kappa_{n}^{-2n})^{\frac{\epsilon_{n}}{2n}}
B_{n}(u_{n}^{*}) = - Z_{\phi^{2}(n)}^{2}
\kappa_{n} \frac{\partial}{\partial \kappa_{n}}
[\Gamma_{(n)}^{(0,2)}(Q; -Q, \lambda_{n})|_{Q^{2}=\kappa_{n}^{2}}].
\end{equation}
The $\Gamma_{R(n)}^{(N,L)}$ can be generalized to
\begin{equation}
\Gamma_{R (n)}^{(N,L)}(p_{i}, Q_{i},  t, \kappa_{n}) =
\kappa_{n}^{\frac{1}{2} N \gamma_{\phi(n)}^{*} -
L \gamma_{\phi^{2}(n)}^{*}} F_{n}^{(N,L)}(p_{i}, Q_{i},
\kappa_{n} t^{\frac{-1}{\gamma_{\phi^{2} (n)}^{*}}}) .
\end{equation}
The homogeneous part of the solution for $\Gamma_{R(n),h}^{(0,2)}$ is 
temperature dependent and scales at the fixed point as
\begin{equation}
\Gamma_{R (n),h}^{(0,2)}(Q, -Q, t, \kappa_{n}) =
\kappa_{n}^{- 2 \gamma_{\phi^{2}(n)}^{*}} F_{n}^{(0,2)}(Q,- Q,
\kappa_{n} t^{\frac{-1}{\gamma_{\phi^{2} (n)}^{*}}}) .
\end{equation}
This vertex function is going to be identified with the specific heat at
zero external momentum insertion $Q=0$. Use of dimensional analysis yields 
the result:
\begin{equation}
\Gamma_{R (n),h}^{(0,2)}(Q, -Q, t, \kappa_{n}) =
\rho_{n}^{n[\frac{m_{n}}{n} - 4] + 2\gamma_{\phi^{2} (n)}^{*}}
\Gamma_{R (n),h}^{(0,2)}(\rho_{n}^{-1}Q, - \rho_{n}^{-1}Q,
\rho_{n}^{-2n} t,\rho_{n}^{-1} \kappa_{n}) .
\end{equation}
Substituting this equation in the solution at the fixed point leads to
\begin{equation}
\Gamma_{R (n),h}^{(0,2)}(Q, -Q, t, \kappa_{n}) =
\rho_{n}^{n [\frac{m_{n}}{n} - 4] + 2\gamma_{\phi^{2} (n)}^{*}}
\kappa_{n}^{- 2 \gamma_{\phi^{2}(n)}^{*}} F_{n}^{(0,2)}(\rho_{n}^{-1} Q,
-\rho_{n}^{-1} Q,
\rho_{n}^{-1} \kappa_{n}(\rho_{n}^{-2n} t)^{\frac{-1}{\gamma_{\phi^{2} (n)}^{*}}}) .
\end{equation}
The choice
$\rho_{n} = \kappa_{n} (\frac{t}{\kappa_{n}^{2n}})^{\frac{1}{\theta_{n} + 2n}}$ can be made. Substitution of this choice into the last equation in 
the limit $Q \rightarrow 0$ and identifying the power of $t$ with the 
specific heat exponent $\alpha_{n}$, we obtain:
\begin{equation}
\alpha_{n} = 2 - m_{n} \nu_{n} .
\end{equation}
The inhomogeneous part can be found by taking $Q=0$ and choosing a 
particular solution in the standard way. Therefore, the general solution 
at the fixed point is given by
\begin{equation}
\Gamma_{R (n)}^{(0,2)} = (\kappa_{n}^{-2 n})^{\frac{\epsilon_{n}}{2 n}}
(C_{n} (\frac{t}{\kappa_{n}^{2 n}})^{- \alpha_{n}} +
\frac{\nu_{n}}{\nu_{n} m -2} B_{n}(u_{n}^{*})).
\end{equation}

Next, let us concentrate ourselves in the scaling relations when the 
system is below the Lifshitz critical temperature $T<T_{L}$.
The relation among the renormalized magnetic field, the renormalized vertex
parts for $t<0$ and the magnetization $M$ is given by
\begin{equation}
H_{(n)}(t, M, u_{n}, \kappa_{n}) = \sum_{N=1}^{\infty} \frac{1}{N!} M^{N}
\Gamma_{R(n)}^{(1+N)}(k_{i} = 0; t, u_{n}, \kappa_{n}).
\end{equation}
The RG equation satisfied by the magnetic field is:
\begin{equation}
(\kappa_{n} \frac{\partial}{\partial \kappa_{n}} +
\beta_{n}\frac{\partial}{\partial u_{n}}
- \frac{1}{2} N \gamma_{\phi (n)}(u_{n})(N + M \frac{\partial}{\partial M}) +
\gamma_{\phi^{2} (n)} t \frac{\partial}{\partial t})
H_{(n)}(t, M, u_{n}, \kappa_{n}) = 0 .
\end{equation}
The solution of the equation of state at the fixed point has the following 
property
\begin{equation}
H_{(n)}(t, M, \kappa_{n}) = \kappa_{n}^{\frac{\eta_{n}}{2}}
h_{n}(\kappa_{n} M^{\frac{2}{\eta_{n}}}, \kappa_{n} t^{\frac{-1}
{\gamma_{\phi^{2} (n)}}}).
\end{equation}
The scale change in the magnetic field followed by a flow in the external 
momenta can be written in the form:
\begin{equation}
H_{(n)}(t, M, \kappa_{n}) = \rho_{n}^{n [\frac{m_{n}}{2n} + 1]}\nonumber\\
\;\; H_{n}(\frac{t}{\rho_{n}^{2n}}, \frac{M}{\rho_{n}^{n[\frac{m_{n}}{2n} - 1]}},
\frac{\kappa_{n}}{\rho_{n}}) .
\end{equation}
The flow parameter $\rho_{n}$ is chosen as a power of $M$ such that:
\begin{equation}
\rho_{n} = \kappa_{n} [\frac{M}{\kappa_{n}^{[\frac{m_{n}}{2} - n]}}]^{\frac{2}
{m_{n} - 2n + \eta_{n}}} ,
\end{equation}
and from the scaling form of the equation of state
$H_{(n)}(t, M) = M^{\delta_{n}} f_{(n)}(\frac{t}{M^{\frac{1}{\beta_{n}}}})$, we
obtain the following scaling relations:
\begin{mathletters}
\begin{eqnarray}
\delta_{n} &=& \frac{m_{n} + 2n - \eta_{n}}{m_{n} - 2n + \eta_{n}}, \\
\beta_{n} &=& \frac{1}{2} \nu_{n} (m_{n} - 2n + \eta_{n}),
\end{eqnarray}
\end{mathletters}
which imply the Widom $\gamma_{n} = \beta_{n} (\delta_{n} -1)$ and
Rushbrooke $\alpha_{n} + 2 \beta_{n} + \gamma_{n} = 2$ relations. 

Except for some minor modifications, the renormalization group treatment 
of each isotropic behavior is equivalent to treat separately each competing 
subspace appearing in the most general anisotropic behavior. It is easy to 
see that all scaling laws reduce to those from the usual critical behavior 
described by a $\phi^{4}$ field theory for $n=1$. Then, the usual critical 
behavior is actually a first character isotropic Lifshitz critical behavior. 
For $n=2$, they easily reproduce those associated to the second character 
Lifshitz point \cite{Leite2}. Therefore, a new nomenclature emerges from the 
study of these higher character Lifshitz critical behaviors: the number of 
neighbors coupled through competing interactions is a fundamental parameter, 
generalizing the concept of universality class. Thus, the universality 
classes of isotropic behaviors $(d=m_{n})$ are characterized by $(N,d,n)$. 
These statements will be put on a firmer ground when we calculate the critical 
exponents, as we shall see in the next sections. 

The isotropic behaviors are calculated using the generalized orthogonal 
approximation as well as exactly, without the resource of any approximation. 
The approximation is useful to complete the unified analytical description of 
the higher character Lifshitz critical behavior in its full generality, 
at least at the loop order considered here. On the other hand, the analytic 
exact (perturbative) solution is a conceptual step forward towards a better 
comprehension of this sort of system. At this point, the reader 
should consult the Appendix B in order to access the computation of the 
Feynman integrals using the generalized orthogonal approximation and 
Appendix C to see the exact computation required to finding the critical 
exponents.

\section{Critical exponents for the anisotropic behaviors}

In this section we compute the critical exponents using the generalized 
orthogonal approximation with the results derived in Appendix A. First, 
we attack the problem using normalization 
conditions. Second, the results are checked using a variant of the minimal 
subtraction scheme first developed in \cite{Leite2} for the anisotropic 
second character $m_{2}$-fold Lifshitz point.

\subsection{Normalization conditions and critical exponents}

The bare coupling constants and renormalization functions were defined in 
section II. They are given by
\begin{mathletters}
\begin{eqnarray}
&& u_{on} = u_{n} (1 + a_{1n} u_{n} + a_{2n} u_{n}^{2}) ,\\
&& Z_{\phi (n)} = 1 + b_{2n} u_{n}^{2} + b_{3n} u_{n}^{3} ,\\
&& \bar{Z}_{\phi^{2} (n)} = 1 + c_{1n} u_{n} + c_{2n} u_{n}^{2} ,
\end{eqnarray}
\end{mathletters}
where the constants $a_{in}, b_{in}, c_{in}$ depend on Feynman
diagrams computed at suitable symmetry points. The critical exponents 
associated to correlations perpendicular or parallel to the arbitrary 
competing $m_{n}$-dimensional subspace can be calculated through the 
specification of the corresponding symmetry point.

In terms of the constants defined above, the beta functions and 
renormalization constants can be cast in the form:
\begin{mathletters}
\begin{eqnarray}
&& \beta_{n}  =  -n \epsilon_{L}u_{n}[1 - a_{1n} u_{n}
+2(a_{1n}^{2} -a_{2n}) u_{n}^{2}],\\
&& \gamma_{\phi (n)} = -n \epsilon_{L}u_{n}[2b_{2n} u_{n}
+ (3 b_{3n}  - 2 b_{2n} a_{1n}) u_{n}^{2}],\\
&& \bar{\gamma}_{\phi^{2} (n)} = n \epsilon_{L}u_{n}[c_{1n}
+ (2 c_{2n}  - c_{1n}^{2} - a_{1n} c_{1n})u_{n}    ].
\end{eqnarray}
\end{mathletters}

The above coefficients can be figured out as functions of
the integrals calculated at the symmetry points. We find
\begin{mathletters}
\begin{eqnarray}
&& a_{1n} = \frac{N+8}{6 \epsilon_{L}}[1 + h_{m_{L}} \epsilon_{L}] ,\\
&& a_{2n} = (\frac{N+8}{6 \epsilon_{L}})^{2}
+ [\frac{(N+8)^{2}}{18}h_{m_{L}} - \frac{(3N+14)}{24}]
\frac{1}{\epsilon_{L}} ,\\
&& b_{2n} = -\frac{(N+2)}{144 \epsilon_{L}}[1 +
(2 h_{m_{L}} + \frac{1}{4}) \epsilon_{L}], \\
&& b_{3n} = -\frac{(N+2)(N+8)}{1296 \epsilon_{L}^{2}} +
\frac{(N+2)(N+8)}{108 \epsilon_{L}}(-\frac{1}{4} h_{m_{L}} +
\frac{1}{48}), \\
&& c_{1n} = \frac{(N+2)}{6 \epsilon_{L}}[1 + h_{m_{L}} \epsilon_{L}], \\
&& c_{2n} = \frac{(N+2)(N+5)}{36 \epsilon_{L}^{2}}
+ \frac{(N+2)}{3 \epsilon_{L}}[\frac{(N+5)}{3} h_{m_{L}} - \frac{1}{4}].
\end{eqnarray}
\end{mathletters}
These expressions are sufficient to find out the fixed points at 
$O(\epsilon_{L}^{2})$, which are defined by $\beta_{n}(u_{n}^{*}) = 0$. 
As was seen in the last section, every integral computed at arbitrary 
symmetry points $SP_{1}, ..., SP_{L}$ gives the same result, 
irrespective of the considered subspace. 
The overall factor of $n=1,...,L$ in the $\beta_{n}$ functions drops 
out at the fixed points,
such that the renormalization group transformations realized
over $\kappa_{1}$,..., and $\kappa_{L}$ will flow to the same fixed
point given by ($u_{1}^{*} = ... = u_{L}^{*} \equiv u^{*}$)
\begin{equation}
u^{\ast}=\frac{6}{8 + N}\,\epsilon_L\Biggl\{1 + \epsilon_L
\,\Biggl[ - h_{m_{L}} + \frac{(9N + 42)}{(8 + N)^{2}}\Biggr]\Biggr\}\;\;.
\end{equation}
It is instructive to separate 
explicitly the noncompeting and competing subspaces. The functions 
$\gamma_{\phi (1)}$ and $\bar{\gamma}_{\phi^{2} (1)}$ read 
\begin{eqnarray}
&& \gamma_{\phi (1)} = \frac{(N+2)}{72} [1 + (2 h_{m_{L}} + \frac{1}{4})
\epsilon_{L}]u_{1}^{2} - \frac{(N+2)(N+8)}{864} u_{1}^{3}, \\
&& \bar{\gamma}_{\phi^{2} (1)} = \frac{(N+2)}{6} u_{1}[1
+ h_{m_{L}} \epsilon_{L} - \frac{1}{2} u_{1}].
\end{eqnarray}
When the value of the fixed point is substituted into these equations, 
using the relation among these functions and the critical exponents 
$\eta_{1}(\equiv \eta_{L2})$ and $\nu_{1}(\equiv \nu_{L2})$, we find:
\begin{eqnarray}
&& \eta_{1}= \frac{1}{2} \epsilon_{L}^{2}\,\frac{N + 2}{(N+8)^2}
[1 + \epsilon_{L}(\frac{6(3N + 14)}{(N + 8)^{2}} - \frac{1}{4})] ,\\
&& \nu_{1} =\frac{1}{2} + \frac{(N + 2)}{4(N + 8)} \epsilon_{L}
+  \frac{1}{8}\frac{(N + 2)(N^{2} + 23N + 60)} {(N + 8)^3} \epsilon_{L}^{2}.
\end{eqnarray}
The coefficient of each power of $\epsilon_{L}$ is the same
as that coming from the second character Lifshitz behavior 
$m_{3}=...=m_{L}=0$. Consequently, the reduction to the Ising-like 
universality class $m_{2}=0$ case is warranted. The several beta functions 
corresponding to distinct competing axes satisfy the property 
$\beta_{n} = n \beta_{1}$. This implies that 
$\gamma_{\phi (n)}= n \gamma_{\phi (1)}$ and
$\bar{\gamma}_{\phi^{2} (n)} = n \bar{\gamma}_{\phi^{2} (1)}$. Then, 
we have 
\begin{eqnarray}
&& \eta_{n}= \frac{n}{2}(\epsilon_{L}^{2}\,\frac{(N + 2)}{(N+8)^2}
[1 + \epsilon_{L}(\frac{6(3N + 14)}{(N + 8)^{2}} - \frac{1}{4})]) ,\\
&& \nu_{n} = \frac{1}{n} (\frac{1}{2} + \frac{(N + 2)}{4(N + 8)} \epsilon_{L}
+  \frac{1}{8}\frac{(N + 2)(N^{2} + 23N + 60)} {(N + 8)^3} \epsilon_{L}^{2}).
\end{eqnarray}
At $O(\epsilon_{L}^{3})$, the relation $\eta_{n} = n \eta_{1}$
is satisfied, whereas at $O(\epsilon_{L}^{2})$,  the relation
$\nu_{n} = \frac{1}{n} \nu_{1}$ holds. Strong anisotropic
scale invariance\cite{He} is {\it exact} to the perturbative order
considered here, and within the generalized orthogonal approximation 
it is expected to hold at arbitrary higher loop order. Differently from 
the critical indices $\eta_{n}$ and $\nu_{n}$ which depend explicitly on 
the $m_{n}$-dimensional subspace under consideration, the other exponents 
take the same value in each subspace even though they are obtained through 
independent scaling relations along the distinct competing axes. They are 
given by
\begin{eqnarray}
&& \gamma_{L} = 1 + \frac{(N+2)}{2(N + 8)} \epsilon_{L}
+\frac{(N + 2)(N^{2} + 22N + 52)}{4(N + 8)^{3}} \epsilon_{L}^{2}, \\
&& \alpha_{L} = \frac{(4 - N)}{2(N + 8)} \epsilon_{L}
- \frac{(N + 2)(N^{2} + 30N + 56)}{4(N + 8)^{3}} \epsilon_{L}^{2} ,\\
&& \beta_{L} = \frac{1}{2} - \frac{3}{2(N + 8)} \epsilon_{L}
+ \frac{(N + 2)(2N + 1)}{2(N + 8)^{3}} \epsilon_{L}^{2} ,\\
&& \delta_{L} = 3 + \epsilon_{L}
+ \frac{(N^{2} + 14N + 60)}{2(N + 8)^{2}} \epsilon_{L}^{2}.
\end{eqnarray}
The exponents correctly reduce to those from the second character behavior 
\cite{Leite2,Leite4}, with a further reduction to the Ising-like case 
when $m_{2}=0$. In fact the universality classes reduction of generic 
higher character anisotropic Lifshitz points to that from Ising-like 
critical points is manifest in {\it all} critical exponents. Hence, 
this universality class reduction is a generic property of arbitrary 
competing systems. 

To check the correctness of these exponents, it is convenient to
calculate them in another renormalization procedure. Let us check these 
results using minimal subtraction of dimensional poles, as we are going to 
show next.

\subsection{Minimal subtraction and critical exponents}

In the minimal subtraction renormalization scheme, the common situation is 
to have just one momenta scale $\mu$\cite{Si} which is called $\kappa$ in 
the present paper. Nevertheless, the dimensional redefinitions performed 
over the external momenta characterizing arbitrary types of competing axes 
permit a picture of the anisotropic cases with $L$ independent momenta scales. 

The calculation of the critical exponents along an arbitrary kind of 
competition subspace can be done, provided all external momenta not 
belonging to that subspace are set to zero. Then, we define 
$\kappa_{j}$ to be the typical scale parameter of the $j$th 
subspace, calculate the renormalization functions for arbitrary 
external momenta along the $m_{j}$th space directions and require 
minimal subtraction of dimensional poles. This procedure is inspired in 
the method formerly discussed in the second character anisotropic 
Lifshitz behaviors \cite{Leite2}. Thus, although the external momentum 
associated to the competing subspace under consideration is kept arbitrary 
in all stages of the computation, the same is not true for all other external 
momenta corresponding to distinct competing subspaces. This restriction on the 
values of all the external momenta is the price to be paid in order to 
describe independently the scale transformations of each inequivalent 
subspace. This is the main difference of this minimal subtraction scheme 
with several independent momentum scales from the conventional method with 
just one momentum scale. 

Here we will content ourselves in showing that the diagrammatic procedure 
to calculate the fixed point using minimal subtraction results in the same 
functions $\gamma_{\phi (n)}$ and  $\bar{\gamma}_{\phi^{2} (n)}$ at the 
fixed point as those coming from normalization conditions. This is 
equivalent to prove the renormalization scheme independence of all 
critical indices.

In minimal subtraction, the dimensionless bare coupling constants and the 
renormalization functions are defined by
\begin{mathletters}
\begin{eqnarray}
&& u_{0n} = u_{n}[1 + \sum_{i=1}^{\infty} a_{in}(\epsilon_{L})
u_{n}^{i}], \\
&& Z_{\phi (n)} = 1 + \sum_{i=1}^{\infty} b_{in}(\epsilon_{L})
u_{n}^{i},\\
&& \bar{Z}_{\phi^{2} (n)} = 1 + \sum_{i=1}^{\infty} c_{in}(\epsilon_{L})
u_{n}^{i}.
\end{eqnarray}
\end{mathletters}
The renormalized vertex parts
\begin{mathletters}
\begin{eqnarray}
&& \Gamma_{R (n)}^{(2)}(k_{(n)}, u_{n}, \kappa_{n}) = Z_{\phi (n)}
\Gamma_{(n)}^{(2)}(k_{n}, u_{0n}, \kappa_{n}), \\
&& \Gamma_{R (n)}^{(4)}(k_{i(n)}, u_{n}, \kappa_{n}) = Z_{\phi (n)}^{2} \Gamma_{(n)}^{(4)}(k_{i (n)}, u_{0n}, \kappa_{n}), \\
&& \Gamma_{R (n)}^{(2,1)}(k_{1(n)}, k_{2(n)}, p_{(n)};
u_{n}, \kappa_{n}) = \bar{Z}_{\phi^{2} (n)}
\Gamma_{(n)}^{(2,1)}(k_{1(n)}, k_{2(n)}, p_{(n)}, u_{0n}, \kappa_{n}),
\end{eqnarray}
\end{mathletters}
are finite by construction when $\epsilon_{L} \rightarrow 0$, 
order by order in $u_{n}$. One should bear in mind that the external 
momenta in the bare vertices are mutiplied by $\kappa_{n}^{-1}$. Since 
$k_{i(1)} = p_{i}$
are the external momenta perpendicular to the competing axes, whereas
$k_{i(n)} = k'_{i(n)}$ are the external momenta parallel to the
$m_{n}$-dimensional type of competing subspace, the coefficients 
$a_{in}(\epsilon_{L}),
b_{in}(\epsilon_{L})$ and $c_{in}(\epsilon_{L})$ are obtained
by requiring that the poles in $\epsilon_{L}$ be minimally subtracted.
The bare vertices are written in the form
\begin{mathletters}
\begin{eqnarray}
&& \Gamma_{(n)}^{(2)}(k_{(n)}, u_{0n}, \kappa_{n}) =
k_{(n)}^{2n}(1- B_{2n} u_{0n}^{2} + B_{3n}u_{0n}^{3}), \\
&& \Gamma_{(n)}^{(4)}(k_{i(n)}, u_{0n}, \kappa_{(n)}) =
\kappa_{n}^{n \epsilon} u_{0n}
[1- A_{1n} u_{0n}
+ (A_{2n}^{(1)} + A_{2n}^{(2)})u_{0n}^{2}], \\
&& \Gamma_{(n)}^{(2,1)}(k_{1(n)}, k_{2(n)}, p_{(n)};
u_{0n}, \kappa_{n}) = 1 - C_{1n} u_{0n}
+ (C_{2n}^{(1)} + C_{2n}^{(2)}) u_{0n}^{2}.
\end{eqnarray}
\end{mathletters}
Remember that $B_{2n}$ is proportional to the integral $I_{3}$ and 
$B_{3n}$ is proportional to $I_{5}$ which are calculated with all external 
momenta not belonging to the $m_{n}$-dimensional subspace set to zero. 

The coefficients are expressed explicitly by the following integrals:
\begin{mathletters}
\begin{eqnarray}
&& A_{1n} = \frac{(N+8)}{18}[ I_{2}(\frac{k_{1(n)} + k_{2(n)}}
{\kappa_{n}}) +  I_{2}(\frac{k_{1(n)} + k_{3(n)}}
{\kappa_{n}}) + I_{2}(\frac{k_{2(n)} + k_{3(n)}}
{\kappa_{n}})] ,\\
&& A_{2n}^{(1)} = \frac{(N^{2} + 6N + 20)}{108}
[I_{2}^{2}(\frac{k_{1(n)} + k_{2(n)}}{\kappa_{n}})
+  I_{2}^{2}(\frac{k_{1(n)} + k_{3(n)}}
{\kappa_{n}}) + I_{2}^{2}(\frac{k_{2(n)} + k_{3(n)}}
{\kappa_{n}})] ,\\
&& A_{2n}^{(2)} = \frac{(5N + 22)}{54}
[I_{4}( \frac{k_{i(n)}}{\kappa_{n}}) + 5   \;permutations] ,\\
&& B_{2n} = \frac{(N+2)}{18}I_{3}(\frac{k_{(n)}}{\kappa_{n}}) ,\\
&& B_{3n} =
\frac{(N+2)(N+8)}{108}I_{5}(\frac{k_{n}}{\kappa_{n}}) ,\\
&& C_{1n} = \frac{N+2}{18}[ I_{2}(\frac{k_{1(n)} + k_{2(n)}}
{\kappa_{n}}) +  I_{2}(\frac{k_{1(n)} + k_{3(n)}}
{\kappa_{n}}) + I_{2}(\frac{k_{2(n)} + k_{3(n)}}
{\kappa_{n}})] ,\\
&& C_{2n}^{(1)} = \frac{(N+2)^{2}}{108}
[I_{2}^{2}(\frac{k_{1(n)} + k_{2(n)}}{\kappa_{n}})
+  I_{2}^{2}(\frac{k_{1(n)} + k_{3(n)}}
{\kappa_{n}}) + I_{2}^{2}(\frac{k_{2(n)} + k_{3(n)}}
{\kappa_{n}})] ,\\
&& C_{2n}^{(2)} = \frac{N+2}{36}[I_{4}(\frac{k_{i(n)}}{\kappa_{n}}) + 5   \;permutations].
\end{eqnarray}
\end{mathletters}

We have at hand all we need to determine the normalization constants 
at least at two-loop order. Requiring minimal subtraction for the 
renormalized vertex parts above listed, it can be verified that all 
the logarithmic integrals depending upon each arbitrary 
(nonvanishing) external momenta subspace appearing in $I_{2}, I_{3}, I_{4}$,
and $I_{5}$ cancell out. This leads to the following expressions for the 
normalization functions and coupling constants:
\begin{mathletters}
\begin{eqnarray}
&& u_{0n} = u_{n}(1 + \frac{(N+8)}{6 \epsilon_{L}} u_{n}
+ [\frac{(N+8)^{2}}{36 \epsilon_{L}^{2}} - \frac{(3N+14)}{24
\epsilon_{L}}] u_{n}^{2}), \\
&& Z_{\phi (n)} = 1 - \frac{N+2}{144 \epsilon_L} u_{n}^{2}
+ [-\frac{(N+2)(N+8)}{1296  \epsilon_{L}^{2}} + \frac{(N+2)(N+8)}{5184
\epsilon_{L}}] u_{n}^{3}, \\
&& \bar{Z}_{\phi^{2} (n)} = 1 + \frac{N+2}{6 \epsilon_L} u_{n}
+ [\frac{(N+2)(N+5)}{36 \epsilon_{L}^{2}} - \frac{(N+2)}{24
\epsilon_{L}}] u_{n}^{2}).
\end{eqnarray}
\end{mathletters}

Using the renormalization constants we obtain:
\begin{eqnarray}
&& \gamma_{\phi (n)} = n [\frac{(N+2)}{72}u_{n}^{2}
- \frac{(N+2)(N+8)}{1728}u_{n}^{3}],\\
&& \bar{\gamma}_{\phi^{2} (n)} = n \frac{(N+2)}{6} u_{n}
[ 1 - \frac{1}{2} u_{n}].
\end{eqnarray}

The fixed points are defined by $\beta_{n}(u_{n}^{*}) =
0$. Then,  we learn that the fixed points generated by
renormalization group transformations over $\kappa_{1}$,..., and 
$\kappa_{L}$ are the same, namely
\begin{equation}
u_{n}^{\ast}=\frac{6}{8 + N}\,\epsilon_L\Biggl\{1 + \epsilon_L
\,\Biggl[\frac{(9N + 42)}{(8 + N)^{2}}\Biggr]\Biggr\}\;\;.
\end{equation}

When this result is replaced in the renormalization constants
at the fixed point it yields $\gamma_{\phi (n)}^{*}= \eta_{n}$,
where $\eta_{n}$ are given by Eqs. (71) and (73). In addition,
we have
\begin{equation}
\bar{\gamma^{*}}_{\phi^{2} (n)} = n \frac{(N+2)}{(N+8)} \epsilon_{L}
[ 1 + \frac{6(N+3)}{(N+8)^{2}} \epsilon_{L}].
\end{equation}
The reader can verify that the exponents $\nu_{n}$ encountered by 
using the last equation are the same as those obtained via normalization 
conditions Eqs. (72) and (74). This proves the consistency of either 
renormalization scheme for the anisotropic Lifshitz critical behaviors.

\subsection{Discussion}

The generic $L$th character Lifshitz critical behavior naturally 
extends the comprehension of the usual second character Lifshitz 
criticality. The latter is characterized by one 
noncompeting subspace and only one 
competing subspace, whereas the former is characterized by several types 
of competing subspaces. The expressions for the critical exponents can be 
analysed to extract further information concerning those systems. For 
instance, when $m_{3}\neq 0$ and $m_{2}=m_{4}=...=m_{L}=0$, the third 
character behavior is recovered and correctly reduces to the Ising-like 
behavior for $m_{3}=0$. The main characteristic of generic third character 
behavior is that there are  $m_{2}, m_{3}\neq 0$ competing axes with 
$m_{4}=...=m_{L}=0$, and so on. From a phenomenological perspective in 
magnetic systems, may be it is worthy to assemble all magnetic 
materials presenting Lishitz critical behavior and analyse their critical 
exponents. Choosing those alloys in the same conjectured universality 
class, the greater the difference in their critical exponents the more 
likely they are in an alternative universality class contained in the CECI 
model analysed here.

It is interesting to note that exact strong anisotropic scale invariance 
\cite{He} is valid in the CECI model as a result of the generalized 
orthogonal approximation. Since the model describes the physics of 
short ranged competing systems, the limit $L \rightarrow \infty$ in the 
$Lth$ charater Lifshitz point should not be taken, since it would describe 
{\it long range} competing systems. Nevertheless, since no restriction on 
$L$ was made in the beginning of the discussion, the CECI model can be viewed 
as describing a particular type of long range competing interactions. If we 
go on and take this limit in the expression of the critical exponents we 
find that $\eta_{L}$ tends to infinity, whereas $\nu_{L}\rightarrow 0$ as a 
consequence of the strong anisotropic scale invariance. This implies that 
close to the Lifshitz critical temperature the correlation length 
$\xi_{L}$ does not diverge in that limit, instead of having a 
usual power law divergence. This fact is a nonperturbative result valid 
to all orders in perturbation theory within the context of the 
$\epsilon_{L}$-expansion.

Another feature emerges from this limit by 
looking at the critical dimension $d_{c}=4+m_{\infty}$. (Recall that in the 
$L$th anisotropic character critical behavior the system has $m_{L}$ 
competing axes and $d-m_{L}$ noncompeting space directions, i.e., 
$m_{2}=...=m_{L-1}=0$.) This limit yields a mechanism which is a natural new 
way to study extra dimensions without destroying the renormalizability of 
the corresponding field theory, while retaining the nontrivial aspect of 
the fixed point. In spite of the divergence of the anomalous dimension and 
the vanishing of the correlation length exponent along the $m_{L}$ competing 
directions in this limit, all other exponents are well behaved and have the 
same value of those corresponding to space directions without competition, 
as can be seen explicitly by the expressions for the exponents. This is so 
since the anisotropic scaling laws only contain the safe combination 
$L \nu_{L}$ which is always finite in the limit $L \rightarrow \infty$. 
Once more, note that $m_{\infty}\rightarrow 0$ 
reduces to the previous $\phi^{4}$ universality classes. Recall that the 
isotropic case is characterized by $d=m_{L}=4L$. Some care 
care must be taken in the interpretation of this case in the limit 
$L\rightarrow \infty$, for the approach from the anisotropic 
($0\leq m_{L}<4L$) to the isotropic case takes infinite steps, which 
is not as simple as in the case when $L$ is finite. This point deserves its 
own analysis for future work.

On the other hand, the last few years have witnessed some ideas in 
quantum field theories that resemble very much issues contained in 
the analysis of Lifshitz critical phenomena. The closest analogy with 
the Lifshitz field theoretic tools presented here is the very recent 
idea of ghost condensation producing a consistent modification of 
gravity in the infrared (long distance limit) \cite{AH,Tsu-Sa,Amen}. 
In this framework, gravity is modified to have attractive {\it as well as} 
repulsive components. The latter mimics a kind of dark energy \cite{KNg}, 
whose ghost condensate is a physical fluid arising from a theory where a real 
scalar field changes with a constant velocity. The ghost condensate 
appears as the physical scalar excitation around the background generated 
by the scalar field whose vacuum expectation value is defined by a constant 
value of its time derivative. Consequently, the effective action for the field 
representing the ghost condensate has kinetic terms with quadratic time 
derivatives and quartic space derivatives of the field, therefore breaking 
Lorentz invariance \cite{AH}. The instability in momentum space is manifest 
in the absence of kinetic terms quadratic in the 
space derivatives of the ghost condensate. This 
characteristic is a result of the competing nature between the attractive 
and repulsive components of the gravitational force. This is a precise 
analogy with the Lifshitz critical region in the case of the usual second 
character behavior included in the discussion given in the present paper. The 
utilization of an analogous reasoning leads us to conclude that 
the CECI model can be used to extract further insights from these new 
effective quantum field theories where Lorentz invariance is broken. It 
permits generalizations for the ghost condensate when higher 
powers in space derivatives of this field are present in its corresponding 
effective action. For example, at large distances the case of quadratic time 
derivatives and 6th space derivatives in the kinetic term would correspond 
to a gravitational interaction with attractive/repulsive/attractive competing 
components and so on. Further analysis of the model might be helpful to 
address the perturbative calculations regarding the ghost condensate in that 
scalar field background. 

Numerical methods to probe the results obtained here are in their infancy 
for the CECI model. Earlier Monte Carlo simulations for a uniaxial third 
character behavior were performed \cite{Se1} and the existence of the 
corresponding Lifshitz point at nonzero temperature was established. 
Unfortunately, no critical exponent was determined for the third character 
Lifshitz point. Since then, perhaps due to the fact that these higher 
character criticalities were not well understood theoretically, these 
methods are still waiting for more investigations. Recently, a model with 
antiferromagnetic couplings between nearest neighbors as well as 
antiferromagnetic exchange interactions between third neighbors was studied 
using Monte Carlo simulations and some quantum properties were 
investigated \cite{Ca-Sa}. No ferromagnetic couplings appear among arbitrary 
neighbors. Even though there is a quantum Lifshitz point there, it pertains 
to a universality class which might be different from that representing 
third character critical points discussed in the present paper. Further 
numerical studies motivated by the CECI model are necessary to understand 
completely the classical and quantum properties of its critical regions.

\section{Isotropic critical exponents in the orthogonal approximation}

We now address the calculation of critical exponents using the generalized 
orthogonal approximation from the results obtained in Appendix B. As we have 
seen in the calculation of Feynman 
integrals, this problem can also be tackled without performing any 
approximation during all steps of the calculation. This 
technique shall be postponed until next section. In a way or another, 
the approach is equivalent to treat each competing subspace separately. 
Though very similar, the framework of this section turns out to be more 
economical than that reported in the anisotropic cases. 

\subsection{Critical exponents in normalization conditions}

The basic definitions of the bare coupling constants and renormalization 
functions were encountered before and are given by
\begin{mathletters}
\begin{eqnarray}
&& u_{0n} = u_{n} (1 + a_{1n} u_{n} + a_{2n} u_{n}^{2}) ,\\
&& Z_{\phi (n)} = 1 + b_{2n} u_{n}^{2} + b_{3n} u_{n}^{3} ,\\
&& \bar{Z}_{\phi^{2} (n)} = 1 + c_{1n} u_{n} + c_{2n} u_{n}^{2} ,
\end{eqnarray}
\end{mathletters}
where the constants $a_{in}, b_{in}, c_{in}$ depend on Feynman
integrals at the symmetry point called $SP_{n}$. Let 
$\kappa_{n}$ denote the competing $m_{n}$-dimensional subspace in the 
isotropic cases.

The beta-function and renormalization constants can be expressed in 
the form 
\begin{mathletters}
\begin{eqnarray}
&& \beta_{n}  =  - \epsilon_{n}u_{n}[1 - a_{1n} u_{n}
+2(a_{1n}^{2} -a_{2n}) u_{n}^{2}],\\
&& \gamma_{\phi (n)} = - \epsilon_{n}u_{n}[2b_{2n} u_{n}
+ (3 b_{3n}  - 2 b_{2n} a_{1n}) u_{n}^{2}],\\
&& \bar{\gamma}_{\phi^{2} (n)} = \epsilon_{n}u_{n}[c_{1n}
+ (2 c_{2n}  - c_{1n}^{2} - a_{1n} c_{1n})u_{n}    ].
\end{eqnarray}
\end{mathletters}
The coefficients above obtained as functions of the 
integrals calculated at the symmetry point read
\begin{mathletters}
\begin{eqnarray}
&& a_{1n} = \frac{N+8}{6 \epsilon_{n}}[1 + \frac{1}{2n} \epsilon_{n}] ,\\
&& a_{2n} = (\frac{N+8}{6 \epsilon_{n}})^{2}
+ [\frac{2N^{2} + 23N + 86}{72n \epsilon_{n}}] ,\\
&& b_{2n} = -\frac{(N+2)}{144n \epsilon_{n}}[1 + \frac{5}{4n} \epsilon_{n}], \\
&& b_{3n} = -\frac{(N+2)(N+8)}{1296n \epsilon_{n}^{2}} -
5 \frac{(N+2)(N+8)}{5184n^{2} \epsilon_{n}}, \\
&& c_{1n} = \frac{(N+2)}{6 \epsilon_{n}}[1 + \frac{1}{2n} \epsilon_{n}], \\
&& c_{2n} = \frac{(N+2)(N+5)}{36 \epsilon_{n}^{2}}
+ \frac{(N+2)(2N+7)}{72n \epsilon_{n}}.
\end{eqnarray}
\end{mathletters}
The equation $\beta_{n}(u_{n}^{*}) = 0$ defines the fixed point. Thus, 
we find
\begin{equation}
u_{n}^{\ast}=\frac{6}{8 + N}\,\epsilon_{n}\Biggl\{1 + \epsilon_{n}
\frac{1}{n}\Biggl[ - \frac{1}{2} + \frac{(9N + 42)}{(8 + N)^{2}}\Biggr]\Biggr\}\;\;.
\end{equation}
We stress that this fixed point is different from that arising in the
anisotropic behavior and cannot be obtained from it within the 
$\epsilon_{L}$-expansion described above. The 
functions $\gamma_{\phi (n)}$ and $\bar{\gamma}_{\phi^{2} (n)}$ are found 
to be
\begin{eqnarray}
&& \gamma_{\phi (n)} = \frac{(N+2)}{72n} [1 + \frac{5}{4n}
\epsilon_{n}]u_{n}^{2} - \frac{(N+2)(N+8)}{864 n^{2}} u_{n}^{3}, \\
&& \bar{\gamma}_{\phi^{2} (n)} = \frac{(N+2)}{6} u_{n}[1
+ \frac{1}{2n} \epsilon_{n} - \frac{1}{2n} u_{n}].
\end{eqnarray}
When the fixed point is replaced inside these equations, using the
relation among these functions and the critical exponents $\eta_{n}$ and
$\nu_{n}$, we find:
\begin{eqnarray}
&& \eta_{n}= \frac{1}{2n} \epsilon_{n}^{2}\,\frac{N + 2}{(N+8)^2}
[1 + \epsilon_{n}\frac{1}{n}(\frac{6(3N + 14)}{(N + 8)^{2}} - \frac{1}{4})] ,\\
&& \nu_{n} =\frac{1}{2n} + \frac{(N + 2)}{4 n^{2}(N + 8)} \epsilon_{n}
+  \frac{1}{8 n^{3}}\frac{(N + 2)(N^{2} + 23N + 60)} {(N + 8)^3} \epsilon_{n}^{2}.
\end{eqnarray}
The coefficient of the $\epsilon_{n}^{2}$ term in the
exponent $\eta_{n}$ is positive, consistent with its counterpart in the
anisotropic cases as well as in the Ising-like case. In the generalized 
orthogonal approximation the competing momenta are not sufficient to 
induce its change of sign. 

Now using the scaling relations derived for the isotropic case we obtain
immediately
\begin{eqnarray}
&& \gamma_{n} = 1 + \frac{(N+2)}{2n (N + 8)} \epsilon_{n}
+\frac{(N + 2)(N^{2} + 22N + 52)}{4 n^{2}(N + 8)^{3}} \epsilon_{n}^{2}, \\
&& \alpha_{n} = \frac{(4 - N)}{2n (N + 8)} \epsilon_{n}
- \frac{(N + 2)(N^{2} + 30N + 56)}{4 n^{2} (N + 8)^{3}} \epsilon_{n}^{2} ,\\
&& \beta_{n} = \frac{1}{2} - \frac{3}{2n (N + 8)} \epsilon_{n}
+ \frac{(N + 2)(2N + 1)}{2 n^{2} (N + 8)^{3}} \epsilon_{n}^{2} ,\\
&& \delta_{n} = 3 + \frac{1}{n} \epsilon_{n}
+ \frac{(N^{2} + 14N + 60)}{2 n^{2} (N + 8)^{2}} \epsilon_{n}^{2}.
\end{eqnarray}
The explicit dependence on the number of neighbors coupled through 
competing interactions is manifest in the above exponents. Hence, the 
universality classes for an arbitrary isotropic ($d=m_{n}$) competing 
system are determined by $(N,d,n)$. The interesting fact is that the 
results for the ordinary critical behavior are correctly recovered in the 
limit $n \rightarrow 1$. 

The orthogonal approximation is a good approximation even for isotropic 
higher character Lifshitz critical behaviors for at least three main 
reasons. It preserves the homogenity of the Feynman integrals in the 
external momenta scales. Second, it indicates which parameters are important 
to describe the universality classes of isotropic systems. And last, 
but not least, it manifests one of the most important properties of 
competing systems, namely, the reduction to the Ising-like universality 
classes in the limit when all interactions beyond first neighbors are 
turned off.

It is important to emphasize a technical detail concerning the 
approximation just employed. The leading singularities 
from the one- and two-loop diagrams contributing to the 
four-point 1PI vertex part do not get modified from the usual 
$\phi^{4}$. On the other hand, the leading singularities of the two- and 
three-loop for the two-point 1PI vertex function do get a factor of 
$\frac{1}{n}$ with respect to those from the usual critical behavior. 
These integrals also do not change signs under the generalized 
orthogonal approximations. The calculation performed without 
approximations shows that the leading singularities of diagrams contributing 
to the 1PI two-point function change in a more complicated way, also 
changing sign depending on the value of $n$. We shall compare the 
differences in the values of the exponents utilizing the orthogonal 
approximation and the exact treatment later on. In particular, we shall 
perform a numerical analysis for the isotropic second character behavior 
to understanding the deviations in both approaches.   

In order to check these results, let us analyse the situation using the 
minimal subtraction scheme.

\subsection{Critical exponents in minimal subtraction}
Minimal subtraction of dimensional poles in the renormalized vertex 
$\Gamma_{R (n)}^{(4)}$ can be used to show that all momentum-dependent 
logarithimic integrals are eliminated in the renormalization process 
leading the bare dimensionless coupling constant to be written as
\begin{equation}
u_{0n} = u_{n}[1 + \frac{(N+8)}{6 \epsilon_{n}} u_{n} +
(\frac{(N+8)^{2}}{36 \epsilon_{n}^{2}} - \frac{(3N+14)}{24n \epsilon_{n}})
u_{n}^{2}].
\end{equation}
The fixed point can be found out
\begin{equation}
u_{n}^{*} = \frac{6}{(N+8)} \epsilon_{n} +
\frac{18(3N+14)}{n (N+8)^{3}} \epsilon_{n}^{2}.
\end{equation}
In addition, the normalization constants are given by:
\begin{eqnarray}
&& Z_{\phi (n)} = 1 - \frac{(N+2)}{144n \epsilon_{n}} u_{n}^{2} \nonumber\\
&& + [-\frac{(N+2)(N+8)}{1296n \epsilon_{n}^{2}} + \frac{(N+2)(N+8)}{5184 n^{2}
\epsilon_{n}}] u_{n}^{3},\\
&& \bar{Z}_{\phi^{2} (n)} = 1 + \frac{(N+2)}{6 \epsilon_{n}} u_{n} \nonumber\\
&& + [\frac{(N+2)(N+5)}{36 \epsilon_{n}^{2}} - \frac{(N+2)}{24n \epsilon_{n}}]
u_{n}^{2}.
\end{eqnarray}

Consequently, the functions $\gamma_{\phi (n)}$ and 
$\bar{\gamma}_{\phi^{2} (n)}$ can be obtained directly
\begin{mathletters}
\begin{eqnarray}
&& \gamma_{\phi (n)} = \frac{(N+2)}{72n} u_{n}^{2} - \frac{(N+2)(N+8)}{1728 n^{2}} u_{n}^{3},\\
&& \gamma_{\phi^{2} (n)} = \frac{(N+2)}{6}(u_{n} - \frac{1}{2n} u_{n}^{2}).
\end{eqnarray}
\end{mathletters}

Substitution of these expressions in the function 
$\gamma_{\phi (n)}^{*}$ at the fixed point, one obtains the value 
of $\eta_{n}$ as obtained in (94). The function
$\bar{\gamma}_{\phi^{2} (n)}^{*}$ at the fixed point reads
\begin{equation}
\bar{\gamma}_{\phi^{2} (n)}^{*} = \frac{(N+2)}{(N+8)} \epsilon_{n} [1 +
\frac{6(N+3)}{n (N+8)^{2}} \epsilon_{n}],
\end{equation}
that is the same as the one obtained in the fixed point using normalization
conditions. Therefore, it yields the same critical exponent $\nu_{n}$ 
from (95) as can be easily cheched. This constitutes the equivalence 
between the two renormalization schemes.

\section{Isotropic critical exponents in the exact calculation}
We now turn our attention to the calculation of the isotropic critical 
exponents exactly, i.e., not using the orthogonal approximation. The 
algorithm we need to employ to obtain the critical indices is pretty 
much the same as that used in the orthogonal approximation, since the 
renormalization program and the scaling laws are approximation independent. 
The difference is that one should replace the Feynman integrals by their 
exact values already calculated in Appendix C. As we are going to discuss 
the case $n=2$ in the next section using normalization conditions and minimal 
subtraction, we shall take normalization conditions in this section. 
Nevertheless, with the resources furnished in the text, the reader should 
be able to check the exponents using minimal subtraction as well.  

Using the definitions given for the power series of the bare 
dimensionless coupling constant and renormalization functions in terms 
of the dimensionless renormalized coupling constant, we find the following 
values for the coefficients of each power of $u_{n}$: 
\begin{mathletters}
\begin{eqnarray}
&& a_{1n} = \frac{N+8}{6 \epsilon_{n}}[1 + D(n) \epsilon_{n}] ,\\
&& a_{2n} = (\frac{N+8}{6 \epsilon_{n}})^{2}
+ \frac{1}{\epsilon_{n}}[\frac{(N^{2} + 21N + 86) D(n)}{18}  
- \frac{5N+22}{18} 
- \frac{(N+2) (-1)^{n} \Gamma(2n)^{2}}{36 \Gamma(n+1) \Gamma(3n)}\nonumber\\
&& -\frac{5N+22}{18}(\sum_{p=1}^{2n-2} \frac{1}{2n-p} 
- 2 \sum_{p=1}^{n-1} \frac{1}{n-p})],\\
&& b_{2n} = (-1)^{n} \frac{(N+2) \Gamma(2n)^{2}}
{72 \Gamma(n+1) \Gamma(3n) \epsilon_{n}}[1 + (D(n) + \frac{3}{4} 
- \frac{3}{2} \sum_{p=2}^{2n-1} \frac{1}{p} \nonumber\\
&& + \frac{1}{2} \sum_{p=1}^{n-1} \frac{1}{n-p} 
+ \frac{3}{2} \sum_{p=3}^{3n-1} \frac{1}{p})\epsilon_{n}], \\
&& b_{3n} = (-1)^{n}\frac{(N+2)(N+8)\Gamma(2n)^{2}}{108 \Gamma(n+1) \Gamma(3n)}[\frac{1}{6 \epsilon_{n}^{2}} 
+ \frac{1}{\epsilon_{n}}(\frac{D(n)}{3} + \frac{1}{24}\nonumber\\
&& - \frac{1}{12} \sum_{p=2}^{2n-1} \frac{1}{p} 
- \frac{1}{12} \sum_{p=1}^{n-1} \frac{1}{n-p} 
+\frac{1}{12} \sum_{p=3}^{3n-1} \frac{1}{p})], \\
&& c_{1n} = \frac{(N+2)}{6 \epsilon_{n}}[1 + D(n) \epsilon_{n}], \\
&& c_{2n} = \frac{(N+2)(N+5)}{36 \epsilon_{n}^{2}}
+ \frac{(N+2)}{6\epsilon_{n}}[ \frac{(N+8)}{3} D(n) -\frac{1}{2}(1 + D(n) 
+ \sum_{p=1}^{2n-2} \frac{1}{2n-p} - 2 \sum_{p=1}^{n-1} \frac{1}{n-p})].
\end{eqnarray}
\end{mathletters}
Here $D(n)= \frac{1}{2} \psi(2n) - \psi(n) + \frac{1}{2} \psi(1)$.
The fixed point at two-loop level is defined as the zero of the 
$\beta$-function and it is given by the following expression:
\begin{eqnarray}
&& u_{n}^{*} = \frac{6 \epsilon_{n}}{(N+8)}[1 + 
\epsilon_{n}[\frac{1}{(N+8)^{2}}
(\frac{(-1)^{n} \Gamma(2n)^{2} (2N+4)}{\Gamma(n+1) \Gamma(3n)} + 
(20N+88)(1-D(n)) - D(n) \nonumber\\
&& + \frac{(20N+88)}{(N+8)^{2}}(\sum_{p=1}^{2n-2} \frac{1}{2n-p} 
- 2 \sum_{p=1}^{n-1} \frac{1}{2n-p})]].
\end{eqnarray}
The renormalization functions $\gamma_{\phi (n)}(u_{n})$ and 
$\bar{\gamma}_{\phi^{2} (n)}(u_{n})$ can be expressed in the 
following simple manner in terms of $u_{n}$:
\begin{eqnarray}
&&\gamma_{\phi(n)}(u_{n}) = (-1)^{n+1} 
\frac{(N+2) \Gamma(2n)^{2}}{36 \Gamma(n+1) \Gamma(3n)}[1 
+ (D(n) + \frac{3}{4} - \frac{3}{2} \sum_{p=2}^{2n-1} \frac{1}{p} 
+ \frac{1}{2} \sum_{p=1}^{n-1} \frac{1}{n-p} \nonumber\\
&& + \frac{3}{2} \sum_{p=3}^{3n-1} \frac{1}{p})\epsilon_{n}]u_{n}^{2} 
+ (-1)^{n+1} \frac{(N+2)(N+8)\Gamma(2n)^{2}}{216 \Gamma(n+1) \Gamma(3n)}
[ - \frac{1}{2} + \sum_{p=2}^{2n-1} \frac{1}{p}
- \sum_{p=1}^{n-1} \frac{1}{n-p} - \sum_{p=3}^{3n-1} \frac{1}{p}]u_{n}^{3};
\end{eqnarray}
\begin{eqnarray}
\bar{\gamma}_{\phi^{2}(n)}(u_{n}) = \frac{(N+2)}{6}u_{n}[1 + 
D(n) \epsilon_{n}  -D(n) u_{n}].
\end{eqnarray}
Substitution of the fixed point in the first equation gives directly the 
anomalous dimensions $\eta_{n}$. Using the scaling law relating the 
second expression with the exponents $\eta_{n}$ and $\nu_{n}$, one 
obtains the exponent $\nu_{n}$. Thus, we get 
\begin{eqnarray}
\eta_{n} = (-1)^{n+1} 
\frac{(N+2)\Gamma(2n)^{2}}{(N+8)^{2} \Gamma(n+1) \Gamma(3n)} \epsilon_{n}^{2} 
+(-1)^{n+1}\frac{(N+2)\Gamma(2n)^{2} F(N,n)}{(N+8)^{2} \Gamma(n+1) \Gamma(3n)} 
\epsilon_{n}^{3};
\end{eqnarray}
where 
\begin{eqnarray}
&&F(N,n)=[((-1)^{n} \frac{\Gamma(2n)^{2} (4N+8)}{\Gamma(n+1) 
\Gamma(3n)} + (40N+176)D(n)) \frac{1}{(N+8)^{2}}\nonumber\\
&&- \frac{3}{4} 
- \sum_{p=1}^{2n-1} \frac{1}{p} + \frac{1}{2} \sum_{p=1}^{n-1} 
\frac{1}{p} + \frac{1}{2} \sum_{p=1}^{3n-1} \frac{1}{p}].
\end{eqnarray}
Analogously,
\begin{eqnarray}
&& \nu_{n} = \frac{1}{2n} + \frac{(N+2)}{4n^{2}(N+8)}\epsilon_{n} 
+ \frac{(N+2)}{4n^{2}(N+8)^{3}} \epsilon_{n}^{2}
[(-1)^{n} (N-4) \frac{\Gamma(2n)^{2}}{\Gamma(n+1) \Gamma(3n)} 
+ \frac{(N+2)(N+8)}{2n} \nonumber\\
&& + (14N+40)D(n)].
\end{eqnarray}

Now, we use the scaling relations to obtain the remaining critical 
exponents. We find:
\begin{eqnarray}
&&\gamma_{n}= 1 + \frac{(N+2)}{2n(N+8)}\epsilon_{n} 
+ \frac{(N+2)}{4n^{2}(N+8)^{3}}\epsilon_{n}^{2}[(-1)^{n}\frac{2n(2N+4) 
\Gamma(2n)^{2}}{\Gamma(n+1) \Gamma(3n)} + (N+2)(N+8)\nonumber\\
&& + 2n(14N+40)D(n)],
\end{eqnarray}
\begin{eqnarray}
&&\alpha_{n}= \frac{(4-N)}{2n(N+8)}\epsilon_{n} 
- \frac{(N+2)}{4n^{2}(N+8)^{3}}\epsilon_{n}^{2}[(-1)^{n}\frac{4n(N-4) 
\Gamma(2n)^{2}}{\Gamma(n+1) \Gamma(3n)} + (N-4)(N+8)\nonumber\\
&& + 4n(14N+40)D(n)],
\end{eqnarray}
\begin{eqnarray}
&&\beta_{n}= \frac{1}{2} - \frac{3}{2n(N+8)}\epsilon_{n} 
+ \frac{(N+2)}{4n^{2}(N+8)^{3}}\epsilon_{n}^{2}[(-1)^{n+1}
\frac{12n \Gamma(2n)^{2}}{\Gamma(n+1) \Gamma(3n)} -3(N+8)\nonumber\\
&& + n(14N+40)D(n)],
\end{eqnarray}
\begin{eqnarray}
&&\delta_{n} = 3 + \frac{\epsilon_{n}}{n} 
+ \frac{\epsilon_{n}^{2}}{2 n^{2} (N+8)^{2}}[(N+8)^{2} + (-1)^{n} \frac{4n(N+2)\Gamma(2n)^{2}}{\Gamma(n+1) \Gamma(3n)}].
\end{eqnarray}

Two important features emerge from this exact picture of the critical 
exponents shown above. It is instructive to discuss the exponent $\eta_{n}$. 
First, the sign of the lowest ($O(\epsilon_{n}^{2})$) 
contribution to the exponent $\eta_{n}$ is determined by the value of $n$. 
Remember that this fact merely reflects the change of sign of the two-point 
diagrams depending on the value of $n$. Second, instead of a global factor 
proportional to $n$ the exact solution has a dependence on $n$ coming from 
a product of $\Gamma$ functions having $n$ on their arguments as well as a 
finite sum with terms which depend on $n$. This property is valid for
all critical exponents and appears explicitly at two- and higher-loop 
corrections. This is quite a remarkable new feature of arbitrary isotropic 
competing systems, going beyond the simple polynomial dependence on $n$ 
found using the orthogonal approximation. 

The universality class reduction to that from the Ising-like one in the 
limit $n \rightarrow 1$ is obvious. Moreover, the case $n=2$ correctly 
reproduces the solution of an earlier two-loop calculation. Actually, the 
results given above extend the calculation of $\eta_{n}$ for arbitrary 
$n$ to include $O(\epsilon_{n}^{3})$ corrections. In addition, all critical 
exponents presented here at least up to two-loop order generalize all 
previous results for arbitrary higher character isotropic Lifshitz 
points. Therefore, at the loop order considered the results above 
represent the complete solution to the critical exponents of the CECI 
model for arbitrary types of isotropic competing interactions.

A numerical analysis for comparison between the results obtained either 
using the orthogonal approximation or in the exact calculation are 
worthwhile. The case $n=2$ will be analysed later. Here we display the 
numerical results for the cases $n=3,4,5,6$. 

It would be appropriate to calculate the exponents in either approach in  
a particular case in order to see if the difference is meaningful. The 
usual $\epsilon$-expansion is good enough for the numerical estimation of 
critical exponents associated to three-dimensional critical systems even 
though the expansion parameter is not small. We can then ask ouselves if 
the same analogy is valid in order to extract concrete results from a fixed 
value of the space dimension and number of components of the order parameter 
field for arbitrary isotropic higher character Lifshitz points. We shall 
look at space dimensions which yields $\epsilon_{n}=1$ in analogy to the 
method used to extract numerical results from the ordinary 
$\epsilon-$expansion for three-dimensional systems.  

For an eleven-dimensional lattice, take $N=1$ and $n=3$ for the 
isotropic third character behavior. The orthogonal approximation yields 
$\eta_{3}=0.006$, $\nu_{3}=0.177$, $\alpha_{3}=0.046$, $\beta_{3}=0.446$, 
$\gamma_{3}=1.064$ and $\delta_{3}=3.385$. The exact calculation produces 
the exponents $\eta_{3}=0.002$, $\nu_{3}=0.174$, $\alpha_{3}=0.085$, 
$\beta_{3}=0.435$, $\gamma_{3}=1.046$ and $\delta_{3}=3.390$. The maximal 
deviation ($4.1\%$) occurs for the exponent $\alpha_{3}$ followed by the 
deviation in the $\gamma_{3}$ exponent ($1.8\%$) and an error in the 
exponent $\beta_{3}$ around $1\%$. The other exponents have deviations 
smaller than $0.5\%$.

Consider the case $N=1$, $d=15$, $n=4$. The results from the orthogonal 
approximation are: $\eta_{4}=0.005$, $\nu_{4}=0.131$, $\alpha_{4}=0.036$, 
$\beta_{4}=0.459$, $\gamma_{4}=1.046$ and $\delta_{4}=3.279$. The exact 
calculation yields $\eta_{4}=-0.001$, $\nu_{4}=0.129$, $\alpha_{4}=0.058$, 
$\beta_{4}=0.449$, $\gamma_{4}=1.029$ and $\delta_{4}=3.282$. The maximal 
deviations takes place in $\alpha_{4} (2.2\%)$, $\gamma_{4} (1.7\%)$ and 
$\beta_{4} (1\%)$, while the other exponents have errors smaller than 
$0.6\%$.

Let us examine the case $N=1$, $d=19$ and $n=5$. The isotropic exponents 
in the orthogonal approximation are $\eta_{5}=0.004$, $\nu_{5}=0.104$, 
$\alpha_{5}=0.030$, $\beta_{5}=0.467$, $\gamma_{5}=1.036$ and 
$\delta_{5}=3.219$. The exact exponents are $\eta_{5}=0.001$, $\nu_{5}=0.102$, 
$\alpha_{5}=0.064$, $\beta_{5}=0.475$, $\gamma_{5}=1.020$ and 
$\delta_{5}=3.220$. The maximal deviations are in $\alpha_{5} (3.4\%)$, 
and $\gamma_{5} (1.6\%)$, whereas the remaining exponents have errors 
smaller than $0.8\%$\footnote{The maximal deviations also occur for 
$\alpha_{n}$ and $\gamma_{n}$ for $n=6,7,8$ and decreases for increasing $n$. 
For $n=8$, the maximal error for $\alpha_{8}$ is about $2.7\%$, whereas the 
maximal deviation for $\gamma_{8}$ is $1.3\%$.}.  

This analysis leads us to conclude that the orthogonal approximation is 
very precise to predict numerical values in specific situations, since 
the deviations are negligible when compared with the exact calculation. 
Moreover, the above data indicate that the deviations are under control no 
matter how the number of neighbors increases.

The extra insight from the field-theoretic viewpoint is that the 
more neighbors are introduced and coupled trough isotropic competing 
interactions the more the space dimensions seem to split. Then, one 
line with competing interactions up to second neighbors behaves for all 
practical purposes as having 2 dimensions. Pushing the argument further, a 
line with $n$ neighbors interacting through competing interactions seem to 
have $n$ dimensions. This is a striking general property of the 
field theory under consideration: in the massless limit presented here, when 
the free critical propagator is proportional to a $2n$th power of 
momenta each space direction ``splits'' in $n$ dimensions. This is 
a kind of degeneracy in the dimension, which can only be unveiled 
when more participants (neighbors) are allowed to take place in the 
isotropic competing system. This is a new aspect of systems 
whose Lagrangians have kinetic terms described by higher derivatives of 
the field. Further implications of this phenomenon remain to be 
investigated.

Now, let us show that these findings easily 
reproduce and extend the original calculation done by Hornreich, Luban 
and Shtrikman \cite{Ho-Lu-Sh} for the $n=2$ case.

\section{Exact isotropic exponents for the second character case}
In the last section we found the critical exponents for the isotropic case 
when competing interactions among arbitrary neighbors are allowed. We now 
discuss its reduction for the usual second character Lifshitz critical 
behavior. Since we want to compare our results with other which already 
appeared in the literature we shall derive the critical indices using 
normalization conditions and minimal subtraction of poles. 

\subsection{Critical exponents in normalization conditions}
The fixed point can be calculated by replacing $n=2$ in Eq.(107) in 
order to obtain:
\begin{eqnarray}
&&u_{2}^{*} = \frac{6 \epsilon_{2}}{(N+8)}(1 - \frac{1}{3} \epsilon_{2}
[\frac{(41N+202)}{(N+8)^{2}} - \frac{1}{4}]).
\end{eqnarray}
The renormalization functions $\gamma_{\phi (2)}(u_{2})$ and 
$\bar{\gamma}_{\phi^{2} (2)}(u_{2})$ can be expressed in the 
following simple manner in terms of $u_{2}$:
\begin{eqnarray}
&&\gamma_{\phi(2)}(u_{2}) = - \frac{(N+2)}{240}[1 + 
\frac{131}{120} \epsilon_{2}]u_{2}^{2} 
+ \frac{29 (N+2)(N+8)}{28800}u_{2}^{3};
\end{eqnarray}
\begin{eqnarray}
&&\bar{\gamma}_{\phi^{2}(2)}(u_{2}) = \frac{(N+2)}{6}u_{2}[1 - 
\frac{1}{12} \epsilon_{2} + \frac{1}{12}u_{2}].
\end{eqnarray}
Substitution of the fixed point in the first equation gives directly the 
anomalous dimensions $\eta_{2}$. Thus, we get 
\begin{eqnarray}
&&\eta_{2} = - \frac{3(N+2)}{20(N+8)^{2}} \epsilon_{2}^{2} 
+ \frac{(N+2)}{10(N+8)^{2}}[\frac{(41N+202)}{10(N+8)^{2}} + \frac{23}{80}] 
\epsilon_{2}^{3};
\end{eqnarray}
From the scaling law relating $\eta_{2}$ with $\nu_{2}$ we obtain:
\begin{eqnarray}
&&\nu_{2} = \frac{1}{4} + \frac{(N+2)}{16(N+8)}\epsilon_{2} 
+ \frac{(N+2)(15N^{2}+89N+4)}{960(N+8)^{3}} \epsilon_{2}^{2}
\end{eqnarray}
Using the scaling relations to obtain the remaining critical 
exponents, we find:
\begin{eqnarray}
&&\gamma_{2}= 1 + \frac{(N+2)}{4(N+8)}\epsilon_{2} 
+ \frac{(N+2)(15N^{2}+98N+76)}{240(N+8)^{3}}\epsilon_{2}^{2};
\end{eqnarray}
\begin{eqnarray}
&&\alpha_{2}= \frac{(4-N)}{4(N+8)}\epsilon_{2} 
+ \frac{(N+2)(-15N^{2}+62N+952)}{240(N+8)^{3}}\epsilon_{2}^{2};
\end{eqnarray}
\begin{eqnarray}
&&\beta_{2}= \frac{1}{2} - \frac{3}{4(N+8)}\epsilon_{2} 
- \frac{(N+2)(80N+514)}{240(N+8)^{3}}\epsilon_{2}^{2};
\end{eqnarray}
\begin{eqnarray}
&&\delta_{2} = 3 + \frac{1}{2} \epsilon_{2}
+ \frac{(5N^{2}+86N+332)}{40(N+8)^{2}} \epsilon_{2}^{2}.
\end{eqnarray}
Since the earlier calculations in the literature were performed using 
minimal subtraction, we shall discuss it next.   

\subsection{Critical exponents in minimal subtraction}

Now, let us obtain some renormalization functions at the fixed point 
using minimal subtraction. Since these objects are universal, finding 
them is equivalent to find the fixed point. Requiring minimal subtraction 
of the renormalized vertex $\Gamma_{R(2)}^{(4)}$, the cancellations among 
logarithmic integrals for arbitrary external momenta indeed take place as 
expected. It leads to the expression of the bare dimensionless coupling 
constant $u_{02}$ written in terms of the renormalized dimensionless 
coupling constant $u_{2}$, namely
\begin{equation}
u_{02}= u_{2}[1 + \frac{(N+8)}{6 \epsilon_{2}}u_{2} 
+ (\frac{(N+8)^{2}}{36 \epsilon_{2}^{2}} + \frac{(41N+202)}{2160 \epsilon_{2}})
u_{2}^{2}].
\end{equation}
From the definitions of the normalization constants, we can write the 
following expressions
\begin{eqnarray}
&& Z_{\phi (2)} = 1 + \frac{(N+2)}{480 \epsilon_{2}} u_{2}^{2} \nonumber\\
&& + \frac{(N+2)(N+8)}{4320}[\frac{1}{\epsilon_{2}^{2}} - \frac{23}{518400 
\epsilon_{2}}] u_{2}^{3},\\
&& \bar{Z}_{\phi^{2} (2)} = 1 + \frac{(N+2)}{6 \epsilon_{2}} u_{2} \nonumber\\
&& + [\frac{(N+2)(N+5)}{36 \epsilon_{2}^{2}} 
+ \frac{N+2}{144 \epsilon_{2}}]u_{2}^{2}.
\end{eqnarray}
Consequently, the functions $\gamma_{\phi (2)}(u_{2})$ and 
$\bar{\gamma}_{\phi^{2} (2)}(u_{2})$ read
\begin{eqnarray}
&&\gamma_{\phi(2)}(u_{2}) = - \frac{(N+2)}{240}u_{2}^{2} 
+ \frac{23 (N+2)(N+8)}{172800}u_{2}^{3};
\end{eqnarray}
\begin{eqnarray}
&&\bar{\gamma}_{\phi^{2}(2)}(u_{2}) = \frac{(N+2)}{6}u_{2}[1 + 
\frac{1}{12}u_{2}].
\end{eqnarray}
The fixed point can be calculated and shown to be
\begin{equation}
u_{2}^{*} = \frac{6}{(N+8)} \epsilon_{2} -
\frac{(41N+202)}{5 (N+8)^{3}} \epsilon_{2}^{2}.
\end{equation}
Replacement of this expression for the function 
$\gamma_{\phi (2)}(u_{2}^{*})$ gives precisely the exponent $\eta_{2}$ 
of the last section up to $O(\epsilon_{2}^{3})$. The corresponding 
expression for $\bar{\gamma}_{\phi^{2} (2)}(u_{2}^{*})$ is given by
\begin{equation}
\bar{\gamma}_{\phi^{2} (2)}^{*} = \frac{(N+2)}{(N+8)} \epsilon_{2} [1 -
\frac{(N+2)(13N+41)}{15 (N+8)^{2}} \epsilon_{2}].
\end{equation} 
The last expression is actually the same as that coming from 
normalization conditions, therefore leading to the same exponent 
$\nu_{2}$ whereas the remaining exponents are obtained using the scaling 
laws. Thus, the equivalence between the two renormalization schemes is 
complete.

\subsection{Discussion}

First of all, our results for the isotropic $n=2$ case generalizes those 
in the seminal work by Hornreich, Luban and Shtrikman 
\cite{Ho-Lu-Sh}. To see this, we make the identifications 
$\epsilon_{\alpha} \equiv \epsilon_{2}, \eta_{l4} \equiv \eta_{2}$ and 
$\nu_{l4} \equiv \nu_{2}$. The equations (10a,b) in \cite{Ho-Lu-Sh} are 
identical to our results for $\eta_{2}$ and $\nu_{2}$ Eqs.(120)-(121). 
The step forward within our method is the exponent $\eta_{2}$ which is 
obtained up to $O(\epsilon_{2}^{3})$ for the first time. Furthermore, 
using the scaling relations reported in our previous letter \cite{Leite4}, 
we found all critical exponents (Eqs.(122)-(125)) exactly, at least up to 
two-loop order which constitutes another new result for the usual isotropic 
case. 

Next, let us confront the results coming from the generalized 
orthogonal approximation with those from the exact solution at the same 
loop order. Both the exact and the approximated one-loop exponents are the 
same. This can be seen from the $n=2$ particular case or from the generic 
$n$ isotropic criticality by the direct examination of the explicit results 
shown in the present article. Two-loop deviations start at $n=2$ and for 
higher $n$. 

Consider the case $n=2$. Take a magnetic system which has $N=1$ in a 
seven-dimensional lattice and analyse the exponents in each case. We shall 
restrict ourselves to three significative algarisms in our discussion. 
First use the orthogonal approximation. At two-loop order, the correlation 
length exponent yields the result $\nu_{2}=0.276$, and the anomalous dimension 
is given by $\eta_{2}=0.009$. The susceptibility, specific heat, 
magnetization and magnetic field exponents are given by 
$\gamma_{2}=1.103$, $\alpha_{2}=0.061$, $\beta_{2}=0.418$, 
$\delta_{2}=3.616$. Now use the exact two-loop calculation. We obtain the 
following numerical values for the critical indices: $\nu_{2}=0.271$, 
$\eta_{2}=-0.006$, $\gamma_{2}=1.087$, $\alpha_{2}=0.100$, $\beta_{2}=0.406$, 
$\delta_{2}=3.631$. 

Therefore, for all the exponents the difference using either approach 
starts in the second significative algarisms. Specifically, the maximal 
error made by using the orthogonal approximation takes place for the 
specific heat exponent with a deviation of $3.9 \%$ , whereas the minimal 
error occurs in the correlation length exponent whose difference is 
approximately $0.5 \%$ . This is a strong evidence that the 
orthogonal approximation is very good to give reliable information 
for the isotropic case in this specific situation. This feature was 
already encountered for uniaxial anisotropic cases, where the 
approximation showed its usefulness for three-dimensional 
uniaxial systems. We hope that these results may stimulate the search 
for these exponents using Monte Carlo numerical simulations in the 
particular case of second character isotropic Lifshitz critical points. 

The remarkable agreement between the numerical results for the critical 
indices above mentioned either using the orthogonal approximation or the exact 
two-loop calculations corroborates previous conjectures that the orthogonal 
approximation is not only good to describe uniaxial systems pertaining to 
the anisotropic second character Lifshitz critical behaviors but also 
arbitrary isotropic higher character Lifshitz points.

In fact the numerical analysis can be carried out for higher-dimensional 
lattices for higher values of $n$.  Extending this argument, the case 
$L=4n$, $N=1$ and $d=L-1$ yields the same value for the expansion parameter 
and should not deviate very much when both calculations are compared. 
Perhaps the study of the most general arbitrary isotropic points via 
numerical tools might be worthwhile as well, since now we have satisfactory 
numerical results coming from a purely analytical field theoretical 
investigation.

\section{Conclusions and perspectives}

In this paper we have discussed the field theoretic description of the most 
general competing system, which has a simple lattice model representation 
named CECI model. It consists of a modified Ising model presenting the most 
general type of competing exchange couplings among arbitrary neighbors and 
includes other models previously reported. We have derived explicitly the 
critical exponents in the anisotropic as well as in the isotropic situations 
at least up to two-loop level. The CECI model and its field theory 
representation generalize the description of the second character Lifshitz 
universality classes in a nontrivial way. In particular, strong anisotropic 
scale invariance is exact up to the loop order considered here. The 
universality class reduction is a general property of both anisotropic and 
isotropic critical behaviors. It implies that when the interactions beyond 
first neighbors are turned off, the Ising-like universality classes are 
recovered. This feature is manifest in all exponents.

The anisotropic exponents were calculated by using the generalized 
orthogonal approximation. The calculation of loop integrals is 
consistent, since it is rooted in the physical property of homogeneity. 
It is required for a satisfactory renormalization group treatment with 
several independent relevant length scales represented by each correlation 
length associated to the several competing axes. The fixed point is the 
same irrespective of the competing axes under consideration. In close 
analogy to the second character case, this result is expected to hold in 
arbitrary perturbative orders. The second character 
Lifshitz exponents are easily recovered as a particular case of the generic 
anisotropic situation described in the paper. Although it is desirable to 
have an approach that does not require the use of approximations for the 
anisotropic behaviors, in the present moment it is not obvious. It is a 
consequence of the appearance of many competing subspaces simultaneously 
which makes the exact calculation (if not impossible) very difficult. We 
hope the techniques developed in the present work shed light on a quest 
for a solution without the necessity of approximations for the anisotropic 
cases.

The isotropic behaviors were calculated using two different trends to 
evaluate the Feynman diagrams. The first of them makes use of the orthogonal 
approximation. It can be noticed that the isotropic behavior cannot be 
obtained from any type of anisotropic behavior within the framework of the 
$\epsilon_{L}$-expansions developed in the present work even though the same 
kind of approximations are employed in both cases. This generalizes the 
previous situation taking place for second character Lifshitz points. Next, we 
attacked the diagrams without making any approximation. The result is 
that deviations in the calculation of the critical exponents start at 
two-loop level in comparison to the outcome provenient of the orthogonal 
approximation. We obtain as a particular case of the generic isotropic 
behavior the second character isotropic behavior, which extend 
the results first derived by Hornreich, Luban and Shtrikman 
\cite{Ho-Lu-Sh}. In this way, we obtain $\eta_{2}$ up to three-loop 
order and $\nu_{2}$ at two-loops as well as the remaining exponents via the 
scaling relations derived previously in \cite{Leite4}, a result which was 
lacking since the discovery of the second character isotropic Lifshitz point.

The most immediate application of the formalism just presented is the 
calculation of all universal amplitude ratios of certain quantities close 
to generic higher character Lifshitz points. It is amazing that a detailed 
scale theory for these quantities is lacking in the literature, even for 
the simple second character Lifshitz points. In fact, some results were 
presented for the susceptibility \cite{Leite5} and specific heat 
\cite{Leite6} amplitude ratios at one-loop order. The latter amplitude 
ratio proved to be reliable to explain the experimental result associated 
to the magnetic material $MnP$ \cite{Be}. Nevertheless, a thorough 
renormalization group analysis is necessary in order to have a better 
comprehension of the several scale transformations in each competing 
subspace and the role played by them in the treatment of generic amplitude 
ratios. In principle we can extend the formalism presented here for the 
calculations of amplitude ratios including the most general competing 
system. 
 
The treatment of finite-size effects for the most general Lifshitz 
critical behavior can be developed as a direct extension of the approach 
to the Ising-like critical behavior \cite{Bre-Zinn,Nemi}. The applications 
might include systems which are finite (or semi-infinite) 
along one (or several) of their dimensions, but of infinite 
extent in the remaining directions. It would be interesting to see how 
different competing axes alter the approach to the bulk criticalities, for 
example in parallel plate geometries. The utilization of different boundary 
conditions in a layered geometry would be particularly simple and 
instructive to see how the generalization works for arbitrary competing 
systems. It could be used to investigate how amplitude ratios change with 
different boundary conditions with respect to the situation occurring in bulk 
systems \cite{Leite7}.  

Typical examples are systems which are finite in all directions, such as 
a (hyper) cube of size $L$, and systems
which are of infinite size in $d'=d-1$ dimensions but are either of
finite thickness $L$ along the remaining direction ($d$-dimensional layered
geometry) or of a semi-infinite extension. The presence of geometrical
restrictions on the domain of systems also requires the introduction of
boundary conditions (periodic, antiperiodic, Dirichlet and Neumann)
satisfied by the order parameter on the surfaces. In particular, the
validity limits of the $\epsilon_{L}$-expansion for these systems and the
approach to bulk criticality in a layered geometry can be studied 
\cite{Leite7}.

One interesting aspect of the generalized orthogonal approximation is 
that it can actually address the problem of calculating Feynman 
integrals originating in field theories in the massless limit with odd 
(greater than 2) powers of momentum in the propagator as well. This subject 
goes beyond the realm of critical phenomena in competing systems. It might 
be useful for treating perturbatively a recent proposal of an effective 
quantum field theory with cubic kinetic terms \cite{MP}. In addition, the 
general framework can be used to treat perturbatively other effective 
quantum field theories with higher derivative kinetic terms which break 
Lorentz invariance in the infrared regime as the effect of combined 
gravitational attraction and repulsion \cite{AH}. This type of theory 
resembles very much the second character Lifshitz critical behavior 
above discussed. One can expect that introducing more and more 
competing gravitational interactions in the infrared (long distance) 
limit higher powers will appear in the kinetic terms of the 
corresponding effective field theory. The perturbative analysis of this 
system would be quite analogous to the generic higher character discussed 
in the present work. 

To summarize, we have described the generic higher character 
Lifshitz critical behaviors. New field theoretical tools were exposed 
resulting in new analytical expressions for all the critical
indices in the isotropic as well as in the anisotropic cases at least at
$O(\epsilon_{L}^{2})$. Other aspects like crossover phenomena and 
tricritical behavior for this model remain to be studied. New anisotropic 
behaviors in the absence of a uniform ordered phase for the CECI model 
are under current investigation.

\section{Acknowledgments}
I thank kind hospitality at the 
Instituto Tecnol\'ogico de Aeron\'autica where this work has begun and 
T. Frederico for discussions. I wish to thank J. X. de Carvalho for the 
preparation of the figures, V. O. Rivelles for conversations and 
N. Berkovits for useful discussions on an earlier version of this work. 
I would like to acknowledge financial support from FAPESP, grant number 
00/06572-6.

\appendix 
\section{Feynman integrals for anisotropic behaviors}

In general the critical exponents and other universal ammounts are 
independent of the renormalization group scheme employed. The explicit 
calculations in this section are presented in such a way that the results 
can be checked using more than one renormalization procedure. We shall now 
describe the generalized orthogonal approximation (GOA) for the integrals 
appearing in the anisotropic cases.

In order to accomplish the task of calculating the critical indices at least 
up to two-loop level, we need a minimal set of Feynman diagrams to work with. 
The one- two- and  three-loop integrals we shall need to determine are given 
by the following expressions

\begin{equation}
I_2 =  \int \frac{d^{d-\sum_{n=2}^{L} m_{n}}q \Pi_{n=2}^{L} d^{m_{n}}k_{(n)}}
{[\bigl(\sum_{n=2}^{L}(k_{(n)} + K_{(n)}^{'})^{2}\bigr)^{n} +
(q + P)^{2}] \left(\sum_{n=2}^{L} (k_{(n)}^{2})^{n} + q^{2}  \right)}\;\;\;,
\end{equation}

\begin{eqnarray}
I_{3} =&& \int \frac{d^{d-\sum_{n=2}^{L} m_{n}}{q_{1}}d^{d-\sum_{n=2}^{L} m_{n}}q_{2} \Pi_{n=2}^{L} d^{m_{n}}k_{1 (n)}\Pi_{n=2}^{L} d^{m_{n}}k_{2 (n)}}
{\left( q_{1}^{2} + \sum_{n=2}^{L}(k_{1 (n)}^{2})^{n} \right)
\left( q_{2}^{2} +  \sum_{n=2}^{L}(k_{2 (n)}^{2})^{n}\right)}\nonumber\\
&& \qquad\qquad \times \frac{1}{[(q_{1} + q_{2} + P)^{2} + \bigl(\sum_{n=2}^{L}(k_{1 (n)} + k_{2 (n)} + K_{(n)}^{'})^{2}\bigr)^{n}]}\;\;,
\end{eqnarray}

\begin{eqnarray}
I_{4} =&& \int \frac{d^{d-\sum_{n=2}^{L} m_{n}}{q_{1}}d^{d-\sum_{n=2}^{L} m_{n}}q_{2}\Pi_{n=2}^{L} d^{m_{n}}k_{1 (n)}\Pi_{n=2}^{L} d^{m_{n}}k_{2 (n)}}
{\left ( q_{1}^{2} + \sum_{n=2}^{L}(k_{1 (n)}^{2})^{n}\right)
\left( (P - q_{1})^{2} + \sum_{n=2}^{L} \bigl((K'_{(n)} - k_{1 (n)})^{2}\bigr)^{n}  \right)}\nonumber\\
&&\qquad \times \frac{1}
{\left( q_{2}^{2} +  \sum_{n=2}^{L}(k_{2 (n)}^{2})^{n}\right)[(q_{1} - q_{2} + p_{3})^{2} + \sum_{n=2}^{L}\bigl((k_{1 (n)} - k_{2 (n)} + k_{3 (n)}')^{2}\bigl)^{n}]}\;\;,
\end{eqnarray}

\begin{eqnarray}
I_{5} &=&
\int \frac{d^{d-\sum_{n=2}^{L} m_{n}}q_{1} d^{d-\sum_{n=2}^{L} m_{n}}q_{2} d^{d-\sum_{n=2}^{L} m_{n}}q_{3} \Pi_{n=2}^{L} d^{m_{n}}k_{1 (n)}}
{\left( q_{1}^{2} + \sum_{n=2}^{L}(k_{1 (n)}^{2})^{n} \right)
\left( q_{2}^{2} + \sum_{n=2}^{L}(k_{2 (n)}^{2})^{n}\right)
\left( q_{3}^{2} + \sum_{n=2}^{L}(k_{3 (n)}^{2})^{n}\right)}\nonumber\\
&& \qquad\qquad\qquad \times \frac{\Pi_{n=2}^{L} d^{m_{n}}k_{2 (n)} \Pi_{n=2}^{L} d^{m_{n}}k_{3 (n)}}{[ (q_{1} + q_{2} - p)^{2} + \sum_{n=2}^{L} \bigl((k_{1(n)} + k_{2(n)} - k'_{(n)})^{2}\bigr)^{n}]}\nonumber\\
&& \qquad\qquad\qquad \times \frac{1}{[(q_{1} + q_{3} - p)^{2} + \sum_{n=2}^{L}\bigl((k_{1(n)} + k_{3(n)} -
  k'_{(n)})^{2}\bigr)^{n}]}.
\end{eqnarray}
We stress that in the above expressions $P, p_{3}$ and $p$ are external 
momenta perpendicular to the various competing axes, whereas 
$K'_{(n)}, k'_{3(n)}$ and $k'_{(n)}$ are external momenta characterizing 
the $n$th competing axes, respectively.     
For arbitrary values of the external momenta, these integrals can be 
calculated by making use of approximations very similar to those 
first developed to describe second character Lifshitz points \cite{Leite2}. 
As a matter of fact, the generalized dissipative approximation was 
formerly used to compute the critical exponents out of the anomalous 
dimension and correlation length exponents corresponding to space directions 
perpendicular to the competing axes for this model. Indeed, the 
generalized orthogonal approximation was figured out using a similar 
analogy. Nevertheless, since the dissipative approximation cannot approach 
the isotropic cases, we shall not describe it in this paper. 
Instead, we shall make use of the generalized orthogonal approximation, 
for it can approach both anisotropic and isotropic behaviors.

Certain useful identities will be derived in order to solve the integrals 
above. Let us proceed to find out them. Our starting point is the integral 
derived in \cite{Leite2}, namely
\begin{equation}
\int_{-\infty}^{\infty} dx_{1}...dx_{m} exp(- a(x_{1}^{2} + ...+x_{m}^{2})^{n})
= \frac{1}{2n} \Gamma(\frac{m}{2n}) a^{\frac{-m}{2n}} S_{m}.
\end{equation}
After choosing $r^{2}= x_{1}^{2} + ... + x_{m}^{2}$, one can take 
$y=r^{n}$ to write last equation in the form
\begin{equation}
\int_{0}^{\infty} dy y^{\frac{m}{n} - 1} exp(-ay^{2})
= \frac{1}{2} a^{\frac{-m}{2n}} \Gamma(\frac{m}{2n}).
\end{equation}

The integral 
\begin{equation}
\int_{-\infty}^{\infty} exp(-ax^{2k} - b(x + x_{0})^{2k}) dx 
\end{equation}
cannot be solved exactly for all $k$. In fact for generic $k \geq 2$ it 
has no elementary primitive. Nevertheless, one can select only the 
homogeneous function of $a$ by using the following approximation. First, 
make the change of variables $y=x^{k}$. Second perform the approximation 
$(x+x_{0})^{k} \cong x^{k} + x_{0}^{k}$. Thus, this integral becomes:
\begin{equation}
\int_{-\infty}^{\infty} exp(-ax^{2k} - b(x + x_{0})^{2k}) dx = 
exp(-by_{0}^{2}) \frac{2}{k}\int_{0}^{\infty} exp(-(a+b)y^{2} - 2byy_{0}) 
y^{\frac{1}{k} -1}dy
\end{equation}     
Next, perform another change of variables, namely, 
$y' = y + \frac{by_{0}}{a+b}$. We then obtain:
\begin{eqnarray}
&\int_{-\infty}^{\infty} exp(-ax^{2k} - b(x + x_{0})^{2k}) dx = 
exp(-\frac{ab}{a+b} y_{0}^{2}) \frac{2}{k} [\int_{0}^{\infty} 
exp(-(a+b)y'^{2}) (y' -\frac{by_{0}}{a+b})^{\frac{1}{k}-1}dy' \nonumber\\
& - \int_{0}^{\frac{by_{0}}{a+b}} exp(-(a+b)y'^{2}) (y' -\frac{by_{0}}{a+b})^{\frac{1}{k}-1}dy'].
\end{eqnarray}
Since the integrals are convergent, we can make use of the
approximation $(y'-\frac{b}{2a})^{\frac{1}{k}-1} = y'^{\frac{1}{k}-1} + ...$. 
The elipsis stands for the remaining terms that will be subtracted from the 
last integral, producing a type of error function. The original integral is 
then approximated by its leading contribution, i.e.
\begin{equation}
\int_{-\infty}^{\infty} exp(-ax^{2k} - b(x + x_{0})^{2k}) dx \cong 
exp(-\frac{ab}{a+b} x_{0}^{2k}) \frac{1}{k}
\Gamma(\frac{1}{2k}) (a+b)^{-\frac{1}{2k}}.
\end{equation}
It can be easily shown that the generalization for the $m$-sphere yields
\begin{eqnarray}
&\int_{-\infty}^{\infty} exp[-a(x_{1}^{2} + ... + x_{m}^{2})^{k}
- b((x_{1}+x_{01})^{2} + ... + (x_{m}+x_{0m})^{2})]^{k} dx_{1}...dx_{m}  
\cong \\
&exp(-\frac{ab}{a+b} x_{0}^{2k}) \frac{1}{2k} S_{m} \Gamma(\frac{m}{2k})
(a+b)^{-\frac{m}{2k}}\nonumber.
\end{eqnarray}
We found appropriate to name this approximation the generalized orthogonal 
approximation, for it is a natural generalization of that discussed in the 
usual second character case \cite{Leite2}.

Let us now start our calculation of loop integrals given by Eqs.(A1)-(A4). 
We can calculate the one-loop integral using two Schwinger parameters. Using 
the formula derived above, the integration over quadratic momenta can be 
shown to be given by  
\begin{eqnarray}
&& I_2= \frac{1}{2} S_{(d-\sum_{n=2}^{L}m_{n})}
\Gamma(\frac{(d-\sum_{n=2}^{L}m_{n})}{2}) \int^{\infty}_{0}\int^{\infty}_{0}
d\alpha_{1}d\alpha_{2}\,\exp(- \frac{\alpha_{1}
\alpha_{2}P^{2}}{\alpha_{1} + \alpha_{2}})\nonumber\\
&& \quad\times(\alpha_{1} + \alpha_{2})^{- \bigl(\frac{d-\sum_{n=2}^{L}m_{n}}{2} \bigr)}
\int (\Pi_{n=2}^{L}d^{m_{n}}k_{(n)}) \exp(-\alpha_{1}\sum_{n=2}^{L}(k_{(n)}^{2})^{n} - \alpha_{2} \sum_{n=2}^{L}((k_{(n)} + K_{(n)}^{'})^{2})^{n}).
\end{eqnarray}
We can now expand the argument of the last exponentials. Notice that now we 
have a product of independent integrals corresponding to the momentum 
components along arbitrary competing subspaces. Those integrals
cannot be performed analytically. If we use the homogeneity property of the 
remaining integrals in the arbitrary competing external momenta scales, we 
can simplify the calculation by utilizing the generalized orthogonal 
approximation. In fact, taking 
$(k_{(n)} + K'_{(n)})^{n} \cong k_{(n)}^{n} + K_{(n)}^{' n}$ and using (A11), 
we obtain:
\begin{eqnarray}
&\int d^{m_{n}}k_{(n)} 
exp(-\alpha_{1}k_{(n)}^{2n} -\alpha_{2}(k_{(n)}+ K'_{(n)})^{2n}) = \frac{1}{2n} 
S_{m_{n}} \nonumber\\
& \times (\alpha_{1} + \alpha_{2})^{-\frac{m_{n}}{2n}} exp(-\frac{\alpha_{1} \alpha_{2} K_{(n)}^{' 2n}}{\alpha_{1} + \alpha_{2}}) \Gamma(\frac{m_{n}}{2n}) 
\end{eqnarray}

Therefore, we can write the integral as
\begin{eqnarray}
&& I_2= \frac{1}{2} S_{(d-\sum_{n=2}^{L}m_{n})} 
\Gamma(\frac{(d-\sum_{n=2}^{L}m_{n})}{2})(\Pi_{n=2}^{L} 
\frac{S_{m_{n}} \Gamma(\frac{m_{n}}{2n})}{2n})
\int^{\infty}_{0}\int^{\infty}_{0}d\alpha_{1}d\alpha_{2}\nonumber\\
&&\quad\times \,\exp(- \frac{\alpha_{1}
\alpha_{2}(P^{2} + \sum_{n=2}^{L}((K'_{(n)})^{2})^{n}}{\alpha_{1} + \alpha_{2}})\;
(\alpha_{1} + \alpha_{2})^{- \frac{\bigl(d -\sum_{n=2}^{L}\frac{(n-1)m_{n}}{n}\bigr)}{2}}.
\end{eqnarray}
Take $x=\alpha_{1} (P^{2} + \sum_{n=2}^{L}((K'_{(n)})^{2})^{n})$ and
$y = \alpha_{2}(P^{2} + \sum_{n=2}^{L}((K'_{(n)})^{2})^{n})$. Then, define
$v = \frac{x}{x+y}$. Consequently, the parametric integrals can be performed 
with this change of variables. Using the identity
\begin{equation}
\Gamma (a + b x) = \Gamma(a)\,\Bigl[\,1 + b\, x\, \psi(a) + O(x^{2})\,\Bigr],
\end{equation}
and expressing everything in terms of the $\epsilon_{L}$ parameter leads to 
the following expression for $I_{2}$:
\begin{eqnarray}
& I_{2} = \frac{1}{2}[S_{(d-\sum_{n=2}^{L}m_{n})} 
\Gamma(2 - \sum_{n=2}^{L}\frac{m_{n}}{2n})(\Pi_{n=2}^{L} 
\frac{S_{m_{n}} \Gamma(\frac{m_{n}}{2n})}{2n})](1 - \frac{\epsilon_{L}}{2} \psi(2 - \sum_{n=2}^{L}\frac{m_{n}}{2n})) \Gamma(\frac{\epsilon_{L}}{2})\nonumber\\
& \times \; \int_{0}^{1} dv(v(1-v)(P^{2} + \sum_{n=2}^{L}((K'_{(n)})^{2})^{n}))^{\frac{-\epsilon_{L}}{2}}.
\end{eqnarray}
This is a homogeneous function (with the same homogeneity degree)
in $(P,K'_{(n)})$ . The factor 
$[S_{(d-\sum_{n=2}^{L}m_{n})} 
\Gamma(2 - \sum_{n=2}^{L}\frac{m_{n}}{2n})(\Pi_{n=2}^{L} 
\frac{S_{m_{n}} \Gamma(\frac{m_{n}}{2n})}{2n})]$ is going to 
be absorbed in a redefinition of the coupling constant after performing 
each loop integral and shall be omited henceforth. We can follow two 
routes from last equation. The first one is to perform the $v$ integral in 
terms of products of the Euler $\Gamma$ functions. It will be useful in the 
calculation of higher-loop integrals since we need the subdiagrams of this 
type in order to compute the complete integral. Then, we obtain
\begin{equation}
{I}_{2}(P,K'_{(n)}) = (P^{2} + \sum_{n=2}^{L}((K'_{(n)})^{2})^{n}))^{\frac{-\epsilon_{L}}{2}} 
\frac{1}{\epsilon_{L}}\biggl(1 + h_{m_{L}} \epsilon_{L} \biggr)\;\;,
\end{equation}
This is appropriate to 
calculate the critical exponents only using normalization conditions. But we 
would like the solution in a form suitable for minimal subtraction as well. 
The second possibility convenient for both types of renormalization schemes 
is to expand the last integral as
\begin{equation}
\int_{0}^{1} dv (v(1-v)(P^{2} + \sum_{n=2}^{L}((K'_{(n)})^{2})^{n}))^{\frac{-\epsilon_{L}}{2}} =
1 - \frac{\epsilon_{L}}{2}L(P,K'_{(n)}),
\end{equation}
where
\begin{equation}
L(P,K'_{(n)}) =
\int_{0}^{1} dv   \;\;ln[v(1-v)(P^{2} +\sum_{n=2}^{L}((K'_{(n)})^{2})^{n} )].
\end{equation}
Hence, this integral reads
\begin{equation}
{I}_{2}(P,K'_{(n)}) =
\frac{1}{\epsilon_{L}}\biggl(1 + (h_{m_{L}} - 1) \epsilon_{L}
-\frac{\epsilon_{L}}{2} L(P,K'_{(n)}) \biggr)\;\;,
\end{equation}
where 
$h_{m_{L}}= 1 + \frac{(\psi(1) - \psi(2- \sum_{n=2}^{L}\frac{m_{n}}{2n}))}
{2}$. 
Notice that whenever $m_{3} = ...=m_{L}=0$, $h_{m_{2}}=[i_{2}]_{m}$, and the 
usual anisotropic Lifshitz critical behavior is trivially obtained from this 
more general competing situation. This form is convenient for the 
renormalization using minimal subtraction. Instead, for normalization 
conditions we have:
\begin{equation}
{I}_{2 SP_{1}} = ...={I}_{2 SP_{L}} =
\frac{1}{\epsilon_{L}}\biggl(1 + h_{m_{L}}\,\epsilon_{L} \biggr)\;\;,
\end{equation}
since $L(SP_{1}=...=SP_{L})=-2$, with $SP_{1}\equiv (P^{2}=1,K'_{(n)}=0)$,...,
$SP_{L} \equiv (P=0,(K'_{(L)})^{2}=1)$.

The simplifying condition 
$(k_{(n)}+K'_{(n)})^{n} = k_{(n)}^{n} + K_{(n)}^{' n}$ 
for the one-loop integral can be generalized to the higher-loop 
graphs. It is translated in the statement that {\it the loop momenta 
characterizing a certain competition subspace in a given bubble (subdiagram) 
do not mix to all loop momenta not
belonging to that bubble}. The simplest practical application of this 
principle can be viewed in the calculation of the ``sunset'' two-loop 
integral $I_{3}$ contributing to the two-point function
\begin{eqnarray}
I_{3} =&& \int \frac{d^{d-\sum_{n=2}^{L} m_{n}}{q_{1}}d^{d-\sum_{n=2}^{L} m_{n}}q_{2} \Pi_{n=2}^{L} d^{m_{n}}k_{1 (n)}\Pi_{n=2}^{L} d^{m_{n}}k_{2 (n)}}
{\left( q_{1}^{2} + \sum_{n=2}^{L}(k_{1 (n)}^{2})^{n} \right)
\left( q_{2}^{2} +  \sum_{n=2}^{L}(k_{2 (n)}^{2})^{n}\right)}\nonumber\\
&& \qquad\qquad \times \frac{1}{[(q_{1} + q_{2} + P)^{2} + \bigl(\sum_{n=2}^{L}(k_{1 (n)} + k_{2 (n)} + K_{(n)}^{'})^{2}\bigr)^{n}]}\;\;,
\end{eqnarray}

Defining $K^{''}_{(n)}= k_{1(n)} + K^{'}_{(n)}$ and using the condition 
$k_{2(n)}.K''_{(n)}=0$, one can solve the integral over $q_{2},k_{2(n)}$ 
first, picking out only the homogeneous part of each individual integral. 
The remaining parametric integrals contains the divergence 
(pole in $\epsilon_{L}$) and can be solved as before. Using 
Eq.(A17), we obtain:
\begin{eqnarray}
I_{3}(P, K') =&& \frac{1}{\epsilon_{L}}(1 +h_{m_{L}})  \nonumber\\
&&\qquad \times \int \frac{d^{d-\sum_{n=2}^{L} m_{n}}q_{1} \Pi_{n=2}^{L} 
d^{m_{n}}k_{1 (n)}}{ \left( q_{1}^{2} + \sum_{n=2}^{L}(k_{1 (n)}^{2})^{n} \right) 
[(q_{1} + P)^{2} + \sum_{n=2}^{L} \bigl ((k_{1 (n)} + K'_{(n)})^{2}\bigr)^{n}]^{\frac{\epsilon_{L}}{2}}}. 
\end{eqnarray}
Using Feynman parameters, integrating the loop momenta along with the 
remaining parametric integrals, and expanding the resulting $\Gamma$ functions in $\epsilon_{L}$ we find:
\begin{equation}
I_{3}(P, K') = (P^{2} + \sum_{n=2}^{L}K^{' 2n}_{n})\frac{-1}{8 \epsilon_{L}}
(1+ 2h_{m_{L}} \epsilon_{L} -\frac{3}{4} \epsilon_{L} -
2 \epsilon_{L} L_{3}(P, K'_{(n)})),
\end{equation}
where
\begin{equation}
L_{3}(P, K') = \int_{0}^{1} dx (1-x) ln[(P^{2} + \sum_{n=2}^{L}K^{' 2n}_{n}) x(1-x)].
\end{equation}
At the symmetry points $SP_{n}$, it can be rewritten as
\begin{equation}
I_{3 SP_{1}} =...= I_{3 SP_{L}} = \frac{-1}{8 \epsilon_{L}}
(1+ 2 h_{m_{L}} \epsilon_{L} + \frac{5}{4} \epsilon_{L}).
\end{equation}
From the above equation we can derive the expressions:
\begin{equation}
I_{3 SP_{1}}^{'} (\equiv \frac{\partial I_{3 SP_{1}}}{\partial P^{2}})=...=
I_{3 SP_{L}}^{'} (\equiv \frac{\partial I_{3 SP_{2}}}{\partial K^{' 2L}_{(L)}}) =
\frac{-1}{8 \epsilon_{L}}
(1+2h_{m_{L}} \epsilon_{L} + \frac{1}{4} \epsilon_{L}).
\end{equation}
To complete our description of the 1$PI$ two-point vertex parts, 
consider the integral
\begin{eqnarray}
I_{5} &=&
\int \frac{d^{d-\sum_{n=2}^{L} m_{n}}q_{1} d^{d-\sum_{n=2}^{L} m_{n}}q_{2} d^{d-\sum_{n=2}^{L} m_{n}}q_{3} \Pi_{n=2}^{L} d^{m_{n}}k_{1 (n)}}
{\left( q_{1}^{2} + \sum_{n=2}^{L}(k_{1 (n)}^{2})^{n} \right)
\left( q_{2}^{2} + \sum_{n=2}^{L}(k_{2 (n)}^{2})^{n}\right)
\left( q_{3}^{2} + \sum_{n=2}^{L}(k_{3 (n)}^{2})^{n}\right)}\nonumber\\
&& \qquad\qquad\qquad \times \frac{\Pi_{n=2}^{L} d^{m_{n}}k_{2 (n)} \Pi_{n=2}^{L} d^{m_{n}}k_{3 (n)}}{[ (q_{1} + q_{2} - p)^{2} + \sum_{n=2}^{L} \bigl((k_{1(n)} + k_{2(n)} - k'_{(n)})^{2}\bigr)^{n}]}\nonumber\\
&& \qquad\qquad\qquad \times \frac{1}{[(q_{1} + q_{3} - p)^{2} + \sum_{n=2}^{L}\bigl((k_{1(n)} + k_{3(n)} -
  k'_{(n)})^{2}\bigr)^{n}]},
\end{eqnarray}
which is the three-loop diagram contributing to the two-point vertex
function. Incidentally, there is a symmetry in the dummy loop
momenta  $q_{2} \rightarrow q_{3}$ and $k_{2(n)} \rightarrow k_{3(n)}$.
Concerning the integrations either over $q_{2}, k_{2(n)}$ or 
$q_{3}, k_{3(n)}$, we use the condition $(k_{2(n)}+(k_{1} - K'))^{n} = 
k_{2(n)}^{n} + (k_{1(n)} - K'_{(n)})^{n}$ when the integration is performed
over $k_{2}$ as well as $(k_{3(n)}+(k_{1(n)} - K'_{(n)}))^{n} = 
k_{3(n)}^{n} + (k_{1} - K'_{(n)})^{n}$ when the integral over
$k_{3}$ is realized. The two internal bubbles, which are represented by the
integrals over $(q_{2}, k_{2(n)})$ and $(q_{3}, k_{3(n)})$, respectively,
are actually the same, resulting in 
$I_{2}((q_{1} - P), (k_{1(n)} - K'_{(n)}))$. 
Next take $P\rightarrow -P$, $K'_{(n)}\rightarrow -K'_{(n)}$. 
Using a Feynman parameter and proceeding in close analogy to the calculation
of $I_{3}$ we find:
\begin{equation}
I_{5}(P, K'_{(n)}) = (P^{2} + \sum_{n=2}^{L}K^{' 2n}_{n}) \frac{-1}{6 \epsilon_{L}^{2}}
(1+3h_{m_{L}} \epsilon_{L} - \epsilon_{L} - 3 \epsilon_{L} L_{3}(P, K'_{(n)})),
\end{equation}
At the symmetry points $SP_{1}$, ..., $SP_{L}$ one obtains:
\begin{equation}
I_{5 SP_{1}}^{'} (\equiv \frac{\partial I_{5 SP_{1}}}{\partial P^{2}})=...=
I_{5 SP_{L}}^{'} (\equiv \frac{\partial I_{5 SP_{2}}}{\partial K^{'2L}_{(L)}}) =
\frac{-1}{6 \epsilon_{L}^{2}}
(1+3h_{m_{L}} \epsilon_{L} + \frac{1}{2} \epsilon_{L}).
\end{equation}
Finally we compute one of the two-loop diagrams
contributing to the four-point function, namely
\begin{eqnarray}
I_{4} =&& \int \frac{d^{d-\sum_{n=2}^{L} m_{n}}{q_{1}}d^{d-\sum_{n=2}^{L} m_{n}}q_{2}\Pi_{n=2}^{L} d^{m_{n}}k_{1 (n)}\Pi_{n=2}^{L} d^{m_{n}}k_{2 (n)}}
{\left ( q_{1}^{2} + \sum_{n=2}^{L}(k_{1 (n)}^{2})^{n}\right)
\left( (P - q_{1})^{2} + \sum_{n=2}^{L} \bigl((K'_{(n)} - k_{1 (n)})^{2}\bigr)^{n}  \right)}\nonumber\\
&&\qquad \times \frac{1}
{\left( q_{2}^{2} +  \sum_{n=2}^{L}(k_{2 (n)}^{2})^{n}\right)[(q_{1} - q_{2} + p_{3})^{2} + \sum_{n=2}^{L}\bigl((k_{1 (n)} - k_{2 (n)} + k_{3 (n)}')^{2}\bigl)^{n}]}\;\;.
\end{eqnarray}
Recall that $P= p_{1} + p_{2}$, $p_{i}$ ($i=1,...,3$) are external
momenta perpendicular to the competing axes. On the other hand, 
$K'_{(n)}= k'_{1(n)} + k'_{2(n)}$,
and $k'_{i(n)}$ ($i=1,...,3$) are the external momenta along arbitrary 
competition directions. We can integrate first over the bubble 
$(q_{2}, k_{2(n)})$. It is convenient to choose Schwinger parameters in 
the calculation. Then, we use two Feynman parameters and solve for the loop 
momenta to obtain the following parametric form
\begin{eqnarray}
I_{4}&=& \frac{1}{2} f_{m}(\epsilon_{L})\,
\frac{\Gamma(\epsilon_L) \Gamma(2 - \sum_{n=2}^{L}\frac{m_{n}}{2n} -\frac{\epsilon_{L}}{2}) S_{(d-\sum_{n=2}^{L} m_{n})}}{\Gamma\biggl(\frac{\epsilon_L}{2}\biggr)}(\Pi_{n=2}^{L} \frac{\Gamma(\frac{m_{n}}{2n}) S_{m{n}}}{2n})\nonumber\\ 
&&\, \times \int_0^1 dy\, y\,(1-y)^{\frac{1}{2}\epsilon_L-1}
 \int_0^1 dz
\biggl[ yz(1-yz)(P^{2} + \sum_{n=2}^{L} K^{'2n}_{(n)})
+y(1-y)(p_{3}^{2} + \sum_{n=2}^{L} k^{'2n}_{3(n)})\nonumber\\
&& -2yz(1-y)(p_3.P + \sum_{n=2}^{L} (-1)^{n} k^{'n}_{3(n)} K^{'n}_{(n)})\biggr]^{-\epsilon_L}.
\end{eqnarray}
The integral  over $y$ is singular at $y=1$ when
$\epsilon_L=0$. Replace the value $y=1$ inside the
integral over $z$ \cite{Am,Leite2}, integrate over $y$ and expand the Gamma 
functions in $\epsilon_{L}$. This implies that
\begin{equation}
I_{4} = \frac{1}{2 \epsilon_{L}^{2}} \Bigl(1 +
2\;h_{m_{L}} \epsilon_{L} -\frac{3}{2} \epsilon_{L} - \epsilon_{L} L(P,K')\Bigr).
\end{equation}
This form is particularly suitable for the renormalization procedure using
minimal subtraction. For the purpose of normalization conditions, the 
value of this integral at the symmetry points discussed before is given by
\begin{equation}
I_{4 SP_{1}} = ... = I_{4 SP_{L}} = \frac{1}{2 \epsilon_{L}^{2}} \Bigl(1 +
2\;h_{m_{L}} \epsilon_{L} +\frac{1}{2} \epsilon_{L}\Bigr).
\end{equation}
The method proposed here is equivalent to a new regularization procedure to
calculate Feynman integrals whose propagators have any combination of even 
powers of momenta. We can define the measure of the $m_{n}$-dimensional
sphere in terms of a half integer measure. In fact, taking $k=p^{2n}$,
one has
$d^{m_{n}}k \equiv d^{\frac{m_{n}}{2n}}p = 
\frac {1}{2n} p^{\frac{m_{n}}{2n}-1}dp d\Omega_{m_{n}}$.
Hence, the approximation required to solve the integrals results that 
the new ``measure''$d^{\frac{m_{n}}{2n}}p$ is invariant under 
translations $p'=p+a$. This is a simple generalization of the same property 
valid for the usual $m_{2}$-fold Lifshitz behaviors. 

\section{Isotropic diagrams in the generalized orthogonal approximation}

The computation of the Feynman integrals using the generalized orthogonal 
approximation is simpler in the isotropic cases, since there is only one 
subspace to be integrated over. At the Lifshitz point
$\delta_{0n} = \tau_{nn'}= 0$ and solely the $2L$th power of momentum 
appears in the propagator for the case of $L$th character isotropic 
critical point.
The isotropic analogous of the one-loop integral contributing to the 
four-point vertex part is
\begin{equation}
I_2 =  \int \frac{d^{m_{n}}k}{\bigl((k + K^{'})^{2}\bigr)^{n}  (k^{2})^{n} }\;\;\;.
\end{equation}
We can use two Schwinger parameters and the
orthogonality condition $(k+K')^{n} \cong k^{n} + K^{'n}$, resulting in 
the expression
\begin{equation}
I_{2}(K') = \int d^{m_{n}}k \int_{0}^{\infty} \int_{0}^{\infty}
d \alpha_{1} d \alpha_{2} e^{-(\alpha_{1} + \alpha_{2})(k^{2})^{n}}
e^{-2\alpha_{2} K'^{n} k^{n}} e^{-\alpha_{2}(K'^{2})^{n}}.
\end{equation}
Turning to polar spherical coordinates, take 
$r^{2}= k_{1}^{2}+...+ k_{m_{n}}^{2}$. Making the transformation $k^{n}=p$ the volume element becomes $d^{m_{n}}k=\frac{1}{n} p^{\frac{m_{n}}{n} -1}dp d\Omega_{m_{n}} \equiv d^{\frac{m_{n}}{n}}p$. The former integral with a $n$th power 
of momenta changes to a quadratic integral over $p$. After discarding the 
infinite terms which change the measure $d^{\frac{m_{n}}{n}}k$
under the translation $y'=y+\frac{b}{2a}$, only the leading contribution
is picked out and we have
\begin{equation}
\int d^{m_{n}}k e^{-a (k^{2})^{n} -b k^{n}} = \int d^{\frac{m_{n}}{n}}p e^{-a p^{2} -b p} \cong a^{-\frac{m_{n}}{2n}} e^{\frac{b^{2}}{4a}} \frac{1}{2n} \Gamma(\frac{m_{n}}{2n}) S_{m_{n}} .
\end{equation}
When this result is replaced into the expression of $I_{2}(K')$, we get to
\begin{equation}
I_{2}(K') = \frac{S_{m_{n}}}{\epsilon_{n}} [ 1 - \frac{\epsilon_{n}}{2n} 
(1+L(K'^{2}))].
\end{equation}
Henceforth we absorb the factor of $S_{m_{n}}$ in this integral through a
redefinition of the coupling constant and shall do so after performing 
each loop integral for arbitrary vertex parts. Note that this absorption 
factor is different from that arising in the anisotropic case in the limit
$d \rightarrow m_{n} =4n -\epsilon_{n}$. Since the geometric angular factor 
coming from the anisotropic cases becomes singular in the above isotropic 
limit the attempt of extrapolating from one case to another is not valid, 
at least within the framework of the $\epsilon_{L}$-expansion presented in 
this work. This is a further technical evidence that the isotropic
and anisotropic cases have to be tackled differently. Thus,
\begin{equation}
I_{2}(K') = \frac{1}{\epsilon_{n}} [ 1 - \frac{\epsilon_{n}}{2n} (1+L(K'^{2}))].
\end{equation}
The suitable symmetry point $(K'^{2})^{n} = 1$ useful for the purpose of 
normalization conditions leads to the following simple outcome
\begin{equation}
I_{2}(K') = \frac{1}{\epsilon_{n}} [1 +\frac{\epsilon_{n}}{2n}].
\end{equation}
The next step is the evaluation of the integral 
\begin{equation}
I_{3} = \int \frac{d^{m_{n}}k_{1} d^{m_{n}}k_{2}}{\bigl((k_{1} + k_{2} + K^{'})^{2}\bigr)^{n}  (k_{1}^{2})^{n}  (k_{2}^{2})^{n}}\;\;\;,
\end{equation}
Integrate first over $k_{2}$. Take $K''= k_{1} + K'$ and use the condition 
$(k_{2}+ K'')^{n} \cong k_{2}^{n} + K^{''n}$ to obtain:
\begin{equation}
I_{3} = \frac{1}{\epsilon_{n}}[1 + \frac{\epsilon_{n}}{2n}]
\int \frac{d^{m_{n}}k_{1}}
{[\bigl((k_{1} + K^{'})^{2}\bigr)^{n}]^{\frac{\epsilon_{n}}{2n}}  (k_{1}^{2})^{n}}\;.
\end{equation}
Utilizing a Feynman parameter, we can integrate over $k_{1}$. After the 
expansion $m_{n}=4n -\epsilon_{n}$ is done inside the argument of the 
resulting $\Gamma$ functions and using the expression
\begin{equation}
\int \frac {d^{\frac{m_{n}}{n}}q}{(q^{2} + 2 k.q + m^{2})^{\alpha}} \cong
\frac{1}{2n} \frac{\Gamma(\frac{m_{n}}{2n}) \Gamma(\alpha - \frac{m_{n}}{2n})
(m^{2} - k^{2})^{\frac{m_{n}}{2n} - \alpha} S_{m_{n}}}{\Gamma(\alpha)},
\end{equation}
the integral $I_{3}$ can be found to be
\begin{equation}
I_{3} = -\frac{(K'^{2})^{n}}{8n \epsilon_{n}}[1 + \epsilon_{n}(\frac{1}{4n}
- \frac{2}{n} L_{3}(K'^{2}))].
\end{equation}
At the symmetry point, this reduces to
\begin{equation}
I_{3} = -\frac{1}{8n \epsilon_{n}}[1 + \frac{9}{4n} \epsilon_{n}],
\end{equation}
implying that
\begin{equation}
\frac{\partial I_{3}}{\partial (K'^{2})^{n}}|_{SP} = I_{3}^{'} =
-\frac{1}{8n \epsilon_{4n}}[1 + \frac{5}{4n} \epsilon_{4n}].
\end{equation}
The 3-loop integral $I_{5}$ is given by
\begin{equation}
I_{5} = \int \frac{d^{m_{n}}k_{1} d^{m_{n}}k_{2}d^{m_{n}}k_{3}}
{\bigl((k_{1} + k_{2} + K^{'})^{2}\bigr)^{n} \bigl((k_{1} + k_{3} + K^{'})^{2}\bigr)^{n} (k_{1}^{2})^{n}  (k_{2}^{2})^{n}  (k_{3}^{2})^{n} }\;\;\;,
\end{equation}
where we took for convenience the redefinition $K'\rightarrow -K'$. The 
integrals over $k_{2}$ and $k_{3}$ are the same. Thus, following analogous 
steps and employing the same 
reasoning as in the calculation of $I_{3}$ we get to
\begin{equation}
I_{5} = -\frac{(K'^{2})^{n}}{6n \epsilon_{n}^{2}}[1 + \epsilon_{n}(\frac{1}{2n}
- \frac{3}{n}L_{3}(K'^{2}))].
\end{equation}
At the symmetry point, the following expression follows trivially
\begin{equation}
\frac{\partial I_{5}}{\partial (K'^{2})^{n}}|_{SP} = I_{5}^{'} =
-\frac{1}{6n \epsilon_{n}^{2}}[1 + \frac{2}{n} \epsilon_{n}].
\end{equation}
The two-loop integral $I_{4}$ in the isotropic behavior is 
\begin{eqnarray}
I_{4}\;\; =&& \int \frac{d^{m_{n}}k_{1}d^{m_{n}}k_{2}}
{(k_{1}^{2})^{n} \bigl((K' - k_{1})^{2}\bigr)^{n}
\left(k_{2}^{n}\right) \bigl((k_{1} - k_{2} + k_{3}')^{2}\bigl)^{n}}\;\;,
\end{eqnarray}
where $K'= k_{1}' + k_{2}'$. The integration can be done along the 
same lines of the computation performed for its anisotropic counterpart. 
It is straightforward to show that
\begin{equation}
I_{4}(K'^{2}) = \frac{1}{2\epsilon_{n}^{2}}
[1 -\frac{\epsilon_{n}}{2n}(1 + 2 L(K'^{2}))].
\end{equation}
At the symmetry point the integral can be rewritten in the form
\begin{equation}
I_{4}(K'^{2}=1) = \frac{1}{2\epsilon_{n}^{2}}
[1 +\frac{3 \epsilon_{n}}{2n}].
\end{equation}
As was shown above, these results are a natural generalization of those 
originally developed for the second character Lifshitz points. It can be 
checked that all integrals reduce to the usual $\phi^{4}$ values for $n=1$ 
and reproduce the results from \cite{Leite2} in case $n=2$. 

\section{Isotropic integrals in the exact calculation}

An interesting feature of the isotropic case is that it can be calculated 
exactly. We now proceed to yield the exact solution to the Feynman diagrams 
without performing approximations.
\begin{equation}
I_2 =  \int \frac{d^{m_{n}}k}{\bigl((k + K^{'})^{2}\bigr)^{n}  (k^{2})^{n} }\;\;\;.
\end{equation}
Using a Feynman parameter and making the continuation 
$d=m_{n}=4n - \epsilon_{n}$ we get 
\begin{equation}
I_{2}(K') = \frac{\Gamma(2n -\frac{\epsilon_{n}}{2}) \Gamma(\frac{\epsilon_{n}}{2}) S_{m_{n}}}
{2 \Gamma(n) \Gamma(n)} [\frac {\Gamma(n) \Gamma(n)}{\Gamma(2n)} 
- \frac{\epsilon_{n}}{2}L_{n}(K')],
\end{equation}
where $L_{n}(K')$ is given by
\begin{equation}
L_{n}(K') = \int_{0}^{1} dx x^{n-1} (1-x)^{n-1} ln[x(1-x)K'^{2}]. 
\end{equation}
This integral is the analogous of the integral $L(K')$ appearing in the 
orthogonal approximation. Here it depends explicitly on $n$, and that is 
the reason we have included a subscript in it to emphasize this dependence. 
The integration over $x$ together with the $\epsilon_{n}$ expansion of the 
Gamma functions results in
\begin{equation}
I_{2}(K')= \frac{S_{m_{n}}}{\epsilon_{n}}[1 
-\frac{\epsilon_{L}}{2}(\psi(2n) - \psi(1) + 
\frac{\Gamma(2n)}{\Gamma(n) \Gamma(n)} L_{n}(K'))].
\end{equation}
As before, we absorb the factor $S_{m_{n}}$ in a redefinition of the 
coupling constant. We need to do this for each loop integral. Thus,
\begin{equation}
I_{2}(K')= \frac{1}{\epsilon_{n}}[1 
-\frac{\epsilon_{n}}{2}(\psi(2n) - \psi(1) + 
\frac{\Gamma(2n)}{\Gamma(n) \Gamma(n)} L_{n}(K'))].
\end{equation}

This is a useful result for doing minimal subtraction. Defining the 
quantity $D(n)=\frac{1}{2} \psi(2n) - \psi(n) + 
\frac{1}{2} \psi(1)$, at the symmetry point $K'^{2}=1$ 
the integral turns out to be
\begin{equation}
I_{2 SP}= \frac{1}{\epsilon_{n}}[1 + D(n) \epsilon_{n}].
\end{equation}

Now, let us calculate the integral $I_{3}.$ As before, take 
$K''= k_{1} + K'$ and solve for the internal bubble $k_{2}$ using 
Feynman parameters solving the momentum independent 
integrals over the Feynman parameters and expanding the Gamma 
functions in $\epsilon_{n}$, we end up with
\begin{equation}
I_{3} = K'^{2n} (-1)^{n} \frac{\Gamma^{2}(2n)}{4 \Gamma(3n) \Gamma(n+1)} 
\frac{1}{\epsilon_{n}}[1 + \epsilon_{n}(B_{n} - \frac{L_{3n}(K')}{A_{n}})],
\end{equation}
where
\begin{equation}
A_{n} = \frac{\Gamma(2n) \Gamma(n)}{\Gamma(3n)},
\end{equation}

\begin{eqnarray}
&& B_{n} = D(n) 
- \frac{1}{2} \sum_{p=1}^{2n-1} \frac{1}{p} 
+ \sum_{p=1}^{n} \frac{1}{p} \nonumber\\ 
&& + 
\frac{\sum_{p=0}^{2n-1} \frac{(2n-1)! (-1)^{p+1}}{2 p! (2n-1-p)! (n+p)^{2}}}{A_{n}},
\end{eqnarray}

\begin{equation}
L_{3n}(K') = \int_{0}^{1} dx x^{2n-1} (1-x)^{n-1}ln[x(1-x)K'^{2}].
\end{equation}
Again, this integral depends explicitly on $n$ and should be compared with 
its counterpart arising in the orthogonal approximation. We then learn that 
for massless propagators with arbitrary power of momenta, the external 
momentum dependent part of the Feynman integrals generalizes the standard 
$\phi^{4}$ in the manner prescribed above.
 
At the symmetry point, the integral $I_{3}$ simplifies to the following 
expression:
\begin{eqnarray}
&& I_{3SP} = (-1)^{n}\frac{\Gamma(2n)^{2}}
{4 \Gamma(n+1) \Gamma(3n) \epsilon_{n}}(1+(D(n) + \frac{3}{4} + \frac{1}{n})\epsilon_{n})\nonumber\\
&& [1 +\frac{3 \epsilon_{n}}{2}
(\sum_{p=3}^{3n-1} \frac{1}{p} - \sum_{p=2}^{2n-1} \frac{1}{p}) 
+ \frac{\epsilon_{n}}{2} \sum_{p=1}^{n-1} \frac{1}{n-p}].
\end{eqnarray}
Notice that while the first term inside the parenthesis contributes 
for arbitrary values of $n$, the last factor of $O(\epsilon_{n})$ 
into the brackets are corrections which contribute solely for $n\geq2$. (A 
similar feature will also take place in the calculation of $I_{4}$ and 
$I_{5}$.)
Therefore, taking the derivative with respect to $K'^{2n}$ at 
the symmetry point produces the result
\begin{eqnarray}
&& I'_{3SP} = (-1)^{n}\frac{\Gamma(2n)^{2}}
{4 \Gamma(n+1) \Gamma(3n) \epsilon_{n}}(1+(D(n) + \frac{3}{4})\epsilon_{n})\nonumber\\
&& [1 +\frac{3 \epsilon_{n}}{2}
(\sum_{p=3}^{3n-1} \frac{1}{p} - \sum_{p=2}^{2n-1} \frac{1}{p}) 
+ \frac{\epsilon_{n}}{2} \sum_{p=1}^{n-1} \frac{1}{n-p}].
\end{eqnarray}

We now calculate the three-loop integral $I_{5}$. Proceeding analogously, 
we can show that it has the solution
\begin{equation}
I_{5} = K'^{2n} (-1)^{n} \frac{\Gamma^{2}(2n)}{3 \Gamma(3n) \Gamma(n+1)} 
\frac{1}{\epsilon_{n}^{2}}
[1 + \epsilon_{n}(C_{n} - \frac{3 L_{3n}(K')}{2A_{n}})],
\end{equation}
where
\begin{eqnarray}
&& C_{n} = 2D(n) 
- \frac{1}{2} \sum_{p=1}^{2n-1} \frac{1}{p} 
+ \frac{3}{2} \sum_{p=1}^{n} \frac{1}{p} \nonumber\\ 
&& + 
\frac{\sum_{p=0}^{2n-1} \frac{(2n-1)! (-1)^{p+1}}{p! (2n-1-p)! (n+p)^{2}}}{A_{n}}.
\end{eqnarray}
 
At the symmetry point this result gets simplified. Taking the derivative 
with respect to the external momenta we get to:
\begin{eqnarray}
&& I'_{5SP} = (-1)^{n}\frac{\Gamma(2n)^{2}}
{3 \Gamma(n+1) \Gamma(3n) \epsilon_{n}^{2}}(1+ D(n) + 1)\epsilon_{L})\nonumber\\
&& [1 + 2 \epsilon_{n}
(\sum_{p=3}^{3n-1} \frac{1}{p} - 2 \sum_{p=2}^{2n-1} \frac{1}{p}) 
+ \epsilon_{n} \sum_{p=1}^{n-1} \frac{1}{n-p}].
\end{eqnarray}
Notice that the $O(\epsilon_{n})$ terms inside the bracket gives a 
nonvanishing contribution only for $n \geq 2$. 

To conclude, let us calculate the integral $I_{4}$. The integral over 
the bubble $k_{2}$ can be solved directly by taking the effective external 
momenta as $K''= -k_{1}-k'_{3}$. It has $I_{2}(K'')$ as a subdiagram. 
Using the information obtained in calculating $I_{2}$ and working out 
the details we find the intermediate result:
\begin{eqnarray}
&& I_{4} = f_{n}(\epsilon_{n}) 
\frac{\Gamma(2n -\frac{\epsilon_{n}}{2}) \Gamma(\epsilon_{n})}
{2 \Gamma(n)^{2} \Gamma(\frac{\epsilon_{n}}{2})} 
\int_{0}^{1} dy y^{2n-1} (1-y)^{\frac{\epsilon_{n}}{2} -1} \nonumber\\  
&& \int_{0}^{1} dz [z(1-z)]^{n-1}[yz(1-yz)K^{'2} + y(1-y)k_{3}^{'2} 
- 2yz(1-y)K'.k'_{3}]^{-\epsilon_{n}}.
\end{eqnarray}
The situation resembles that in the calculation of $I_{4}$ using the 
orthogonal approximation. Again, set $y=1$ in the integral over 
$z$ and carry out the integral over $y$ independently. Performing the 
integral over $y$, we find:
\begin{equation}
I_{4} = f_{n}(\epsilon_{n}) 
\frac{\Gamma(2n -\frac{\epsilon_{n}}{2}) \Gamma(\epsilon_{n}) \Gamma(2n)}
{2 \Gamma(n)^{2} \Gamma(2n + \frac{\epsilon_{n}}{2})} 
\int_{0}^{1}dz [z(1-z)]^{n-1} [z(1-z)K^{' 2}]^{-\epsilon_{n}}.
\end{equation}
Then for the purposes of minimal subtraction, it can be expressed as
\begin{equation}
I_{4} = \frac{1}{2 \epsilon_{n}^{2}}[1 + 
(D(n) -1 - 
\frac{\Gamma(2n) L_{n}}{\Gamma(n)^{2}})\epsilon_{n} - 
\epsilon_{n} \sum_{p=1}^{2n-2} \frac{1}{2n-p}].
\end{equation} 
We emphasize that the last $O(\epsilon_{n})$ term in this expression 
only contributes for $n\geq2$. At the symmetry point, we find 
\begin{equation}
I_{4SP} = \frac{1}{2 \epsilon_{n}^{2}}[1 + 
(D(n) + 1)\epsilon_{n} + 
\epsilon_{n}(\sum_{p=1}^{2n-2} \frac{1}{2n-p} - 2 \sum_{p=1}^{n-1} 
\frac{1}{n-p})].
\end{equation}
Once again, the last two terms of $O(\epsilon_{n})$ containing the sums 
in the above expression correspond to corrections in case $n\geq2$. Since 
the corrections are absent when starting from the scratch for the $n=1$ 
case, by neglecting them in the above expressions it can be easily checked 
that all of these integrals reduce to the values of the ordinary 
$\lambda \phi^{4}$ in the limit $n \rightarrow 1$. We learn that a general 
feature of the exact calculation is that higher loop integrals generally 
receive further contributions to the subleading singularities for 
$n\geq2$. 

In order to make contact with the more concrete case of the usual second 
character Lifshitz point obtained by Hornreich, Luban and Shtrikman 
\cite{Ho-Lu-Sh}, we discuss the particular $n=2$ case next.    

\subsection{The n=2 case}

 Let us now analyse the Feynman integrals involved in the calculation of the 
critical indices of the ordinary second character Lifshitz critical 
behavior. This is a mere particular case of the most general isotropic 
CECI model discussed in the previous subsection. Nevertheless, the discussion 
of this particular case is useful when comparing with the original 
previous result obtained by Hornreich, Luban and Shtrikman about three 
decades ago. Needless to say, both results agree for the exponents $\eta_{L4}$ 
and $\nu_{L4}$, which in the notation of Section IV correspond to $\eta_{2}$ 
and $\nu_{2}$. Moreover, our treatment permits to obtain two new results for 
this behavior: the results for $\eta_{2}$ are extended 
including corrections up to $O(\epsilon_{2}^{3})$ whereas the remaining 
exponents are obtained through the complete set of scaling relations 
derived in \cite{Leite1} up to $O(\epsilon_{2}^{2})$. 

Since the calculation was already indicated in the last subsection, 
we simply quote the results. For calculating the exponents using minimal 
subtraction the most appropriate form of the integrals are given by 

\begin{equation}
I_{2}(K')= \frac{1}{\epsilon_{L}}
[1 - \frac{11 \epsilon_{L}}{12} - 3 \epsilon_{L} L_{2}(K')].
\end{equation}

\begin{equation}
I_{3} = K'^{4} \frac{3}{80 \epsilon_{L}}
[1 - \epsilon_{L}(\frac{17}{120} + 20L_{32}(K'))],
\end{equation}

\begin{equation}
I_{5} = K'^{4} \frac{1}{20 \epsilon_{L}^{2}}
[1 - \epsilon_{L}(\frac{7}{60} + 30L_{32}(K'))],
\end{equation}

\begin{equation}
I_{4} = \frac{1}{2 \epsilon_{L}^{2}}[1 - 
(\frac{23}{12} + 6 L_{12}(K^{' 2}))\epsilon_{L}].
\end{equation}

For the use of normalization conditions, however, it is convenient 
expressing these integrals at their symmetry point. Instead of 
calculating $I_{3}$ and $I_{5}$ at the symmetry point, we need their 
derivatives with respect to $K^{' 4}$ at the symmetry point. Thus, we 
find

\begin{equation}
I_{2SP}= \frac{1}{\epsilon_{L}}[1 
-\frac{\epsilon_{L}}{12}].
\end{equation}

\begin{equation}
I_{3SP}^{'} = \frac{3}{80 \epsilon_{L}}
[1 + \frac{131}{120} \epsilon_{L} ],
\end{equation}

\begin{equation}
I_{5SP}^{'} = \frac{1}{20 \epsilon_{L}^{2}}
[1 + \frac{26}{15}\epsilon_{L}],
\end{equation}

\begin{equation}
I_{4} = \frac{1}{2 \epsilon_{L}^{2}}[1 - \frac{1}{4} \epsilon_{L}].
\end{equation}


\newpage
\begin{center}
FIGURE CAPTIONS
\end{center}

\begin{figure}[htbp]
\caption{The simplest example of the CECI model with uniaxial competing 
interactions between second neighbors as well as uniaxial couplings between 
third neighbors. This system has three independent correlation lengths and 
presents a generic third character anisotropic Lifshitz critical behavior. 
Note that by turning off the interactions among second neighbors leads to 
the simpler third character critical behavior.}
\label{figCECI1}
\end{figure}

\begin{figure}[htbp]
\caption{The phase diagram of a typical uniaxial second character Lifshitz 
critical behavior. The dashed lines indicate a first order transition 
between the uniformly ordered and modulated ordered phase which terminates 
at the Lifshitz point of second character. The parameter $p$ is defined by 
$p=\frac{J_{2}}{J_{1}}$.}
\label{figCECI2}
\end{figure}

\begin{figure}[htbp]
\caption{The superposition of the two two-dimensional independent phase 
diagrams for the simplest CECI model. In this two-dimensional picture, the 
confluence of the two distinct modulated phases, the ferromagnetic and 
the paramagnetic phases occurs at the generic third character Lifshitz point.}
\label{figCECI3}
\end{figure}
\newpage 

\psfig{file=figCECI1.ps, width=15cm}

\psfig{file=figCECI2.ps, width=15cm}

\psfig{file=figCECI3.ps, width=15cm}


\newpage
\begin{references}
\bb{Am}D. J. Amit, {\it in Field Theory, the Renormalization
Group and Critical Phenomena} (World Scientific, Singapore)(1984).
\bb{BLZ}E. Br\'ezin, J. C. Le Guillou, and J. Zinn-Justin, {\it in
Phase Transitions and Critical Phenomena}, edited by C. Domb and M. S.
A
Green (Academic Press, London), vol. 6(1976).
\bb{Ho-Lu-Sh}R. M. Hornreich, M. Luban, and S. Shtrikman
{\it Phys. Rev. Lett.} {\bf 35}, 1678(1975).
\bb{Selke}W. Selke, {\it Phys. Rep.} {\bf 170}, 213(1988); 
W. Selke, {\it in Phase Transitions and
Critical Phenomena}, edited by C. Domb and J. Lebowitz (Academic
Press, London), vol.15 (1998).
\bb{Pleim-Hen}M. Pleimling, and M. Henkel, {\it Phys. Rev. Lett.} {\bf 87},
125702(2001).
\bb{AL1}L. C. de Albuquerque, and M. M. Leite, {\it cond-mat/0006462}; 
L. C. de Albuquerque, and M. M. Leite, {\it J. Phys. A: Math Gen.} {\bf 34}, 
L327(2001).
\bb{Leite1}M. M. Leite, {\it hep-th/0109037}.
\bb{Leite2}M. M. Leite, {\it Phys. Rev.}{\bf B67}, 104415 (2003).
\bb{Mergulho}C. Mergulh\~ao Jr, and C. E. I. Carneiro, {\it Phys. Rev. B}
{\bf 58}, 6047(1998); C. Mergulh\~ao Jr, and C. E. I. Carneiro, 
{\it Phys. Rev. B}{\bf 59}, 13954(1999).
\bb{DS}M. Shpot, and H. W. Diehl, {\it Nucl. Phys.} {\bf B612}(3), 340(2001); 
H. W. Diehl, and M. Shpot, {\it J. Phys. A: Math. Gen.}
{\bf 35}, 6249(2002).
\bb{Se1}W. Selke, {\it Z. Physik}{\bf B 27}, 81 (1977); W. Selke, 
{\it Journ. Magn. Magn. Mat.}{\bf 9}, 7(1978).
\bb{Se2}W. Selke, {\it Phys. Lett.}{\bf A61}, 443(1977).
\bb{NCS}J. F. Nicoll, T. S. Chang, and H. E. Stanley, {\it Phys. Rev.}
{\bf A13},1251 (1976).
\bb{NTCS}J. F. Nicoll, G. F. Tuthill, T. S. Chang, and H. E. Stanley, 
{\it Phys. Lett.}{\bf A58}, 1 (1976). 
\bb{Leite4}M. M. Leite, {\it Phys. Lett.}{\bf A326}, 281 (2004).
\bb{Becerra}C. C. Becerra, Y. Shapira, N. F. Oliveira Jr., and T. S. Chang, 
{\it Phys. Rev. Lett.} {\bf 44}, 1692 (1980); Y. Shapira, C. C. Becerra, 
N. F. Oliveira Jr., and T. S. Chang, {\it Phys. Rev. B} {\bf 24}, 2780 
(1981); Y. Shapira, {\it J. Appl. Phys.} {\bf 53}, 1914 (1982).
\bb{Yokoi}C. S. O. Yokoi, M. D. Coutinho-Filho, and S. R. Salinas, 
{\it Phys. Rev. B} {\bf 24}, 5430 (1981); {\bf 29}, 6341 (1984).
\bb{Fra-Hen}S. Redner, and H. E. Stanley, {\it J. Phys. C}{\bf 10}, 4765 
(1977); L. Frachebourg, and M. Henkel, {\it Physica A}{\bf 195}, 577 (1993).
\bb{AH}N. Arkani-Hamed, H. Cheng, M. A. Luty, and S. Mukohyama, 
{\it JHEP} {\bf 0405}, 074 (2004); N. Arkani-Hamed, P. Creminelli, 
S. Mukohyama, and M. Zaldarriaga, {\it JCAP} {\bf 0404}, 001 (2004). 
\bb{Wilson} K. G. Wilson, {\it Rev. Mod. Phys.} {\bf 47}, 773 (1975).
\bb{Lev}A. P. Levanyuk, Soviet Physics-J. Exptl. Theoret. Phys. {\bf 36}, 
571 (1959).
\bb{Ginz}V. L. Ginzburg, Soviet Physics-Solid State {\bf 2}, 1824 (1960).
\bb{Amit1}D. J. Amit, {\it J. Phys. C: Solid State}{\bf 7}, 3369 (1974).
\bb{He}M. Henkel, {\it Phys. Rev. Lett.} {\bf 78}, 1940 (1997).
\bb{Si}W. Siegel {\it in Fields, hep-th /9912205}.
\bb{Tsu-Sa}S. Tsujikawa, M. Sami {\it hep-th/0409212}.
\bb{Amen}L. Amendola, {\it hep-th/0409224}.
\bb{KNg}A. Krause, and S. P. Ng, {\it hep-th/0409241}.
\bb{Ca-Sa}L. Capriotti, and S. Sachdev, {\it cond-mat/0409519}.
\bb{Leite5}M. M. Leite, {\it Phys. Rev. B} {\bf 61}, 14691(2000).
\bb{Leite6}M. M. Leite, {\it Phys. Rev.}{\bf B 68}, 052408 (2003).
\bb{Be}V. Bindilatti, C. C. Becerra, and N. F. Oliveira Jr., {\it Phys. Rev.} 
{\bf B40}, 9412(1989).
\bb{Bre-Zinn} E. Br\'ezin, and J. Zinn-Justin, {\it Nucl. Phys.} {\bf B257}, 
867(1985).
\bb{Nemi} A. M. Nemirovsky, and K. F. Freed,{\it Nucl. Phys.} {\bf B270}, 
423(1986); {\it J. Phys. A: Math. Gen.} {\bf 18}, L319(1985).
\bb{Leite7} M. M. Leite, A. M. Nemirovsky, and M. D. Coutinho-Filho, 
{\it J. Magn. Magn. Mater.} {\bf 104-107}, 181(1992); M. M. Leite, 
M. Sardelich and M. D. Coutinho-Filho, {\it Phys. Rev. E} {\bf 59}, 
2683(1999).
\bb{MP}R. C. Myers, and M. Pospelov, {\it Phys. Rev. Lett.}{\it 90}, 
211601 (2003).
\end{references}
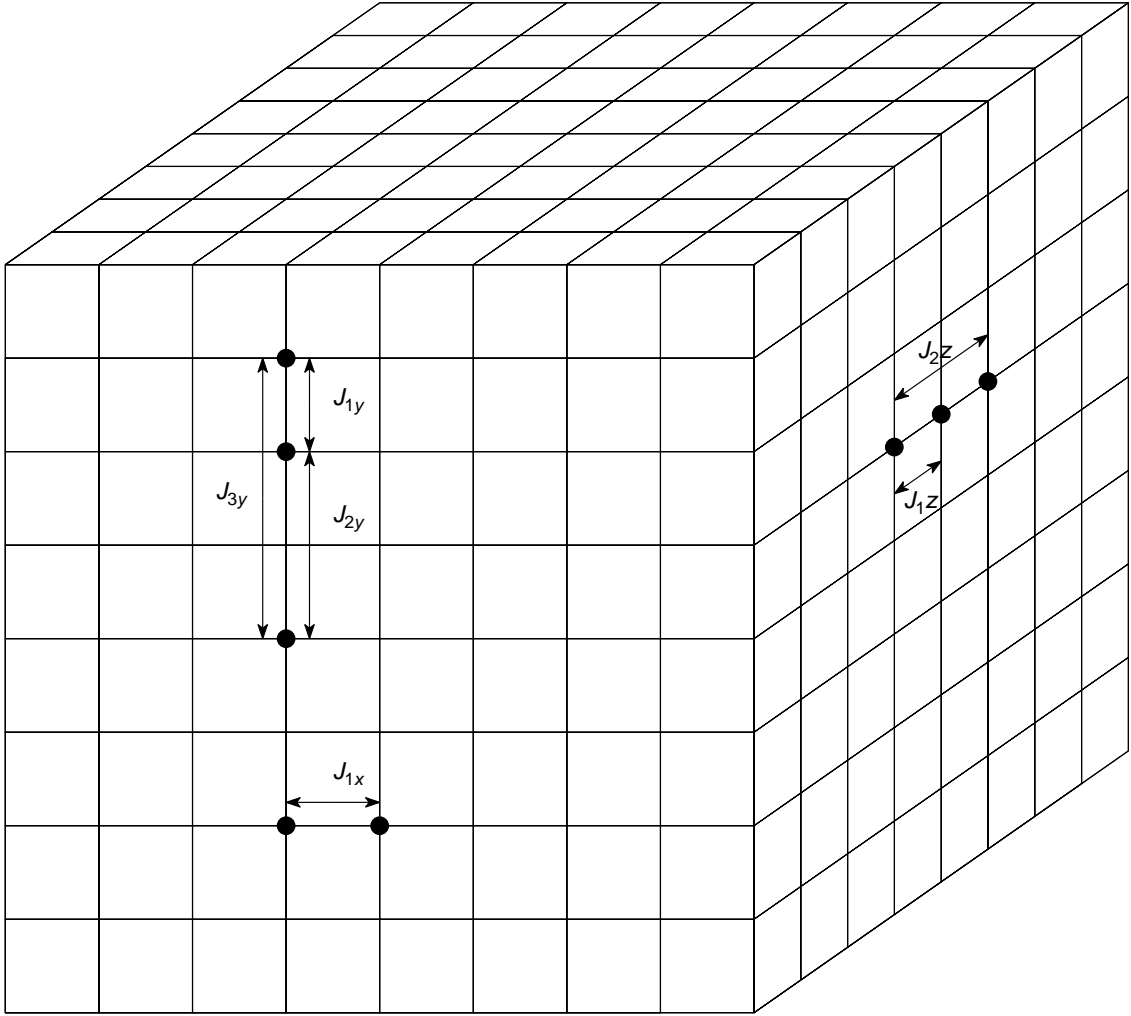
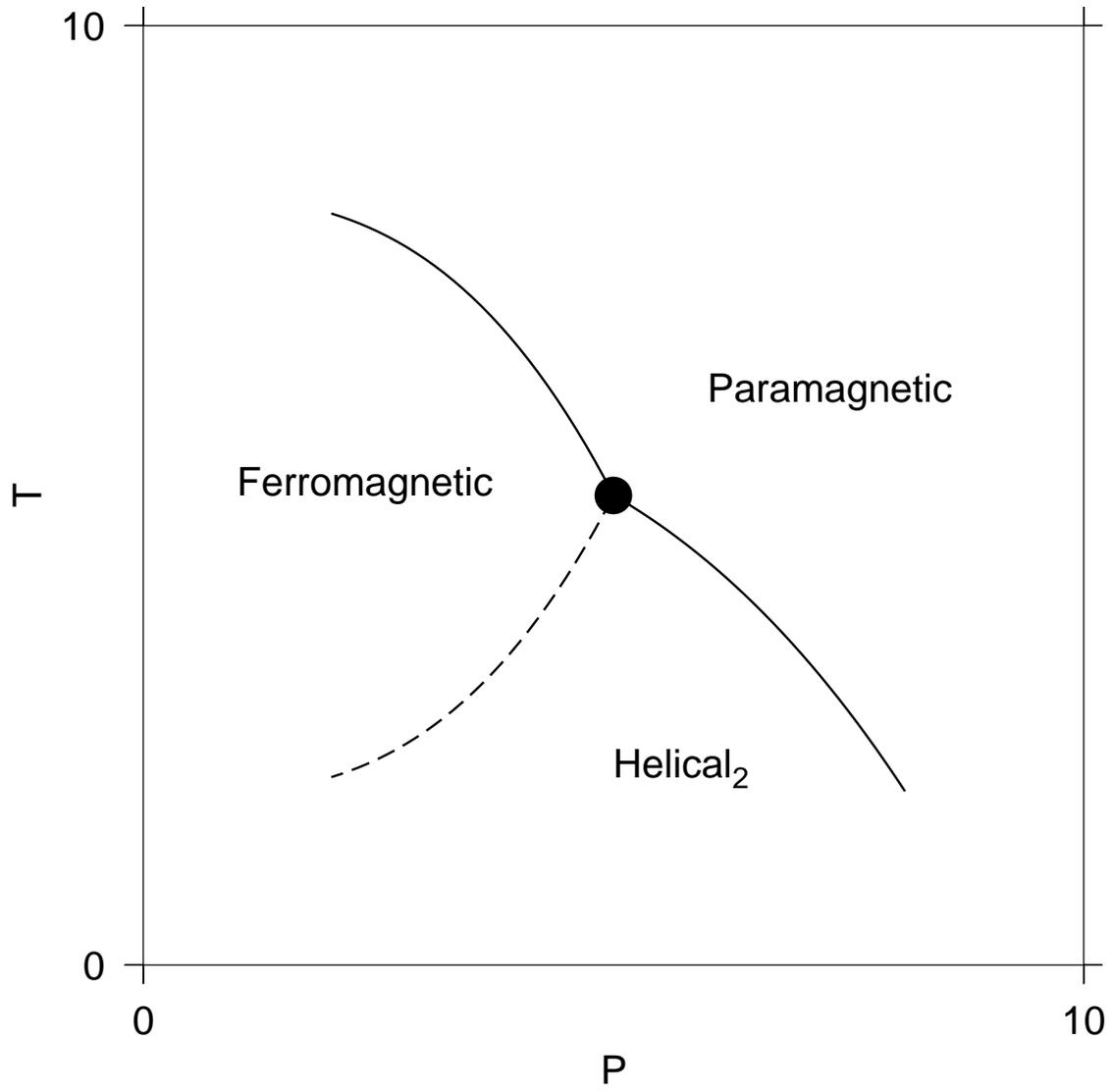
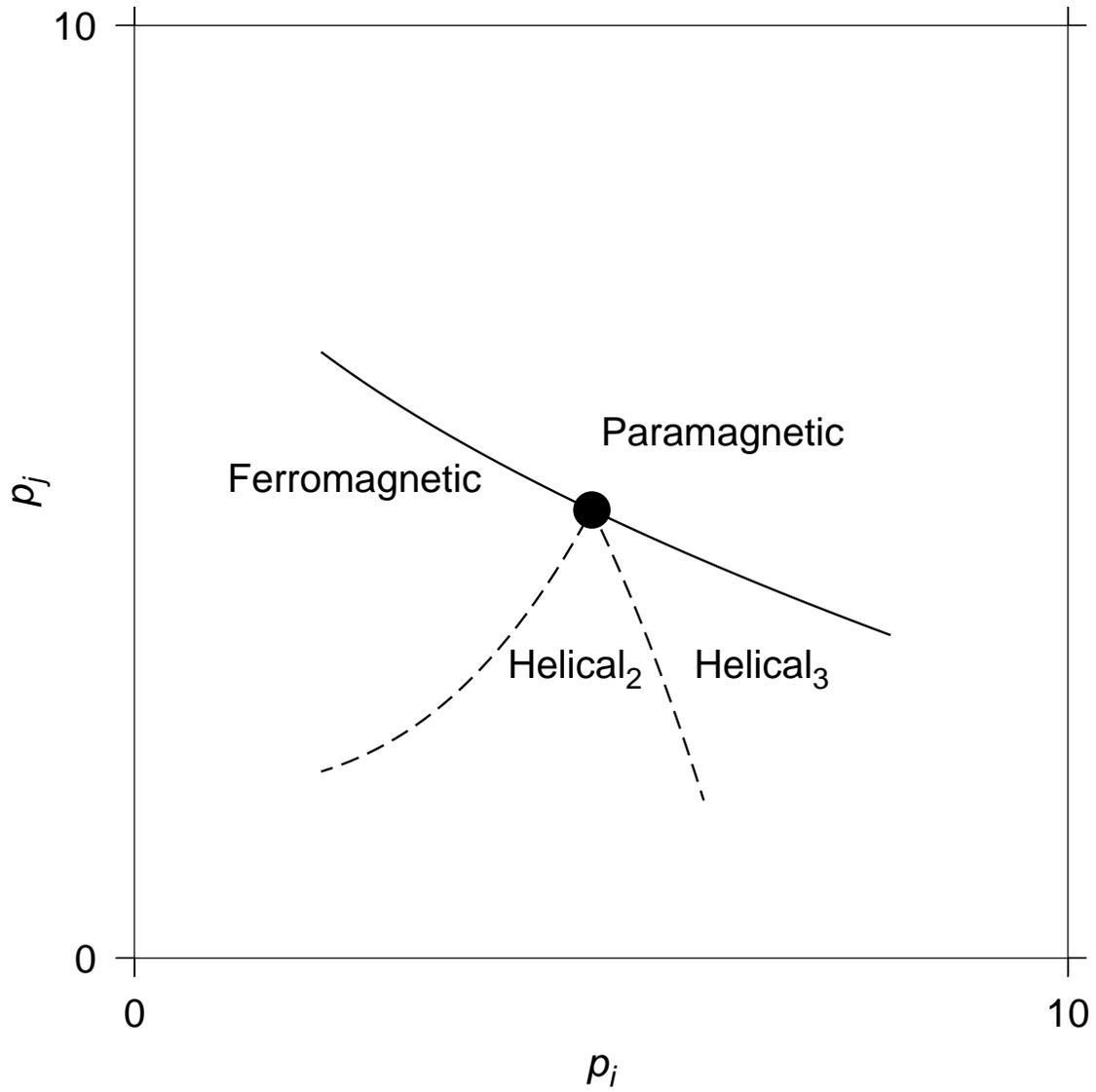
\end{document}